\documentclass[%
 reprint,
superscriptaddress,
 amsmath,amssymb,
 aps,
]{revtex4-2}

\usepackage{graphicx}% Include figure files
\usepackage{dcolumn}% Align table columns on decimal point
\usepackage{bm}% bold math
\newcommand{\eqlab}[1]{\label{eq:#1}}
\renewcommand{\eqref}[1]{Eq.~(\ref{eq:#1})}
\newcommand{\eqsref}[2]{Eqs.~(\ref{eq:#1}) and~(\ref{eq:#2})}

\newcommand{\figref}[1]{Fig.~\ref{fig:#1}}

\newcommand{\figlab}[1]{\label{fig:#1}}

\newcommand{\secref}[1]{Section~\ref{sec:#1}}

\newcommand{\seclab}[1]{\label{sec:#1}}

\newcommand{\appref}[1]{Appendix~\ref{app:#1}}
\newcommand{\appsref}[2]{Appendices~\ref{app:#1} and~\ref{app:#2}}
\newcommand{\applab}[1]{\label{app:#1}}

\newcommand{\equal}{\!=\!}
\newcommand{\minus}{\!-\!}
\newcommand{\plus}{\!+\!}

\newcommand{\ket}[1]{|{#1}\rangle}
\newcommand{\bra}[1]{\langle {#1}|}
\newcommand{\expect}[1]{\big \langle {#1} \big \rangle}
\newcommand{\braket}[2]{\langle #1 | #2 \rangle}

\newcommand{\nn}{\nonumber}
\newcommand{\gamsub}[1]{\gamma_{ #1 }}
\newcommand{\tauin}{\tau_{\!\scriptscriptstyle \mathcal{G}}}
\newcommand{\tauG}{\tau_{\!\scriptscriptstyle \mathcal{G}}}
\newcommand{\OmG}{\Omega_{\scriptscriptstyle \mathcal{G}}}
\newcommand{\tauO}{\tau_{\rm{o}}}
\newcommand{\tauI}{T_{\rm{in}}}
\newcommand{\tauenv}{\tau_{ \rm{e}}}
\newcommand{\Tin}{T_{\rm{in}}}
\newcommand{\Tout}{T_{\rm{out}}}
\newcommand{\Tstore}{T_{\rm{store}}}

\newcommand{\msi}[1]{\mbox{\scriptsize\textit{#1}}}

\begin{document}

\newcommand{\msub}[1]{\mbox{\scriptsize #1}}

\preprint{APS/123-QED}

\title{Photon-Photon Interactions in Dynamically Coupled Cavities}% Force line breaks with \\
% \thanks{Footnote to title of article.}

\author{Mikkel Heuck}
 \email{mrheuck@gmail.com}
% \affiliation{DTU Fotonik, Department of Photonics Engineering, Technical University of Denmark, Building 343, 2800 Kgs. Lyngby, Denmark}%
\affiliation{ Department of Electrical Engineering and Computer Science, Massachusetts Institute of Technology,
77 Massachusetts Avenue, Cambridge, Massachusetts 02139, USA}%

\author{Kurt Jacobs}
\affiliation{U.S. Army Research Laboratory, Computational and Information Sciences Directorate, Adelphi, Maryland 20783, USA}%
\affiliation{Department of Physics, University of Massachusetts at Boston, Boston, MA 02125, USA}%
\affiliation{Hearne Institute for Theoretical Physics, Louisiana State University, Baton Rouge, LA 70803, USA}%

\author{Dirk R. Englund}% 
% \email{englund@mit.edu}
\affiliation{ Department of Electrical Engineering and Computer Science, Massachusetts Institute of Technology,
77 Massachusetts Avenue, Cambridge, Massachusetts 02139, USA}%

\date{\today}% It is always \today, today,

\begin{abstract}
We study theoretically the interaction between two photons in a nonlinear cavity. The photons are loaded into the cavity via a method we propose here, in which the input/output coupling of the cavity is effectively controlled via a tunable coupling to a second cavity mode that is itself strongly output-coupled. Incoming photon wave packets can be loaded into the cavity with high fidelity when the timescale of the control is smaller than the duration of the wave packets. Dynamically coupled cavities can be used to avoid limitations in the photon-photon interaction time set by the delay-bandwidth product of passive cavities. Additionally, they enable the elimination of wave packet distortions caused by dispersive cavity transmission and reflection.
% In addition to providing the ability to load and to output arbitrarily shaped photon wave packets, dynamically coupled cavities can be used to avoid limitations in the photon-photon interaction time set by the delay-bandwidth product of passive resonators. Another important application is the elimination of wave packet distortions caused by dispersive cavity transmission and reflection. %We propose a method to achieve an effective dynamical coupling between a waveguide and an otherwise uncoupled cavity mode by controlling its coupling to an auxiliary cavity mode with a strong coupling to the waveguide. 
We consider three kinds of nonlinearities, those arising from $\chi^{\scriptscriptstyle(2)}$ and  $\chi^{\scriptscriptstyle(3)}$ materials and that due to an interaction with a two-level emitter. To analyze the input and output of few-photon wave packets we use a  Schr\"{o}dinger-picture formalism in which travelling-wave fields are discretized into infinitesimal time-bins.  %We believe that dynamically coupled cavities provides a very beneficial tool to improve the performance of quantum devices relying on cavity-enhanced light-matter interactions such as single-photon sources and atom-like quantum memories with photon interfaces.
We suggest that dynamically coupled cavities provide a very useful tool for improving the performance of quantum devices relying on cavity-enhanced light-matter interactions such as single-photon sources and atom-like quantum memories with photon interfaces. As an example, we present simulation results showing that high fidelity two-qubit entangling gates may be constructed using any of the considered nonlinear interactions.

% \begin{description}
% Secondary publications and information retrieval purposes.
% You may use the \texttt{description} environment to structure your abstract;
% use the optional argument of the \verb+\item+ command to give the category of each item. 
% \end{description}
\end{abstract}

\maketitle

% \tableofcontents

%----------------------------------------------------------------
\section{Introduction} \seclab{introduction}
%----------------------------------------------------------------
Photons make excellent flying qubits due to the low decoherence and loss associated with their transport over standard telecommunication fibers. It therefore seems unavoidable that they will play a key role as carriers of quantum information for secure communication networks and distributed quantum computing~\cite{Kimble2008}. The lack of direct interactions between photons makes it very challenging to perform universal quantum information processing using photonic qubits. Indirect interactions may be mediated by materials with optical nonlinearities but these are usually very weak at optical frequencies. Nevertheless, progress in the design and fabrication of nanocavities with very small mode volumes and very large lifetimes~\cite{Hu2016, Choi2017, Hu2018, Zhang2019, Liang2017} has reduced the optical energy required to observe nonlinear interactions close to the single-photon level. To fully exploit the enhanced light-matter interaction inside the cavity, it is necessary for the entire energy of an incoming wave packet to reside in the cavity throughout its lifetime. However, delay-bandwidth trade-offs~\cite{Lenz2001} put bounds on the energy from an incoming wave packet that can reside inside a passive cavity throughout its lifetime. For instance, a rising exponential wave packet may be absorbed completely into a cavity, but only for an infinitesimal time, such that the average energy is smaller than the total incoming energy. The delay-bandwidth limit may be broken using active controls to modify the cavity-waveguide coupling at a timescale smaller than the wave packet temporal width. Such dynamically coupled cavities have been demonstrated in photonic crystals~\cite{Tanaka2007} and ring resonators~\cite{Xu2007}. These demonstrations used short optical pump pulses to generate electric charge carriers in the semiconductor material forming the cavities. The free carrier absorption loss associated with this method degrades the intrinsic quality factor, $Q_L$, which motivates the search for an alternative approach. 

Here, we propose a method to achieve dynamic coupling that uses the parametric nonlinearity of cavity materials ($\chi^{(2)}$ or $\chi^{(3)}$) and therefore avoids loss. 
\begin{figure}[!h] 
\centering
   \includegraphics[angle=0,origin=c,width=8.2cm] {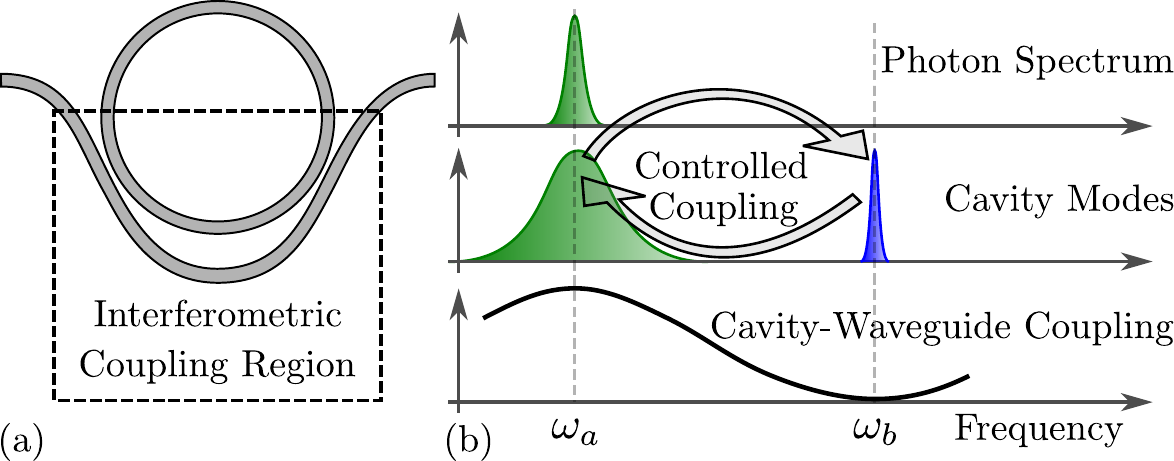}
\caption{(a) Ring resonator interferometrically coupled to a waveguide. (b) Spectra of the incoming photon wave packet (top), cavity resonances coupled via external control fields (center), and cavity-waveguide coupling rate (bottom).}
\figlab{concept figure}
\end{figure} 
Two strong optical control fields may couple two cavity modes via so-called Bragg-scattering four-wave-mixing (FWM) in $\chi^{(3)}$-materials~\cite{McKinstrie2005, Li2016, Heuck2019} and a single control field may do the same in a $\chi^{(2)}$ material~\cite{Guo2016, Zhang2019}, as illustrated with arrows in~\figref{concept figure}b. If the cavity is interferometrically coupled to a waveguide~\cite{Madsen1999} (see~\figref{concept figure}a), one of the cavity modes may be strongly coupled (green mode in~\figref{concept figure}b) whereas the other may be completely decoupled from the waveguide (blue mode in~\figref{concept figure}b). External control over the coupling between the cavity modes therefore introduces a time-dependent effective coupling between the decoupled mode and the waveguide~\cite{Heuck2019}. In other words, photons may be loaded in and out of the decoupled mode via the strongly coupled mode due to their time-dependent mutual coupling.

We succinctly review a Schr\"{o}dinger-picture, discrete-time formalism for treating input and output from optical cavities (equivalent to the well-known Heisenberg-picture  input/output formalism), and show how it can be used to treat the input/output of one- and two-photon wave packets into and out of dynamically coupled nonlinear cavities. %It enables the derivation of input-output relations for continuous-mode photons described by their wave packet amplitudes, $\xi_{\rm{in}}(t)$ and $\xi_{\rm{out}}(t)$. 
We suggest that dynamically coupled cavities would be useful for a range of quantum applications relying on cavity-enhanced light-matter interaction, and specifically use the formalism to calculate the fidelity of two-qubit gates for  travelling-wave photons.\\
% [ref to PRL], but single-qubit operations such as tunable-delay and frequency conversion may also be investigated.\\

% , which is related to overlap integrals between two-time correlation function, $\xi_{\rm{out}}(\tau, t)$ the output
This article is organized as follows:~\secref{discrete time formalism} describes the discrete-time formalism and~\secref{main model} elucidates the Hamiltonians that describe our nonlinear cavity modes.  In \secref{main linear dynamics} we consider the linear regime and examine the dynamics of the cavity modes under the controlled coupling. In~\secref{absorption emission main} we present analytic solutions for the control fields required to absorb and emit wave packets with predefined shapes and consider a specific example in which the wave packets are Gaussian. \secref{nonlinear dynamics main} contains a description of three types of nonlinear interactions, $\chi^{(2)}$, $\chi^{(3)}$, and two-level emitters (TLEs), and considers their application to controlled-phase (c-phase) gates. Finally, we conclude with a discussion of the limitations of our model and suggest other quantum applications that could benefit from dynamically coupled cavities.

%----------------------------------------------------------------

%----------------------------------------------------------------
\section{Discrete-Time Formalism} \seclab{discrete time formalism}
%----------------------------------------------------------------
In our analysis of the dynamics of photons scattering off a system driven by external control fields we discretize the traveling-wave field into time-bins of duration $\Delta t$ as illustrated in~\figref{discrete time illustration}~\cite{Scarani2002, Ciccarello2017, Gross2018}.
\begin{figure}[!h] 
\centering
   \includegraphics[angle=0,origin=c,width=8.2cm] {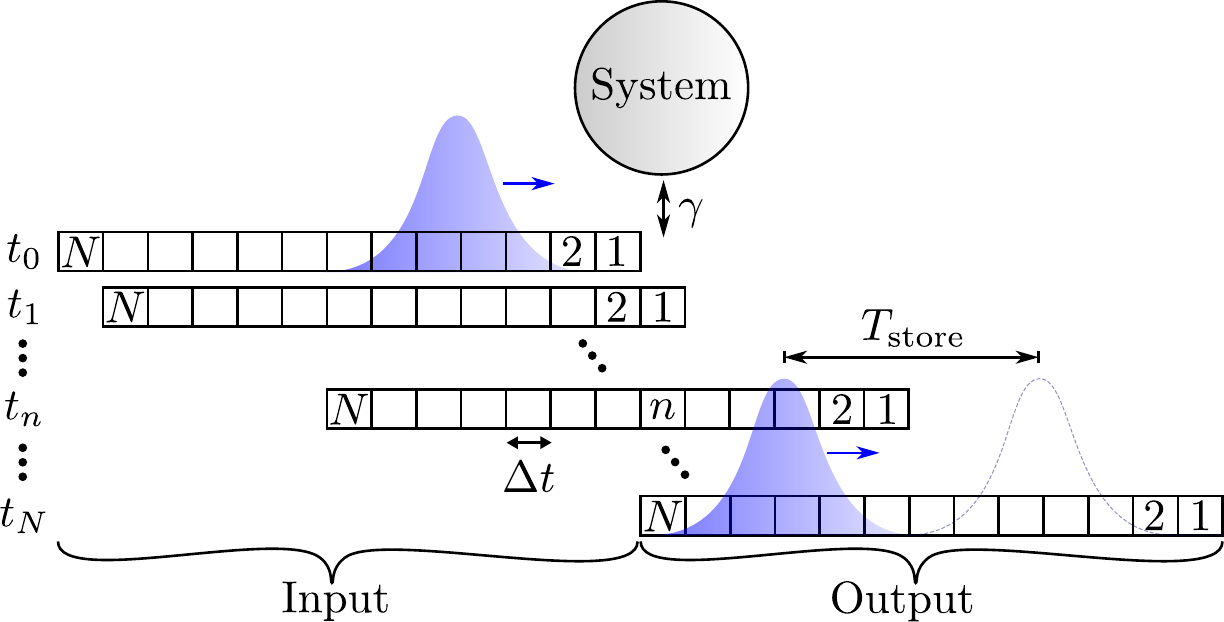}
\caption{Illustration of the discrete time formalism. The time-axis for the travelling-wave field is divided into discrete bins and time evolution is modeled by shifting the time-axis from left to right. The system interacts with one time bin at a time, modelling a  point-interaction with the field as is standard in the input/output formalism for quantum systems.}
\figlab{discrete time illustration}
\end{figure} 
The time-axis may be thought of as a conveyor belt and time evolution corresponds to dragging this conveyor belt past the fixed system one bin at a time. The discretization involves introducing new field operators 
\begin{align}\eqlab{discrete field operator}
   \hat{w}(t_k) = \hat{w}(k\Delta t) \equiv \frac{\hat{w}_k}{\sqrt{\Delta t}}  \;\;\; \mbox{with} \;\;\;  [\hat{w}_j,\hat{w}_k^\dagger] = \delta_{jk} ,  
\end{align}
where $\hat{w}(t_k)$ is the continuous-time annihilation operator that removes a photon from the waveguide at time $t_k$. The operator $\hat{w}_k$ is the discrete-time counterpart of $\hat{w}(t_k)$ that removes a photon from the $k^{\msi{th}}$ time-bin. The factor of $1/\sqrt{\Delta t}$ allows $\hat{w}(t_k)$ to have the canonical commutation relation, $[\hat{w}(t_j), \hat{w}^\dagger(t_k)]\equal \delta(t_j-t_k)$, as $\Delta t \rightarrow 0$. 

For a single-photon input with a wave packet described by $\xi_{\rm{in}}(t)$, the continuous and discrete descriptions are
\begin{align}\eqlab{discrete single photon input state}
    \ket{\psi_{\rm{in}}} = \int_{0}^{T} \!\!dt \xi_{\rm{in}}(t) \hat{w}^\dagger(t) \ket{\emptyset} \approx \sum_{k=1}^N \sqrt{\Delta t} \xi^{\rm{in}}_k  \hat{w}_k^\dagger \ket{\emptyset}, %~\text{with}~ \int_{-\infty}^{\infty} \!\!dt \big|\xi^{\rm{in}}(t)\big|^2 = 1
\end{align}
in which $\int_0^T|\xi_{\rm{in}}(t)|^2 = 1$ so the state is normalized and $\ket{\emptyset}$ denotes the vacuum state of the waveguide. At any time step, $n$ (see~\figref{discrete time illustration}), a photon in bin $k$ is referred to as an \textit{input} photon if $k>n$ and we write the corresponding state of the field as $\hat{w}_k^\dagger \ket{\emptyset}\!\equiv\!\ket{1_k}$. Similarly, if $k\leq n$ the photon is referred to as an \textit{output} photon and we denote the corresponding state of the field by  $\ket{\mathbf{1}_k}$. 

The system depicted in~\figref{discrete time illustration} consists of a nonlinear multimode cavity. We consider up to three cavity modes of which only one will be coupled to the waveguide and another may be coupled to a two-level emitter. Linear coupling between cavity modes will be implemented by nonlinear interactions with  classical control fields. The nonlinear coupling between the photons will originate either from the bulk nonlinearity of the cavity material or an interaction with a TLE. The waveguide-coupled mode is denoted the ``auxiliary'' cavity mode (oscillating at $\omega_a$), and will be used to load and unload photons into and out of the ``primary'' cavity mode (oscillating at $\omega_b$). A third ``tertiary'' cavity mode (oscillating at $\omega_c$), if used, will be coupled to the primary mode and potentially to a TLE. 

We use the Schr\"odinger-picture to derive equations of motion for the time-dependent state coefficients. The unitary time evolution operator describing one time step from $t_{n-1}$ to $t_n$ in~\figref{discrete time illustration} is 
\begin{align}\eqlab{U_n definition}
    \hat{U}_n = \exp \Big( \!-\frac{i}{\hbar} \hat{H}_n \Delta t\Big) = \sum_{m=0}^{\infty} \frac{1}{m!} \Big(\! -\frac{i}{\hbar} \hat{H}_n \Delta t\Big)^{\!m},
\end{align}
such that the updated state is 
\begin{align}\eqlab{time step evolution main}
   \ket{\psi_{n}} = \hat{U}_{n} \ket{\psi_{n-1}},
\end{align}
with $\hat{H}_n$ being the Hamiltonian describing the system and its interaction with the waveguide at time-step $n$. In the next section we explain the model used to describe the system and their interaction with the waveguide and additional loss channels.

% With these four systems the self-energy terms of the Hamiltonian are
% % 
% \begin{align}\eqlab{H0}
%     \hat{H}_0 &= \hbar\omega_a \hat{a}^\dagger\hat{a} +  \hbar\omega_b \hat{b}^\dagger\hat{b}  +  \hbar\omega_c \hat{c}^\dagger\hat{c} + \hbar\omega_e \hat{\sigma}_z,
% \end{align}
% %
% in which $\hat{a}$, $\hat{b}$, and $\hat{c}$ are the annihilation operators for the auxiliary, primary, and tertiary cavity modes, respectively. The operator $\hat{\sigma}_z \!\equiv\! \ket{e}\bra{e}$ where $\ket{e}$ is the excited state of the two-level system. 

% % ------------------------------------------------------------------
% \subsection{Time Evolution}
% % ------------------------------------------------------------------

%------------------------------------------------------------------------------
\section{Model} \seclab{main model}
%------------------------------------------------------------------------------
A model for the complete system consists of a specification of the Hamiltonian in~\eqref{U_n definition}.
It is assumed that the interaction between the system and waveguide occurs at a singular spatial point, which corresponds to interaction only with bin $n$ at time $t_n$. It is therefore convenient to think of $N$ different Hamiltonians,  $\hat{H}_n$, each acting only during the $n^{\msi{th}}$ time step. 

The self-energy terms of the system Hamiltonian in a rotating frame (also know as the interaction picture, see~\appref{rotating frame}) are 
\begin{align}\eqlab{Hamiltonian rotating frame main}
    \hat{H}_{0} = \hbar\delta_a \hat{a}^\dagger\hat{a} +  \hbar\delta_b \hat{b}^\dagger\hat{b}  +  \hbar\delta_c \hat{c}^\dagger\hat{c} + \hbar\delta_e \hat{\sigma}_z,
\end{align} 
where $\hat{a}$, $\hat{b}$, and $\hat{c}$ annihilate, respectively, a photon from the auxiliary cavity mode ($a$), primary mode ($b$), and tertiary mode ($c$). The operator $\hat{\sigma}_z\equal \ket{e}\bra{e}$, with $\ket{e}$ being the excited state of a TLE coupled to mode $c$. The detunings, $\delta_n$, are used to account for discrepancies between energy levels of the system and the incoming photons and control fields as described in~\appref{rotating frame}.
% % 
% \begin{subequations}\eqlab{detunings TLE}
% \begin{align}
%      \text{TLE Material} \qquad \delta_a &\equiv \omega_a - \omega_w \\
%      \delta_b &\equiv \delta_\Lambda + \delta_a \\
%      \delta_c &\equiv \delta_\Pi + \delta_b \\
%      \delta_e &\equiv \omega_e - \omega_c \\
%      \delta_\Lambda &\equiv (\omega_2 - \omega_1) - (\omega_a-\omega_b) \\
%      \delta_\Pi &\equiv (\omega_3 - \omega_1) - (\omega_b-\omega_c).
% \end{align} 
% \end{subequations}
% %

Coupling between the waveguide and the auxiliary cavity mode is described by the Hamiltonian~\cite{Ciccarello2017} %~\cite{Pichler2016}
\begin{align}\eqlab{H cav-wg}
    \hat{H}_n^{\rm{cav-wg}} = i\hbar \sqrt{\frac{\gamma}{\Delta t}} \Big( \hat{a}^\dagger \hat{w}_n - \hat{a}\hat{w}_n^\dagger \Big),
\end{align}
where $\gamma$ is the coupling rate.
% where $\hat{a}$ annihilates a photon from the auxiliary cavity mode, $a$. 
% Note that this is the Hamiltonian in a rotating frame, which is described in~\appref{rotating frame}.
% %-------------------------------------------------------------------------------
% \subsection{Material with Third-order Nonlinearity}
% %-------------------------------------------------------------------------------

As mentioned above, a dynamic cavity-waveguide coupling is established by coupling two cavity modes, one strongly coupled and one decoupled from the waveguide, via nonlinear interactions driven by external control fields. In materials with a third order nonlinearity, $\chi^{(3)}$, the coupling Hamiltonian is 
\begin{align}\eqlab{H cav-cav chi3}
    \hat{H}_n^{\rm{cav-cav}} = \hbar\chi_3 \Big(\hat{p}_1^\dagger\hat{p}_2 \hat{a}^\dagger\hat{b} + \hat{p}_2^\dagger\hat{p}_1 \hat{b}^\dagger\hat{a}\Big),
\end{align}
The operators $\hat{p}_1$ and $\hat{p}_2$ annihilate photons from two pump modes far detuned from modes $a$, $b$, and $c$. The pump fields are treated classically by taking expectation values and making the substitution~\cite{Vernon2016} 
\begin{align}
    \chi_3\langle \hat{p}_2^\dagger \hat{p}_1\rangle \equal \chi_3\alpha_2^*(t_n)\alpha_1(t_n) \equal \Lambda(t_n) ,
\end{align} 
where $\alpha_n$ is the eigenvalue of the annihilation operator $\hat{p}_n$ and $\Lambda(t_n)$ is the complex-valued control field. 
% Here, $\omega_3$ is the frequency difference between the two control fields. The coupling is strongest when $\omega_3$ exactly matches the frequency difference between modes $a$ and $b$. However, in order to study effects of deviations from this perfect energy matching, we keep $\omega_3$ in our analysis. 
With the classical control field, \eqref{H cav-cav chi3} reads 
\begin{align}\eqlab{H cav-cav classical}
    \hat{H}_n^{\rm{cav-cav}} =  \hbar \Big(\Lambda_n^* \hat{a}^\dagger\hat{b} + \Lambda_n \hat{b}^\dagger\hat{a}\Big), 
\end{align}
which now describes a linear coupling between modes $a$ and $b$ driven by the time-dependent control field, $\Lambda(t)$. Note that in the case of a TLE nonlinearity, we introduce a second control field, $\Pi(t)$ that couples modes $b$ and $c$ using pump modes $p_1$ and another mode $p_3$, see~\appref{rotating frame}.  
% That is, once we move to the interaction picture (move into the frame of the self-Hamiltonian), the coupling Hamiltonian becomes 
% % 
% \begin{align}\eqlab{H cav-cav chi3 classical}
%     \hat{H}_n^{\rm{cav-cav, I}} =  \hbar \Big(\Lambda_n^* e^{i\Delta_3 t_n} \hat{a}^\dagger\hat{b} + \Lambda_n e^{-i\Delta_3 t_n}\hat{b}^\dagger\hat{a}\Big), 
% \end{align}
% % 
% in which 
% % 
% \begin{align}
%     \Delta_3 = (\omega_a - \omega_b) - \omega_3 . 
% \end{align} 
% % 
% Alternatively we may use a slightly different interaction picture in which we move into a frame that removes the time-dependent oscillation from the coupling Hamiltonian. For example, if we move into the frame defined by 
% % 
% \begin{align}
%     H_{\msub{f}} = \hbar \omega_a \hat{a}^\dagger \hat{a} + \hbar (\omega_a - \omega_3) \hat{b}^\dagger \hat{b} 
% \end{align} 
% % 
% the full Hamiltonian in the interaction picture is 
% % 
% \begin{align}
%     H_{\msub{I}} = -\hbar\omega_3 \hat{b}^\dagger \hat{b} + \hbar \Big(\Lambda_n^* \hat{a}^\dagger\hat{b} + \Lambda_n \hat{b}^\dagger\hat{a}\Big). 
% \end{align} 
% % 
% Throughout our analysis we will always use an interaction picture in which the coupling terms in the Hamiltonian are time-independent [@Kurt: the coupling terms are still time-dependent, they just don't have the additional time-dependent phase-terms], so that any effective detunings between the cavities that exist after the frequencies of the pumps are included are accounted for by self terms involving frequency differences.  

For $\chi^{\scriptscriptstyle(3)}$ materials, we must also include the cross-phase modulation caused by the pump fields on modes $a$, $b$, and $c$ described by the Hamiltonian
\begin{multline}\eqlab{H xpm pump}
    \hat{H}_n^{X\!P\!M\!,p} = \hbar\chi_3 \sum_{m\equal 1}^2 \hat{p}_m^\dagger\hat{p}_m \Big(\hat{a}^\dagger\hat{a} + \hat{b}^\dagger\hat{b} +\hat{c}^\dagger\hat{c}  \Big)~\rightarrow\\
    2\hbar|\Lambda_n|\Big(\hat{a}^\dagger\hat{a} + \hat{b}^\dagger\hat{b} +\hat{c}^\dagger\hat{c} \Big),
\end{multline}
where we have assumed $\chi_3\langle \hat{p}_2^\dagger \hat{p}_2\rangle \equal \chi_3\langle \hat{p}_1^\dagger \hat{p}_1 \rangle \equal |\Lambda_n|$, which means that the optical energy in each pump mode is identical at all times. \\
% The part of the Hamiltonian that is bi-linear in $\hat{a}$ and $\hat{b}$ is therefore
% % 
% \begin{multline}\eqlab{H lin chi3}
%   \hat{H}_n^{\rm{lin},\chi^{\scriptscriptstyle(3)}} = i\hbar \sqrt{\frac{\gamma}{\Delta t}} \Big( \hat{a}^\dagger \hat{w}_n \minus \hat{a}\hat{w}_n^\dagger \Big) ~+\\
%   \hbar \Big(\Lambda_n^* \hat{a}^\dagger\hat{b} + \Lambda_n\hat{b}^\dagger\hat{a}\Big) + 2\hbar|\Lambda_n|\Big(\hat{b}^\dagger\hat{b} + \hat{a}^\dagger\hat{a} \Big).
% \end{multline}
% % 
% % ------------------------------------------------------------------
% \subsection{Material with Second-order Nonlinearity}
% % ------------------------------------------------------------------

In a $\chi^{\scriptscriptstyle(2)}$ material the cavity-cavity coupling arises from the Hamiltonian
\begin{align}\eqlab{H cav-cav chi2}
    \hat{H}_n^{\rm{cav-cav}} = \hbar\chi_2 \Big(\hat{p}^\dagger \hat{a}^\dagger\hat{b} + \hat{p}\hat{b}^\dagger\hat{a}\Big).
\end{align}
We assume the frequency separation between modes $a$ and $b$ to be in the GHz range and $\hat{p}$ is therefore the annihilation operator of a radio-frequency (RF) electric field that may be applied using electrodes~\cite{Zhang2019}.
% with a carrier frequency $\omega_p\equal \omega_b-\omega_a$. 
Again, we describe it classically by
\begin{align}
    \chi_2 \langle \hat{p} \rangle \equal \chi_2 \alpha_p(t_n) \equal \Lambda_n. 
\end{align}
The coupling Hamiltonian expressed in terms of the classical control field is therefore given by~\eqref{H cav-cav classical} for both second- and third-order nonlinear materials. There is no cross-phase modulation term in the Hamiltonian for a $\chi^{\scriptscriptstyle(2)}$ material (unless a DC electric field is applied), so~\eqref{H xpm pump} does not apply in that case.\\
% % 
% \begin{multline}\eqlab{H lin chi2}
%   \hat{H}_n^{\rm{lin},\chi^{\scriptscriptstyle(2)}} = i\hbar \sqrt{\frac{\gamma}{\Delta t}} \Big( \hat{a}^\dagger \hat{w}_n \minus \hat{a}\hat{w}_n^\dagger \Big) ~+\\
%   \hbar\Big(\Lambda_n^* \hat{a}^\dagger\hat{b} + \Lambda_n \hat{b}^\dagger\hat{a}\Big) .
% \end{multline}
% %

The Hamiltonian describing the three different types of nonlinear materials  are
\begin{subequations}\eqlab{H nonlinear}
\begin{align}
    \hat{H}_{\rm{XPM}} \plus \hat{H}_{\rm{SPM}} &= \hbar\chi_3 \Big[\hat{a}^\dagger\hat{a}\hat{b}^\dagger\hat{b}  +\hat{b}^\dagger\hat{b} \hat{c}^\dagger\hat{c}\Big]~+ \nn\\
    &\hspace{2.2cm} \frac{\hbar\chi_3}{4} \!\sum_{\hat{q}}\!\Big(\! \hat{q}^\dagger\hat{q}\minus 1\!\Big)\hat{q}^\dagger\hat{q} \eqlab{H nonlinear chi3 a}\\
    \hat{H}_{\rm{SHG}}   &= \hbar\chi_2 \Big( \hat{c}\hat{b}^\dagger\hat{b}^\dagger + \hat{c}^\dagger\hat{b}\hat{b}\Big) \eqlab{H nonlinear chi2 a}\\
    \hat{H}_{\rm{TLE}}   &=\hbar  \Big( g\hat{c}^\dagger \hat{\sigma}_{\!-} + g^*\hat{c}\hat{\sigma}_{\!+} \Big) \eqlab{H nonlinear TLE},
\end{align}
\end{subequations}
where $\hat{q}\!\in\!\{\hat{a},\hat{b},\hat{c}\}$ in~\eqref{H nonlinear chi3 a} and $\hat{\sigma}_{\!-} \!\equiv\! \ket{g}\bra{e}$ and $\hat{\sigma}_{\!+} \!\equiv\! \ket{e}\bra{g}$ in~\eqref{H nonlinear TLE} with $\ket{g}$ being the ground state and $\ket{e}$ the excited state of the TLE. Note that not all possible combinations of modes are considered in~\eqref{H nonlinear}, but only those included in the protocols for photon-photon interactions that we consider here.

% We assume the existence of a third cavity mode, $c$, in~\eqsref{H nonlinear chi2 a}{H nonlinear TLE} that couples to $b$ via second-harmonic-generation (SHG) in~\eqref{H nonlinear chi2 a} and to the two-level emitter in~\eqref{H nonlinear TLE}.

%------------------------------------------------------------------------------
\section{Linear Dynamics} \seclab{main linear dynamics}
%------------------------------------------------------------------------------
In this section we derive equations of motion including only the linear dynamics. We start with the simplest case of one photon coupling to one cavity mode to built intuition about the derivation procedure. Then, we consider one photon coupling to a cavity with two modes, and finally two photons coupling to a cavity with two modes.
Having derived equations of motion in the linear regime, it is fairly straight forward to add nonlinear interactions and make the appropriate additions to the equations, which we do in~\secref{nonlinear dynamics main}.

% The unitary time evolution operator for linear dynamics in a $\chi^{(3)}$ material is
% % 
% \begin{multline}
%     \hat{U}_n^{\rm{lin},\chi^{\scriptscriptstyle \!(3)}} \!\!= \hat{\mathbb{I}} + \sqrt{\gamma \Delta t} \Big(\hat{a}^\dagger \hat{w}_n \minus \hat{a} \hat{w}_n^\dagger  \Big) - \frac{\gamma}{2}\Delta t \hat{a}^\dagger \hat{a} \hat{w}_n\hat{w}_n^\dagger  ~-\\
%     i\Delta t\Big(\Lambda_n^* \hat{a}^\dagger\hat{b} + \Lambda_n\hat{b}^\dagger\hat{a}\Big) -i2 |\Lambda_n|\Delta t \Big(\hat{b}^\dagger\hat{b} \plus \hat{a}^\dagger\hat{a} \Big),
% \end{multline} 
% % 
% where the last term would be omitted for a $\chi^{(2)}$ material.

%------------------------------------------------------------------------------
\subsection{One Cavity Mode - One Photon\seclab{one mode one photon}}
%------------------------------------------------------------------------------
Let us begin by considering a single input photon coupling to one cavity mode. The relevant terms of the Hamiltonian are
\begin{align}\eqlab{Hn one mode}
    \hat{H}_n^{(1)} =  \hbar\delta_a\hat{a}^\dagger\hat{a} + i\hbar \sqrt{\frac{\gamma}{\Delta t}} \Big( \hat{a}^\dagger \hat{w}_n - \hat{a}\hat{w}_n^\dagger \Big).
\end{align} 
Keeping only terms to first order in $\Delta t$, the corresponding time-evolution operator is
\begin{multline}\eqlab{Un one mode}
    \hat{U}_n^{(1)} \approx \hat{\mathbb{I}} + \sqrt{\gamma \Delta t} \Big(\hat{a}^\dagger \hat{w}_n - \hat{a} \hat{w}_n^\dagger  \Big) ~-\\
    \frac{\gamma}{2}\Delta t \hat{a}^\dagger \hat{a} \hat{w}_n\hat{w}_n^\dagger -i \delta_a \Delta t \hat{a}^\dagger\hat{a}. 
\end{multline} 
The state at time step $n$ is 
\begin{multline}\eqlab{psi_n one mode one photon main}
    \ket{\psi_n}=  \!\!\sum_{k=n+1}^N \!\!\xi^{\rm{in}}_k  \sqrt{\Delta t} \ket{0} \ket{1_k} ~+\\ 
    \sum_{k=1}^n \xi^{\rm{out}}_k  \sqrt{\Delta t} \ket{0} \ket{\mathbf{1}_k} + \psi_1(n) \ket{1} \ket{\emptyset}, 
\end{multline} 
where $\xi^{\rm{in}}_k\equal \xi_{\rm{in}}(t_k)$ describes the input wave packet. The states $\ket{0} \ket{1_k}$ and $\ket{0} \ket{\mathbf{1}_k}$ correspond to an empty cavity and a photon in bin $k$ on the input ($k> n$) and output ($k\leq n$) side, respectively. The state corresponding to a photon in the cavity has the coefficient $\psi_1(n)$. In~\appref{one photon dynamics one cavity mode} we derive the equation of motion for $\psi_1(t)$ and the input-output relation connecting $\xi_{\rm{out}}(t)$ to $\xi_{\rm{in}}(t)$
\begin{subequations}\eqlab{eom one mode one photon main}
\begin{align}
   \dot{\psi}_1(t) &= -\Big(i\delta_a+\frac{\gamma}{2}\Big)\psi_1(t) + \sqrt{\gamma}\xi_{\rm{in}}(t) \eqlab{eom one mode one photon main psi_1}\\
   \xi_{\rm{out}}(t)    &= \xi_{\rm{in}}(t) - \sqrt{\gamma} \psi_1(t) .
\end{align}
\end{subequations}
These equations have the same form as those derived classically using arguments of energy conservation and time-reversal symmetry~\cite{HermanHaus_book}. They also have the same form as the Heisenberg equations of motion of the usual input-output formalism~\cite{Jacobs14}. 

%------------------------------------------------------------------------------
\subsection{Loss\seclab{loss}}
%------------------------------------------------------------------------------ 
At this stage we consider the effect of loss. It may be conveniently modeled using an additional waveguide with a vacuum input. If the annihilation operator that removes a photon from the loss channel at time $t_n$ is $\hat{l}_n$, then the time-evolution operator has the additional terms
\begin{align}\eqlab{U_n loss}
   \hat{U}_n^{\rm{loss}} = \sum_{\hat{q}}\Big[ \sqrt{\gamma_L\Delta t} \big(\hat{q}^\dagger \hat{l}_n - \hat{q}\hat{l}_n^\dagger \big)    - \frac{\gamma_L}{2}\Delta t\hat{q}^\dagger \hat{q} \hat{l}_n\hat{l}_n^\dagger \Big],
\end{align}
where $\hat{q}$ represents all the cavity modes (we assume they have identical loss rates, $\gamma_L$). If we ignore all states of the loss channel except the vacuum,~\eqref{U_n loss} shows that a term, $-m\gamma_L/2$, is added to all loss terms (with $m$ photons in the cavity mode), such that the loss term in~\eqref{eom one mode one photon main psi_1} would have the coefficient $-(\gamma\plus\gamma_L)/2$. We therefore define the total coupling rate, $\Gamma\equal \gamma\plus\gamma_L$. Noise photons injected into the system from the loss channel due to vacuum fluctuations at finite temperatures is neglected in this treatment. 

Ignoring all states in the loss channel except the vacuum, $\ket{\emptyset}_{_{\!L}}$, our total state is $\ket{\psi}\ket{\emptyset}_{_{\!L}}$. It will not be normalized due to the finite probability of finding photons in the loss channel. We may, however, consider a heralded state, $\ket{\psi_M}$, corresponding to a measurement revealing that the loss channel was, in fact, in the state $\ket{\emptyset}_{_{\!L}}$
% The output states will not be normalized because they only include the first term of the total state 
% % 
% \begin{align}\eqlab{U_n loss}
%   \ket{\Psi} =  \ket{\psi}\otimes \big( \ket{\emptyset}_{_{\!L}} + \ket{\phi}_{_{\!L}}\big),
% \end{align}
% % 
% where the loss channel is in the vacuum state ($\ket{\phi}_{_{\!L}}$ is the part of the loss channel containing photons). If we project the total state onto the subspace with no lost photons, we have 
% 
\begin{align}\eqlab{heralded no loss}
   \ket{\psi_M} =  \frac{\big(\ket{\emptyset}_{_{\!L}}{}_{_{L}}\!\bra{\emptyset}\big) \ket{\Psi}}{\sqrt{\expect{\Psi | \big(\ket{\emptyset}_{_{\!L}}{}_{_{L}}\!\bra{\emptyset}\big) | \Psi }}} = \frac{\ket{\psi}}{\sqrt{1-P_L}},
\end{align}
where $\ket{\Psi}\equal \ket{\psi}\ket{\vartheta}_{_{\!L}}$ is the normalized full state, $\ket{\vartheta}_{_{\!L}}$ is the state of the loss channel, and the probability of losing at least one photon is $P_L \equal 1\minus\big|\expect{\psi|\psi}\big|^2$. The overlap between the output state and some desired state $\ket{\Phi}$ is often used as a metric for the precision with which systems are able to implement desired quantum state transformations. Here, we can define 
\begin{align}\eqlab{complex overlaps Fidelity}
    \braket{\Psi_{\rm{out}}^{(n)}}{\Phi} = \sqrt{F_n} e^{i\theta_n},
\end{align}
where $F_n$ is the state fidelity and $\theta_n$ the phase of the overlap with $n\equal\{1,2\}$ input photons. With the definition of states as superpositions over temporal modes in~\eqref{discrete single photon input state} the overlaps in~\eqref{complex overlaps Fidelity} are
% For applications such as two-qubit logic gates
% If we wish to compare the output state with some desired state $\ket{\Phi}$, we may use the state fidelity as a measure of how close these states are\\
% The important quantities to consider for controlled-phase gates are overlaps of the output states with time-translated versions of the input states
% where $n\equal\{1,2\}$ is the number of input photons and 
% 
\begin{align}\eqlab{complex overlaps xi}
    \braket{\Psi_{\rm{out}}^{(1)}}{ \Phi} &= \int_0^T \!\!\!  \xi_{\rm{out}}(t) \xi_{\Phi} (t)^* dt \\
    \braket{\Psi_{\rm{out}}^{(2)}}{ \Phi} &= \int_0^T \!\!\! \int_0^T \!\!\!  \xi_{\rm{out}}(t_1, t_2) \xi_{\Phi}(t_1)^{\!*} \xi_{\Phi}(t_2)^{\!*}dt_1 dt_2 ,
\end{align}
where we assumed that the desired state for two-photon inputs is a separable state with the same superposition over temporal modes for both photons. Note that, for a single-photon input, 
\begin{align}\eqlab{PL as overlap}
    1 - P_L = \braket{\psi^{(1)}_{\rm{out}}}{\psi^{(1)}_{\rm{out}}} =  \int_0^T \!\!\! | \xi_{\rm{out}}(t)|^2 dt,
\end{align}
which illustrates that when the total state, $\ket{\psi}\ket{\emptyset}_{_{\!L}}$ is not normalized, it means that the integral over $|\xi_{\rm{out}}|^2$ is smaller than one.

% % 
% \begin{align}\eqlab{state fidelity def}
% 	F_s\big(\ket{\Psi}, \ket{\Phi}\big) = \big | \expect{\Psi | \Phi }\big|^2 .
% \end{align}
% % 
If we are only interested in states without lost photons, $\ket{\Phi}\equal \ket{\phi}\ket{\emptyset}_{_{\!L}}$, the fidelity may be written as
\begin{align}\eqlab{state fidelity}
	F_n = \big |  {}_{_{L}}\!\bra{\emptyset} \braket{\psi_{\rm{out}}^{(n)}}{\phi}\ket{\emptyset}_{_{\!L}} \rangle \big|^2 \!= (1\minus P_L) \big | \langle \psi^{(n)}_{{\rm{out}},M} | \phi\rangle \!\big|^2.
\end{align}
\eqref{state fidelity} allows us to define a conditional state fidelity 
\begin{align}\eqlab{conditional state fidelity definition}
	\overline{F}_{n} = \frac{1}{1-P_L}F_n,
\end{align}
which separates the infidelity due to loss from that originating from other sources. This becomes useful later, when we show that dynamically controlled cavities may emit photons into wave packets with a desired shape by reducing the overall emission probability. 

In the following sections, we include the loss term proportional to $\gamma_L$ in all the equations of motion.

%------------------------------------------------------------------------------
\subsection{Two Cavity Modes - One Photon\seclab{two modes one photon}}
%------------------------------------------------------------------------------ 
% 
For two cavity modes and a $\chi^{(3)}$ material, the Hamiltonian describing the linear dynamics is
\begin{multline}\eqlab{Hn two mode}
    \hat{H}_n^{(2)} =  \hbar\delta_a\hat{a}^\dagger\hat{a} \plus \hbar\delta_b\hat{b}^\dagger\hat{b} \plus i\hbar \sqrt{\frac{\gamma}{\Delta t}} \Big( \hat{a}^\dagger \hat{w}_n - \hat{a}\hat{w}_n^\dagger \Big) ~+\\
    \hbar \Big(\Lambda_n^* \hat{a}^\dagger\hat{b} + \Lambda_n \hat{b}^\dagger\hat{a}\Big) + 2\hbar|\Lambda_n|\Big(\hat{a}^\dagger\hat{a} + \hat{b}^\dagger\hat{b} \Big).
\end{multline} 
The corresponding time-evolution operator is
\begin{multline}\eqlab{Un two modes}
    \hat{U}_n^{(2)} \approx \hat{\mathbb{I}} + \sqrt{\gamma \Delta t} \Big(\hat{a}^\dagger \hat{w}_n - \hat{a} \hat{w}_n^\dagger  \Big)  ~-\\
    \frac{\gamma}{2}\Delta t \hat{a}^\dagger \hat{a} \hat{w}_n\hat{w}_n^\dagger - i\Delta t \Big(\Lambda_n^* \hat{a}^\dagger\hat{b} + \Lambda_n\hat{b}^\dagger\hat{a}\Big) ~-\\
    i\Delta t\big(\delta_a+2 |\Lambda_n|\big)\hat{a}^\dagger\hat{a} -  i\Delta t\big(\delta_b+2 |\Lambda_n|\big)\hat{b}^\dagger\hat{b}. 
\end{multline} 
Note that we have omitted the loss terms from~\eqref{U_n loss}, but we will include them in the equations of motion below. The state at time step $n$ is
\begin{multline}\eqlab{psi_n two modes one photon main} 
    \ket{\psi_n}=  \!\!\sum_{k=n+1}^N \!\!\xi^{\rm{in}}_k  \sqrt{\Delta t} \ket{00} \ket{1_k} + \sum_{k=1}^n \xi^{\rm{out}}_k  \sqrt{\Delta t} \ket{00} \ket{\mathbf{1}_k} ~+\\ 
     + \psi_{10}(n) \ket{10} \ket{\emptyset} + \psi_{01}(n) \ket{01} \ket{\emptyset}, 
\end{multline} 
where $\ket{01}\!\equiv\! \ket{0_a}\ket{1_b}$ is the state with one photon in mode $b$. In~\appref{one photon dynamics two cavity modes} we derive the equations of motion for the coefficients $\psi_{10}(t)$ and $\psi_{01}(t)$ along with the input-output relation
\begin{subequations}\eqlab{eoms two modes one photon main}
\begin{align}
    \dot{\psi}_{10}  &=  -\Big(i\delta_a +\frac{\Gamma}{2} + i2|\Lambda| \Big)\psi_{10} - i\Lambda^*\psi_{01} + \sqrt{\gamma}\xi_{\rm{in}} \eqlab{eoms two modes one photon ii 10 main}\\
    \dot{\psi}_{01}  &= -\Big(i\delta_b + \frac{\gamma_L}{2} + i2|\Lambda|\Big)\psi_{01}  -i\Lambda\psi_{10} \eqlab{eoms two modes one photon ii 01 main}\\
     \xi_{\rm{out}}  &= \xi_{\rm{in}} - \sqrt{\gamma}\psi_{10}.
\end{align}
\end{subequations}
Note that we have not explicitly written the time dependence of the functions in~\eqref{eoms two modes one photon main}.

%------------------------------------------------------------------------------
\subsection{Two Cavity Modes - Two Identical Photons\seclab{two modes two photons}}
%------------------------------------------------------------------------------ 
The analysis becomes significantly more complicated for two input photons so we find it beneficial to map out all the different paths they may take from input to output and the different types of states generated in the process, see~\figref{two photons two modes paths}. 
% The state at time step $n$ is  
% % 
% \begin{multline}\eqlab{psi_n two modes two photons main} 
%     \ket{\psi_n}=  \!\!\sqrt{2}\sum_{j>n}^N\sum_{k>j}^N \xi^{\rm{in}}_j\xi^{\rm{in}}_k  \Delta t \ket{00} \ket{1_j 1_k} ~+\\
%     \psi_{10}^{\rm{ii}}(n) \sum_{k>n}^N \xi^{\rm{in}}_k \sqrt{\Delta t} \ket{10}\ket{ 1_k} ~+\\
%     \psi_{01}^{\rm{ii}}(n) \sum_{k>n}^N \xi^{\rm{in}}_k \sqrt{\Delta t} \ket{01}\ket{ 1_k}
% \end{multline} 
% %
%  The time evolution operator therefore given by~\eqref{Un two modes} 
% 
\begin{figure}[htb] 
\centering
   \includegraphics[origin=c,width=8cm] {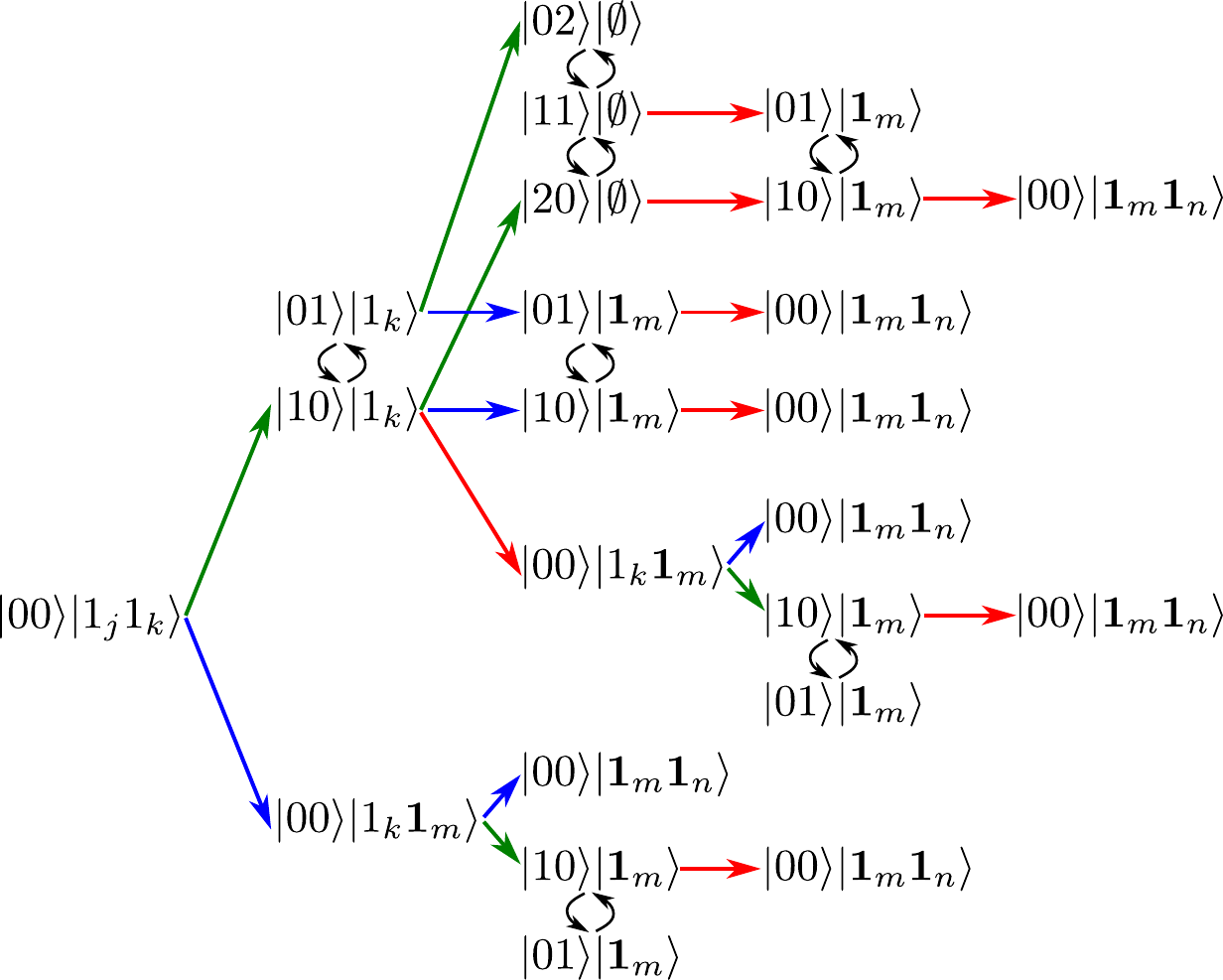}
\caption{Map of states generated with two cavity modes and two input photons and paths from input to output. Green arrows represent absorption of a photon into mode $a$. Red arrows represent emission into the waveguide in bin $m$. Blue arrows represent a photon passing by the system without interacting in time bin $m$. Black arrows indicate the interaction between modes $a$ and $b$ driven by the external control fields. There are five vertical layers going from left to right.}
\figlab{two photons two modes paths}
\end{figure} 
Let us go through the layers of the map from left to right and write down the dynamical equations governing the expansion coefficients of the states in each layer. The first layer only contains the input state
\begin{align}\eqlab{psi_0 two modes two photons main}
    \ket{\psi_0}=  \sqrt{2}\sum_{j=1}^N\sum_{k>j}^N \xi^{\rm{in}}_j \xi^{\rm{in}}_k  \Delta t \ket{00} \ket{ 1_j 1_k}. 
\end{align} 
Note that the summation over $k$ starts at $j$ in~\eqref{psi_0 two modes two photons main} so that the indistinguishable states $\ket{1_j 1_k}$ and $\ket{1_k 1_j}$ are only counted once in the summations. In~\appref{two photon dynamics two cavity modes} we prove that the factor of $\sqrt{2}$ ensures that the state is normalized when the integral of $|\xi_{\rm{in}}(t)|^2$ equals 1. We note that derivations of all the equations of motion for coefficients of the Schr\"{o}dinger picture state in this section may be found in~\appref{two photon dynamics two cavity modes}. 

One of the two photons in layer 1 may be absorbed giving rise to states in layer 2 with one photon in mode $a$ or $b$. The dynamical equations for the coefficients corresponding to these states are 
\begin{subequations}\eqlab{eoms two modes two photons ii main} 
\begin{align}
    \dot{\psi}_{10}^{\rm{ii}} &= -\Big(i\delta_a\plus \frac{\Gamma}{2} \plus i2|\Lambda|\Big)\psi_{10}^{\rm{ii}} - i\Lambda^{\!*}\psi_{01}^{\rm{ii}} + \sqrt{2\gamma}\xi_{\rm{in}} \eqlab{eoms two modes two photons ii 10 main} \\
    \dot{\psi}_{01}^{\rm{ii}} &= -\Big(i\delta_b\plus\frac{\gamma_L}{2}\plus i2|\Lambda|\Big)\psi_{01}^{\rm{ii}} \minus i\Lambda\psi_{10}^{\rm{ii}} \eqlab{eoms two modes two photons ii 01 main},
\end{align} 
\end{subequations}
where we use the superscript $^{\rm{ii}}$ to signify that the driving term in~\eqref{eoms two modes two photons ii 10 main} originates from two input photons, which is why it contains a factor of $\sqrt{2}$ relative to~\eqref{eoms two modes one photon ii 10 main}. A convenient feature of the map in~\figref{two photons two modes paths} is that the couplings represented by black arrows turn up in the equations of motion as coupling terms proportional to the control field, $\Lambda(t)$, and therefore serves to check whether all the dynamics is included. 

The state $\ket{00}\ket{1_k\mathbf{1}_m}$ in layer 2 originates from direct passage of one of the input photons, while in layer 3 it originates from absorption and subsequent emission. If the photon remaining on the input side is later absorbed, it gives rise to states $\ket{10}\ket{\mathbf{1}_m}$ and $\ket{01}\ket{\mathbf{1}_m}$ in layer 3 or 4. The dynamical equations for the coefficients corresponding to these states are
\begin{subequations}\eqlab{eoms two modes two photons i main}
\begin{align}
    \dot{\psi}_{10}^{\rm{i}}(\tau, t)  &\!=\!  -\Big(i\delta_a\plus\frac{\Gamma}{2}\plus i2|\Lambda(t)| \Big)\psi_{10}^{\rm{i}}(\tau, t) ~- \nn\\
    &\hspace{2.6cm} i\Lambda^*\psi_{01}^{\rm{i}}(\tau, t) + \sqrt{\gamma}\xi_{\rm{in}}(t) \eqlab{eoms two modes two photons i 10 main}\\
    \dot{\psi}_{01}^{\rm{i}}(\tau, t)  &\!=\! -\Big(i\delta_b\plus\frac{\gamma_L}{2} \plus i2|\Lambda(t)|\Big) \psi_{01}(\tau, t) ~- \nn\\
    &\hspace{4.4cm}  i\Lambda\psi_{10}^{\rm{i}}(\tau, t),
\end{align} 
\end{subequations}
where the superscript $^{\rm{i}}$ signifies that~\eqref{eoms two modes two photons i 10 main} is driven by a single input photon. The coefficients $\psi_{10}^{\rm{i}}$ and $\psi_{01}^{\rm{i}}$ are functions of two times, $\tau$ being the initial time at which the state $\ket{00}\ket{1_k\mathbf{1}_m}$ was created, and $t\geq \tau$ describing the subsequent evolution of the coefficients. The initial condition of~\eqref{eoms two modes two photons i main} is $\psi_{10}^{\rm{i}}(\tau,\tau) \equal \psi_{01}^{\rm{i}}(\tau,\tau) \equal 0$ since the system is in state $\ket{00}$ at time $\tau$.

States in layer 3 with two photons in the system have coefficients with the following equations of motion
\begin{subequations}\eqlab{ODEs two photon terms main} 
\begin{align}
    \!\dot{\psi}_{20}  &\!= -\big(i2\delta_a\plus \Gamma +i4|\Lambda|\big)\psi_{20} ~-\nn\\
    &\hspace{3.3cm} i\sqrt{2}\Lambda^{\!*}\psi_{11} + \sqrt{2\gamma}\psi_{10}^{\rm{ii}}\xi_{\rm{in}} \\
    \!\dot{\psi}_{11}  &\!= -\Big(i(\delta_a\plus \delta_b) + \frac{\Gamma+\gamma_L}{2}\plus i4|\Lambda|\Big)\psi_{11} ~-\nn\\
    &\hspace{2.2cm} i\sqrt{2}\Big[\Lambda\psi_{20} \plus \Lambda^{\!*}\psi_{02}\Big] \plus \sqrt{\gamma}\psi_{01}^{\rm{ii}}\xi_{\rm{in}} \\
    \!\dot{\psi}_{02}  &\!= \!-\big(i2\delta_b+\gamma_L\plus i4|\Lambda|\big)\psi_{02} \minus i\sqrt{2}\Lambda\psi_{11} . 
\end{align}  
\end{subequations}
The initial conditions are $\psi_{20}(0)\equal \psi_{11}(0)\equal \psi_{02}(0)\equal 0$.

There are other paths leading to the states $\ket{10}\ket{\mathbf{1}_m}$ and $\ket{01}\ket{\mathbf{1}_m}$ than those described by the dynamics in~\eqref{eoms two modes two photons i main}. It could either be from absorption of the first photon followed by direct passage of the second photon or emission from mode $a$ while the state is $\ket{20}\ket{\emptyset}$ or $\ket{11}\ket{\emptyset}$. We use different coefficients for the state originating from these paths because their dynamical equations do not contain driving terms from input photons. The equations are
\begin{subequations}\eqlab{eoms two modes two photons o main} 
\begin{align}
    \dot{\psi}_{10}^{\rm{o}}(\tau, t) &= -\Big(i\delta_a\plus \frac{\Gamma}{2} + i2|\Lambda(t)|\Big)\psi_{10}^{\rm{o}}(\tau, t) ~-\nn\\
    &\hspace{4.0cm} i\Lambda(t)^{\!*}\psi_{01}^{\rm{o}}(\tau, t) \eqlab{SPM 2 photons ODE psi_10^o main} \\
    \dot{\psi}_{01}^{\rm{o}}(\tau, t) &= -\Big(i\delta_b \plus \frac{\gamma_L}{2}+ i2|\Lambda(t)|\Big)\psi_{01}^{\rm{o}}(\tau, t) ~-\nn\\
    &\hspace{4.0cm} i\Lambda(t)\psi_{10}^{\rm{o}}(\tau, t). \eqlab{SPM 2 photons ODE psi_01^o main} 
\end{align} 
\end{subequations}
There are two sets of initial conditions for~\eqref{eoms two modes two photons o main} depending on whether the dynamics originated from the formation of state $\ket{10}\ket{\mathbf{1}_m}$ or $\ket{01}\ket{\mathbf{1}_m}$ at time $\tau\equal m\Delta t$. If the photon started in mode $a$, the initial condition is $\psi_{10}^{\rm{o}}(\tau, \tau)\equal 1$ and $\psi_{01}^{\rm{o}}(\tau, \tau)\equal 0$, and we define $L_{10}\equiv \psi_{10}^{\rm{o}}$ and $L_{01}\equiv \psi_{01}^{\rm{o}}$. If the photon started in mode $b$, the initial condition is $\psi_{10}^{\rm{o}}(\tau, \tau)\equal 0$ and $\psi_{01}^{\rm{o}}(\tau, \tau)\equal 1$, and we define $M_{10}\equiv \psi_{10}^{\rm{o}}$ and $M_{01}\equiv \psi_{01}^{\rm{o}}$. 

\figref{two photons two modes paths} reveals that there are 8 distinct paths from input to output so the coefficient of the output state $\ket{00}\ket{\mathbf{1}_m \mathbf{1}_n}$ should contain 8 terms
\begin{multline}\eqlab{xi^out two modes two photons main} 
    \xi_{\rm{out}}(\tau, t) = \frac{1}{\sqrt{2}}\Big[ \gamma\psi_{11}(\tau) M_{10}(\tau, t) ~+\\
     \sqrt{2}\gamma\psi_{20}(\tau) L_{10}(\tau) - \sqrt{\gamma}\psi_{01}^{\rm{ii}}(\tau)\xi_{\rm{in}}(\tau)M_{10}(\tau, t) ~-\\
     \sqrt{\gamma}\psi_{10}^{\rm{ii}}(\tau)\xi_{\rm{in}}(\tau)L_{10}(\tau, t) -\sqrt{\gamma}\psi_{10}^{\rm{ii}}(\tau)\xi_{\rm{in}}(t)  ~+\\
    \!\gamma\psi_{10}^{\rm{ii}}(\hspace{-0.2mm}\tau\hspace{-0.2mm})\psi_{10}^{\rm{i}}(\tau\!, t) \plus  \xi_{\rm{in}}(\hspace{-0.2mm}\tau\hspace{-0.2mm})\xi_{\rm{in}}(\hspace{-0.2mm}t\hspace{-0.2mm}) \minus \sqrt{2\gamma}\xi_{\rm{in}}(\hspace{-0.2mm}\tau\hspace{-0.2mm})\psi_{10}^{\rm{i}}(\tau\!, t)  \Big], 
\end{multline}
where the first term corresponds to the upper path in~\figref{two photons two modes paths}, the second term to the path immediately below, and so forth. Note that $\tau\leq t$ in~\eqref{xi^out two modes two photons main} and $\xi_{\rm{out}}(\tau, t)\equal \xi_{\rm{out}}(t, \tau)$ follows from the indistinguishability of the photons. The output state is defined as 
\begin{align}\eqlab{output state two modes two photons main} 
    \ket{\psi_{\rm{out}}} \equiv \int_0^T \!\!d\tau \int_0^T  \!\!dt \xi_{\rm{out}}(\tau, t) \hat{w}^\dagger(\tau)\hat{w}^\dagger(t) \ket{\emptyset},
\end{align}
and the integral of $|\xi_{\rm{out}}(\tau,t)|^2$ over $\tau$ and $t$ is 1 (in the absence of loss). To calculate the output state in~\eqref{xi^out two modes two photons main}, we solve the above equations of motion for $N$ different initial conditions corresponding to all the time bins in~\figref{discrete time illustration}.

%-----------------------------------------------------------------------
\section{Absorbing and Emitting Wave Packets via Dynamic Coupling} \seclab{absorption emission main}
%-----------------------------------------------------------------------
In this section we find analytic solutions for the control fields that allow absorption and emission of wave packets with known shapes with arbitrarily high fidelity. We consider a specific example of Gaussian wave packets and show by numerical integration of~\eqref{eoms two modes one photon main} that the fidelity of the absorption and emission process approaches unity very rapidly as the ratio between the cavity-waveguide coupling, $\gamma$, and the wave packet bandwidth, $\OmG$, increases. 
%----------------------------------------------------------------
\subsection{Absorption}
%----------------------------------------------------------------
For the absorption process, the boundary conditions of~\eqref{eoms two modes one photon main} are $\psi_{10}(0)\equal \psi_{01}(0)\equal 0$. We use a subscript $i$ (for ``\emph{in}") on the control function, $\Lambda_i(t)$. The goal is to determine $\Lambda_i(t)$ such that a single incoming photon with wave packet $\xi_{\rm{in}}(t)$ is absorbed into cavity mode $b$. Since $\Lambda_i$ is complex-valued, we write it as $\Lambda_i(t)\equiv |\Lambda_i(t)|\exp[i\phi_i(t)]$. In~\appref{app absorption} we find the solution for a material with a third-order nonlinearity

\begin{subequations}\eqlab{Lambda and phi sol absorption} 
\begin{align}
	 |\Lambda_i(t)|   &= \frac{|f_i(t)| e^{-\frac{\gamsub{L}t}{2}}}{|\xi_{\rm{in}}(t)|\sqrt{ 2\int_0^t\!f_i(s)ds - 4|\xi_{\rm{in}}(t)|^2e^{\gamsub{L}t} } } \eqlab{LAMi main}\\
	 \phi_i(t)        &= -\delta_b t -2\int_0^t \!|\Lambda_i(s)|ds - \arg(\xi_{\rm{in}}) ~+\nn\\
	                & \hspace{1.0cm} \tan^{-1}\!\bigg( \frac{f_i \sin(\theta_i) - g_i \cos(\theta_i)}{f_i \cos(\theta_i) + g_i \sin(\theta_i)}\bigg),
\end{align}
\end{subequations}
%
% % 
% \begin{align}\eqlab{phi sol}
% 	 \phi(t) &= -2\int_0^t \!|\Lambda(s)|ds - \arg(\xi) + \tan^{-1}\!\bigg( \frac{y(t)}{x(t)}\bigg) .
% \end{align}
% % %
% To obtain $x$ and $y$, note that 
% \begin{align} 
% 	x & = \dot{X} = \dot{R}\cos(\theta) - R\sin(\theta)\dot\theta = \frac{f_i \cos(\theta) + g_i \sin(\theta)}{\sqrt{ 2\int\!f_i}} \\ 
% 	y & = \dot{Y} = \dot{R}\sin(\theta) + R\cos(\theta)\dot\theta = \frac{f_i \sin(\theta) - g_i \cos(\theta)}{\sqrt{ 2\int\!f_i}}.
% \end{align} 
% %
% 
where 
\begin{subequations}\eqlab{fi gi definition main}
\begin{align}
	 f_i(t)         &= \Big(\frac{\gamma-\gamsub{L}}{2} \xi_{\rm{in}}(t) - \dot{\xi}(t)\Big)\xi_{\rm{in}}(t)^* e^{\gamsub{L}t} \eqlab{fi definition main} \\
     g_i(t)         &= - 2 |\Lambda_i(t)||\xi_{\rm{in}}(t)|^2 e^{\gamsub{L}t} \eqlab{gi definition main}\\
	 \theta_i(t)    &= -\frac12 \int_0^t\!\! \frac{g_i(s)}{ \int_0^s\! f_i(z)dz}ds \eqlab{theta_i definition main}.
    % x_i(t)            &= \frac{f_i(t) \cos[\theta_i(t)] + g_i(t) \sin[\theta_i(t)]}{\sqrt{2 \int_0^t f_i(s)ds }} .
\end{align} 
\end{subequations}
Note that we have assumed that $\xi_{\rm{in}}$ does not have a time dependent phase, such that $f_i$ and $g_i$ are real functions. It is straight forward to generalize this to chirped pulses with time dependent phase by re-defining $f_i$ and $g_i$. We also assumed $\delta_a\equal 0$ above. \\
% As an example, we consider Gaussian wave packets
% % 
% \begin{align}\eqlab{Gaussian}
% \xi_{\rm{in}}(t) =  \sqrt{\frac{2}{\tauin}} \left(\frac{\text{ln}(2)}{\pi}\right)^{\!\frac{1}{4}} \exp\!\left(\!-2\text{ln}(2)\frac{(t-t_0)^2}{\tauin^2} \right) ,
% \end{align} 
% %
% which are defined such that $|\xi_{\rm{in}}(t)|^2$ has a FWHM of $\tauin$.

% We need an envelope function.
% % 
% \begin{align}\eqlab{envelope function main}
% 	f_{\text{env}}(t) = \bigg[1\plus \frac{1 \plus \sin\!\big( \frac{\pi t}{\tauenv}\big) }{2}  \theta\Big(t \plus \frac{\tauenv}{2}    \Big)  \bigg]  \theta\Big(t\minus \frac{\tauenv}{2}   \Big) ,
% \end{align} 
% %
% where $\theta(t)$ is a step function that equals one when $t\!>\!0$ and zero otherwise. The envelope rises from zero to one in the interval $t\!\in\![-\tauenv/2,~ \tauenv/2]$ as half a period of the sine function. 

% %----------------------------------------------------------------
% \subsubsection{$\chi^{(2)}$ - Medium }\seclab{absorption chi2 main}
% %----------------------------------------------------------------
In the case of a material with a second-order nonlinearity there is no cross-phase modulation from the control field, so $g_i\equal 0$ and the solution reduces to
\begin{subequations}\eqlab{Lambda sol L main chi2} 
\begin{align}
    |\Lambda_i(t)|   &= \frac{|f_i(t)| e^{-\frac{\gamsub{L}t}{2}}}{|\xi_{\rm{in}}(t)|\sqrt{ 2\int_0^t\!f_i(s)ds } }\\
    \phi_i(t)     &=   - \arg(\xi_{\rm{in}}) ,
\end{align} 
\end{subequations}
with $f_i(t)$ still given by~\eqref{fi definition main}.

%----------------------------------------------------------------
\subsection{Emission}\seclab{emission main}
%----------------------------------------------------------------
Without any driving field, the equations of motion are found by setting $\xi_{\rm{in}}\equal 0$ in~\eqref{eoms two modes one photon main}
\begin{subequations}\eqlab{eoms two modes one photon emission main}
\begin{align}
    \dot{\psi}_{10}  &=  \Big(-\frac{\Gamma}{2}-i2|\Lambda_o| \Big)\psi_{10} -i|\Lambda_o|e^{-i\phi_o}\psi_{01}  \eqlab{EOM cp as UL main}\\
    \dot{\psi}_{01}  &= \Big(-\frac{\gamma_L}{2}-i2|\Lambda_o|\Big)\psi_{01}  -i|\Lambda_o|e^{i\phi_o}\psi_{10} \eqlab{EOM cp ai+ UL main}\\
     \xi_{\rm{out}}  &= - \sqrt{\gamma}\psi_{10} .
\end{align} 
\end{subequations}
Note that we use the subscript $o$ (for ``\emph{out}") on the control function in~\eqref{eoms two modes one photon emission main}. The initial condition is $\psi_{10}(0)\equal 0$ and state $\ket{01}\ket{\emptyset}$ has the complex amplitude $\psi_{01}(0)$. The goal is to determine $|\Lambda_o(t)|$ and $\phi_o(t)$ such that $\xi_{\rm{out}}(t)$ equals some desired wave packet, $\xi(t)$. The solution is found in~\appref{app emission}
\begin{subequations}\eqlab{Lambda sol UL main} 
\begin{align}
    |\Lambda_o(t)|  &= \frac{|f_o| e^{-\frac{\gamsub{L}t}{2}}}{|\xi|\sqrt{ \gamma|\psi_{01}(0)|^2 \minus 2\int_0^t\!f_o(s)ds \minus 4|\xi|^2e^{\gamma_Lt} } } \eqlab{LAMo main}\\
    \phi_o(t)     &= -\delta_b  -2\int_0^t |\Lambda_o(s)|ds  - \arg(\xi) ~+\nn\\
    &\hspace{1.0cm} \tan^{-1}\!\bigg( \frac{f_o \cos(\theta_o) - g_o \sin(\theta_o)}{-f_o \sin(\theta_o) - g_o \cos(\theta_o)}\bigg),
\end{align} 
\end{subequations}
where
\begin{subequations}\eqlab{fo go def main}
\begin{align}
f_o(t)     &= \Big(\frac{\Gamma}{2}\xi(t) + \dot{\xi}(t)\Big)\xi(t)^*e^{\gamma_Lt} \eqlab{fo definition main}\\
g_o(t)     &= -2 |\Lambda_o(t)| |\xi(t)|^2 e^{\gamma_Lt}  \eqlab{go definition main}\\
\theta_o(t) &= -\int_0^t\!\! \frac{g_o(s)}{ \gamma |\psi_{01}(0)|^2 - 2\!\int_0^s\! f_o(z)dz} ds.
% x_o(t)      & = \frac{f_o(t) \cos[\theta_o(t)] + g_o (t)\sin[\theta_o(t)]}{\sqrt{ \gamma|\psi_{01}(0)|^2 -  2\int_0^t\!f_o(s)ds}} .
\end{align}
\end{subequations}
Again, we assumed $\delta_a\equal 0$.\\

% %----------------------------------------------------------------
% \subsubsection{$\chi^{(2)}$ - Medium }\seclab{emission chi2 main}
% %----------------------------------------------------------------
The solution simplifies in the case of a material with a second-order nonlinearity
\begin{subequations}\eqlab{Lambda sol UL main chi2} 
\begin{align}
    |\Lambda_o(t)|  &= \frac{|f_o| e^{-\frac{\gamsub{L}t}{2}}}{|\xi|\sqrt{ \gamma|\psi_{01}(0)|^2 - 2\int_0^t\!f_o(s)ds } } \\
    \phi_o(t)     &=   - \arg(\xi) -\frac{\pi}{2},
\end{align} 
\end{subequations}
with $f_o(t)$ still given by~\eqref{fo definition main}.\\

We note that the solutions found in this section correspond to the amplitude and phase inside the cavity modes for the control fields in the case of third-order nonlinear materials. In~\appref{input pumps supp} we derive expressions for the control fields in the waveguide giving rise to these desired cavity-fields.

%----------------------------------------------------------------
\subsection{Gaussian Wave Packet }\seclab{Gaussian main}
%----------------------------------------------------------------
We consider an example of a Gaussian wave packet to investigate how well our absorption and emission technique works. The Gaussian wave packet of the input field is defined as
\begin{align}\eqlab{Gaussian main}
\mathcal{G}(t) &=  \!\sqrt{\frac{2}{\tauG}} \!\left(\frac{\text{ln}(2)}{\pi}\right)^{\!\!\frac{1}{4}} \!\!\exp\!\left(\!-2\text{ln}(2)\frac{t^2}{\tauG^2} \right), 
\end{align} 
where $|\mathcal{G}(t)|^2$ has a full temporal width at half maximum (FWHM) of $\tauG$, spectral width of $\OmG\equal 4\text{ln}(2)/\tauG$, and integrates to 1 (over the infinite interval from $-\infty$ to $\infty$). The input states are characterized by the wave packet $\xi_{\rm{in}}(t)\equal \mathcal{G}(t-\tauI)$ and the ideal output state is characterized by a simple time-translation
\begin{align}\eqlab{ideal Gaussian output state}
\ket{\mathcal{G}_{\rm{out}}} =  \int_0^T \!\! dt \hspace{0.3mm} \mathcal{G}(t-\Tout) \hat{w}^\dagger (t) \ket{\emptyset},
\end{align} 
where $\Tout\equal \tauI\plus \Tstore$. The duration of the entire interaction process, $T\equal \Tout\plus\tauO$, is divided into three time intervals denoted ``absorption", $t\!\in\![0,2\tauI]$, ``storage", $t\!\in\![2\tauI, \Tout\minus \tau_o]$, and ``emission", $t\!\in\![\Tout\minus\tau_o, T]$. Practically, wave packets must have a finite duration and our choice of absorption interval causes a discontinuous jump in $\xi_{\rm{in}}$ from $\xi_{\rm{in}}(0^-)\equal 0$ to $\xi_{\rm{in}}(0^+)\equal \mathcal{G}(-\tauI)$. The field in cavity mode $a$ takes a finite time to build up sufficiently to cause complete destructive interference with the part of the incoming wave packet that did not interact with the cavity. It is therefore impossible to perfectly absorb a wave packet of finite length, but the probability that the photon passes by the cavity without interacting, $P_{\rm{pass}}$, becomes negligible for relatively small values of the ratio $\gamma/\OmG$ as seen below. The problem of absorbing a wave packet of finite length is reflected in the solutions for the control fields in~\eqsref{LAMi main}{LAMo main}, which become imaginary when the terms under the square root in the denominators are negative. As explained in~\appref{Gaussian emission Chi3 app}, we use smoothing functions to avoid divergences and ensure the control functions are zero outside the absorption and emission intervals. The smoothing functions in~\eqref{f up down i app} are parametrized by the on/off duration, $\tauenv$.\\

\figref{chi3 one example} shows an example of the absorption, storage, and emission of a single photon in a Gaussian wave packet. 
\begin{figure}[!h] 
\centering
   \includegraphics[angle=0,origin=c,width=8cm] {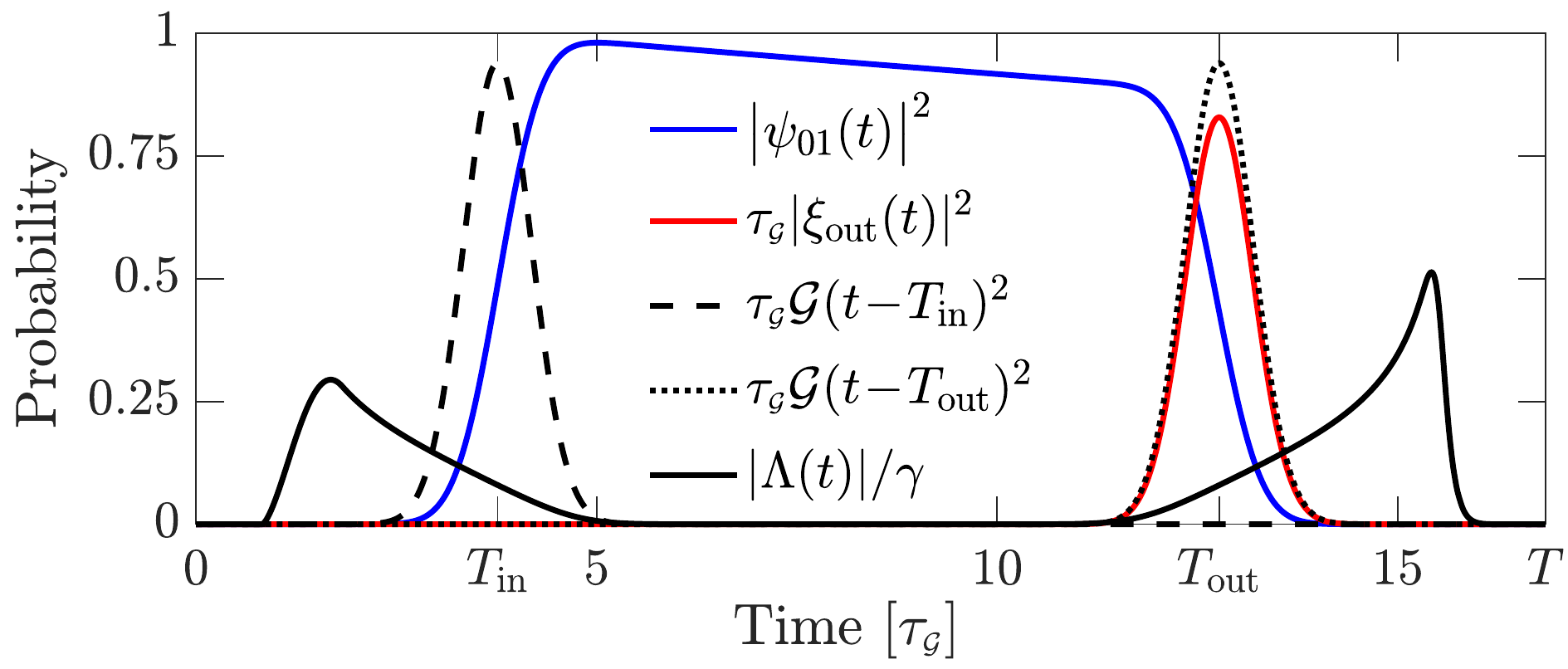}
\caption{Plots of the solution to~\eqref{eoms two modes one photon main} along with the input/output Gaussians and the control field found in~\eqsref{LAMi main}{LAMo main}. Parameters: $\gamma\equal 30\OmG$, $\gamma_L\equal 5\!\times\!10^{-3}\OmG$, $\tauenv\equal \tauG$, $\tau_o\equal 4.08\tauG$, $T_{\rm{store}}\equal 9\tauG$.  }
\figlab{chi3 one example}
\end{figure} 
The control field is given by $\Lambda\equal \Lambda_i\plus \Lambda_o$ since the storage time, $T_{\rm{store}}$, is chosen large enough to avoid overlap between the absorption and emission intervals, $T_{\rm{store}}\!>\! \Tin \plus \tau_o$. Note that the control field responsible for emission is different from a simple time-inversion of the control field responsible for absorption. This is because the presence of loss breaks the time-reversal symmetry of the equations of motion in~\eqref{eoms two modes one photon main}.\\

In the presence of loss, it is possible to emit a wave packet with the desired shape but reduced amplitude, $\xi_{\rm{out}}(t) \!\approx\! \sqrt{\eta} \mathcal{G}(t-\Tout)$, where $\eta$ is a real number smaller than 1. Note, however, that this is only true in the emission interval, $t\!\in\![\Tout\minus\tau_o, T]$, since $\xi_{\rm{out}}(t)$ generally has some small contribution from the absorption interval due to imperfect absorption. The probability that the photon passes by the cavity without being absorbed is
\begin{align}\eqlab{Ppass definition main}
 P_{\rm{pass}} \equiv \int_{0}^{2\tauI} \!|\xi_{\rm{out}}(t)|^2dt.
\end{align} 
The probability of a successful storage process is equal to $\eta$ in the limit $P_{\rm{pass}}\rightarrow 0$. The maximum possible value of $\eta$ can be found by inserting $\xi \equal \sqrt{\eta} \mathcal{G}$ into the denominator of~\eqref{LAMo main} and ensuring that the terms under the square root are positive for all $t$. For the Gaussian in~\eqref{Gaussian main}, we have 
\begin{align}\eqlab{Fo limit main}
\mathcal{F}_o\equiv 2\! \int_{-\infty}^{\infty} \!\!\!f_o(t)dt =  \gamma \exp\!\Big[\gamma_L\Big(\Tout \plus \frac{\gamma_L\tauG^2}{16\ln (2)}\Big)   \Big], 
\end{align} 
and we therefore choose $\eta$ as 
\begin{align}\eqlab{eta main}
\eta = \frac{\gamma|\psi_{01}(0)|^2}{\mathcal{F}_o} \big(1-\epsilon_\eta\big). 
\end{align} 
The value of the small parameter, $\epsilon_\eta$, is optimized to maximize the value of $\eta$ while avoiding divergences in $|\Lambda_o|$. Finite values of $P_{\rm{pass}}$ limits the achievable overlap of the output wave packet with a desired shape, which is seen by calculating the conditional fidelity in~\eqref{conditional state fidelity definition} using $\xi_{\rm{out}}\equal \sqrt{\eta}\mathcal{G}(t\minus\Tout)$ in the emission interval
\begin{align}\eqlab{Fcond approx}
% \overline{F}_{\!s} = \frac{\Big| \int_0^T \!\xi_{\rm{out}}(t)\mathcal{G}(t\minus\Tout)^* dt   \Big|^2 }{ \int_0^T \!|\xi_{\rm{out}}(t)|^2dt \int_0^T \!|\mathcal{G}(t\minus\Tout)|^2dt } \approx
\overline{F}_1 = \frac{\Big| \int_0^T \!\xi_{\rm{out}}(t)\mathcal{G}(t\minus\Tout)^* dt   \Big|^2 }{ \int_0^T \!|\xi_{\rm{out}}(t)|^2dt  } \approx
% ~\approx \\ 
% \frac{\Big| \int_{T-2\tauO}^T \!\xi_{\rm{out}}(t)\mathcal{G}^*(t-\Tout)dt   \Big|^2 }{ \int_{0}^{2\tauI} \!|\xi_{\rm{out}}(t)|^2dt + \int_{T-2\tauO}^T \!|\xi_{\rm{out}}(t)|^2dt} \approx
\frac{\eta}{ P_{\!\rm{pass}} \plus \eta},
\end{align} 
where we changed the lower integration limit from 0 to $\Tout\minus\tauO$ in the numerator since $\mathcal{G}(t\minus\Tout)\!\approx\!0$ outside the emission interval. We also divided the integration of $|\xi_{\rm{out}}|^2$ into intervals $[0,2\tauI]$ and $[\Tout\minus\tauO, T]$ since $|\xi_{\rm{out}}(t)|^2\!\approx\! 0$ in the storage interval.
\begin{figure}[!h] 
\centering
   \includegraphics[angle=0,origin=c,width=7.5cm] {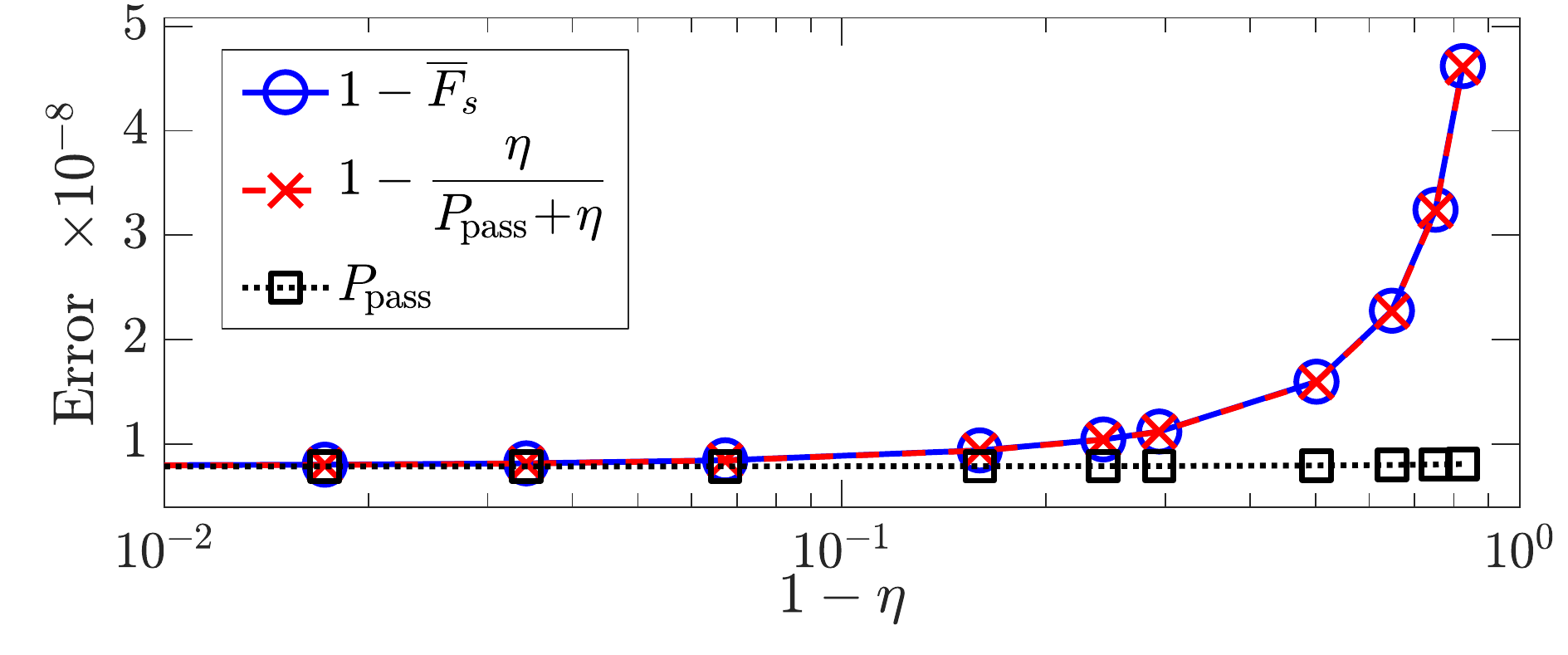}
\caption{Degradation of conditional fidelity in the limit of large loss. Parameters: $\gamma\equal 30\OmG$, $\tauenv\equal \tauin$.   }
\figlab{Fcond degradation}
\end{figure} 
\figref{Fcond degradation} shows a plot of the conditional fidelity using $\xi_{\rm{out}}$ from~\eqref{eoms two modes one photon main} along with the approximation in~\eqref{Fcond approx}. It also shows that $\overline{F}_{\!s}\!\rightarrow\! 1-P_{\!\rm{pass}}$ in the limit where $P_{\!\rm{pass}}\!\ll\!\eta$, which is seen from a Taylor expansion of~\eqref{Fcond approx},
$\overline{F}_{\!s}\approx  1/(1+P_{\!\rm{pass}}/\eta)\approx 1 - P_{\!\rm{pass}}$. It is important to note that~\figref{Fcond degradation} clearly illustrates that very small error in the conditional fidelity is possible even in the case of an efficiency well below unity.\\

The value of $P_{\rm{pass}}$ only depends on the ratio $\gamma/\OmG$ and~\figref{Ppass vs gamma} plots the dependence for both second- and third-order nonlinear materials. 
\begin{figure}[!h] 
\centering
   \includegraphics[angle=0,origin=c,height=3.9cm] {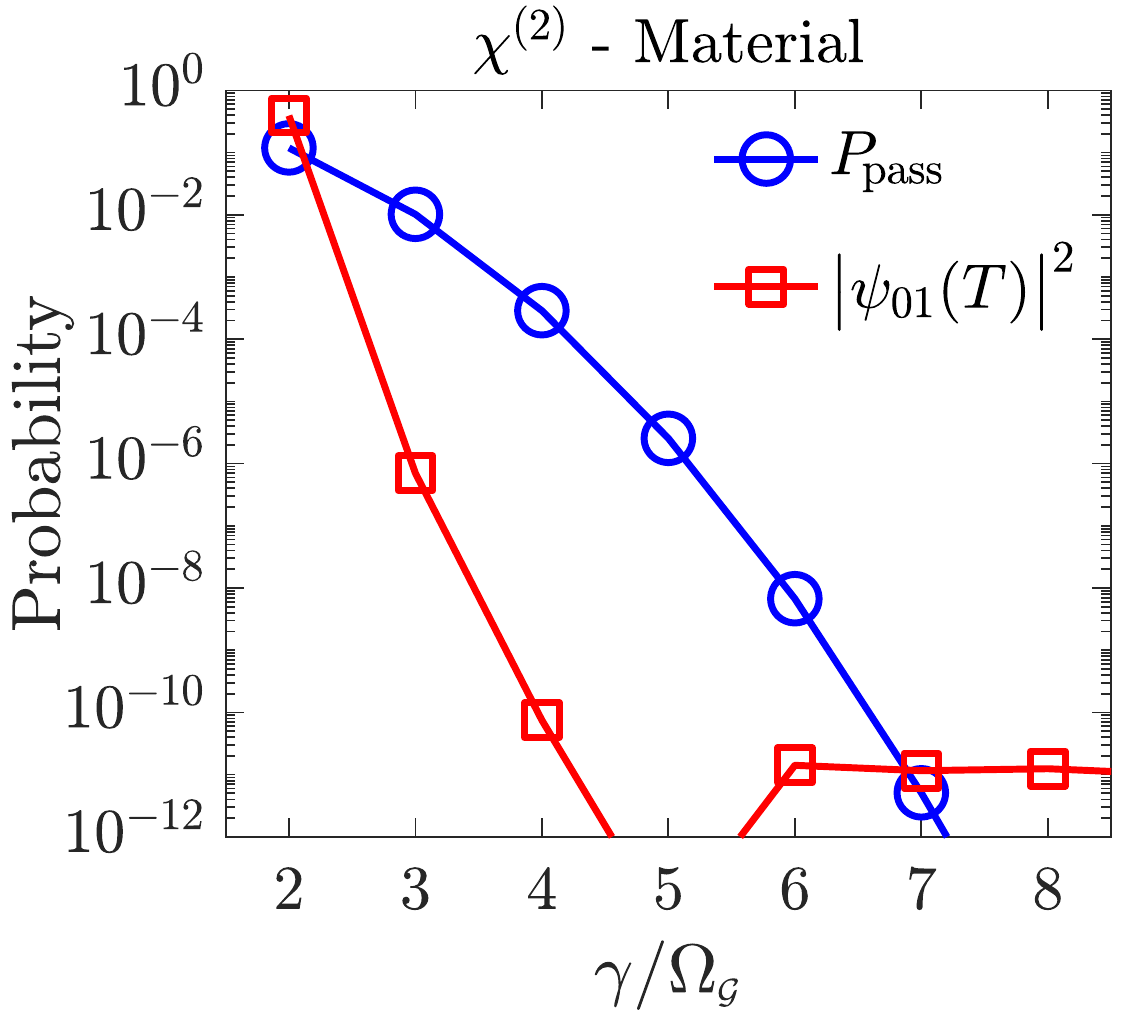}
   \hspace{-0.3cm}
   \includegraphics[angle=0,origin=c,height=3.9cm] {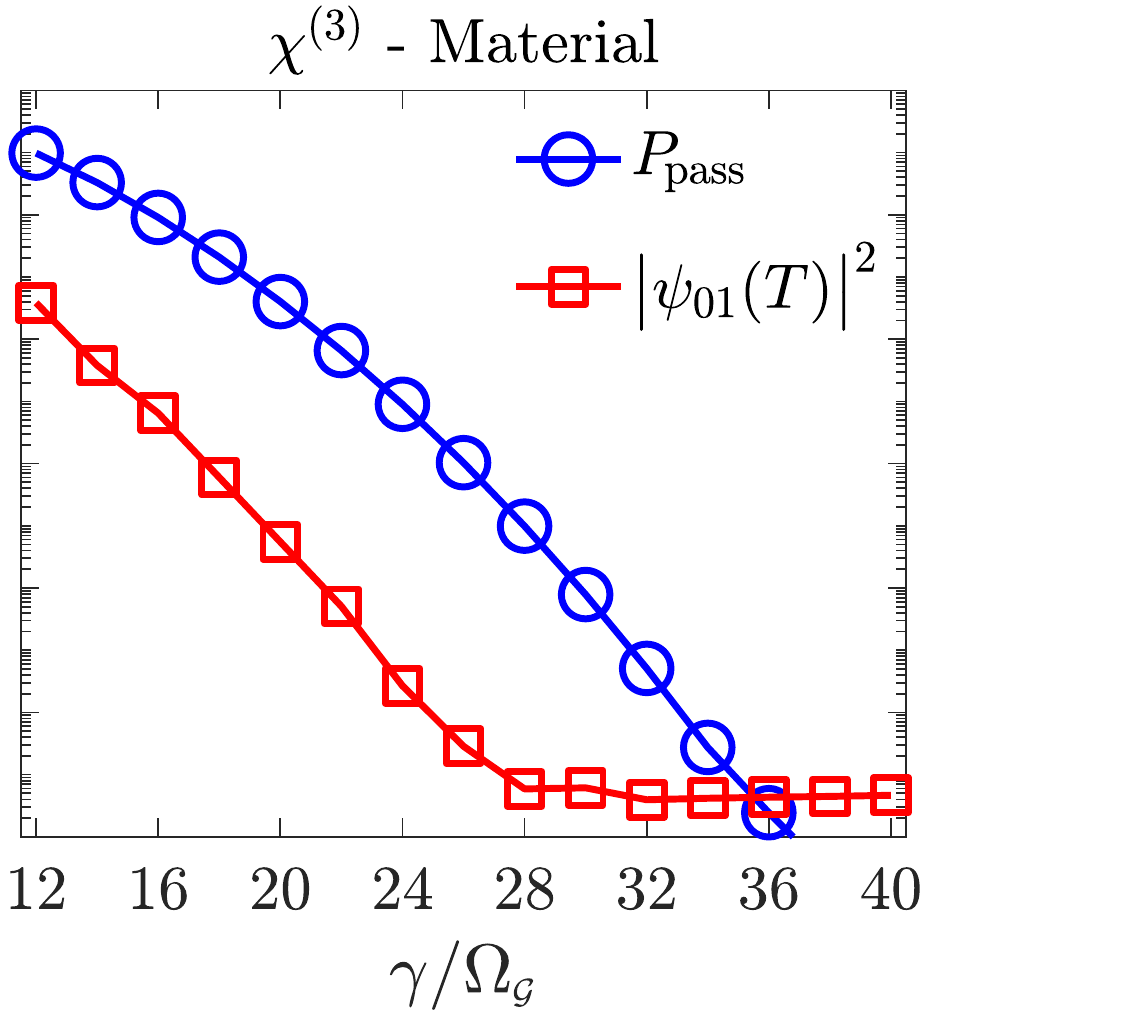}
\caption{$P_{\rm{pass}}$ as a function of $\gamma/\OmG$ for $\chi^{(2)}$ and $\chi^{(3)}$ materials. Parameters: $\gamma_L\equal 0$, $\tauenv\equal \tauG$.  }
\figlab{Ppass vs gamma}
\end{figure} 
It is seen that $P_{\rm{pass}}$ falls off faster for $\chi^{(2)}$ materials due to the absence of cross-phase modulation. In~\appref{when a solution exists} we derive expressions suggesting that a five times larger coupling rate, $\gamma$, is needed for a $\chi^{(3)}$ material, which agrees well with the result in~\figref{Ppass vs gamma}. Importantly,~\figref{Ppass vs gamma} shows that $P_{\rm{pass}}$ approaches zero extremely fast as the ratio $\gamma/\OmG$ increases.\\

%----------------------------------------------------------------
\section{Nonlinear Dynamics} \seclab{nonlinear dynamics main}
%----------------------------------------------------------------
In this section we consider three types of nonlinearities that mediate photon-photon interactions and describe the necessary extensions to the equations of motion in~\secref{main linear dynamics} to account for them. Since we have a particular interest in two-qubit logic gates for quantum information processing, we consider cavity configurations enabling a c-phase gate. Note that we envision a configuration where two identical cavities are placed in between two 50/50 beam-splitters that convert the two-qubit state $\ket{11}$ into $1/\sqrt{2}(\ket{02}+ \ket{20})$~\cite{Knill2001, Nysteen2017}. In this case, the phase $\theta_{n}$ in~\eqref{complex overlaps Fidelity} is important in that $\theta_2\minus 2\theta_1\equal \pi$ is required for the gate transformation $\ket{00}\rightarrow\ket{00}$, $\ket{10}\rightarrow\ket{10}$, $\ket{01}\rightarrow\ket{01}$, $\ket{11}\rightarrow - \ket{11}$.

% In terms of our input/output relations, this corresponds to $\arg[\xi_{\rm{out}}(\tau, t)] - 2\arg[\xi_{\rm{out}}(t)] \equal \pi$, meaning that a two-photon input state acquires an additional $\pi$ phase shift compared to a single-photon input. 
We start by considering a material with a third-order nonlinearity, then we describe second-order nonlinearities, and finally interactions with a two-level emitter.

% The important quantities to consider for controlled-phase gates are overlaps of the output states with time-translated versions of the input states
% % 
% \begin{align}\eqlab{complex overlaps}
%     \braket{\psi_{\rm{out}}^{(n)}}{ \mathcal{G}_{\rm{out}}} = \sqrt{F_n} e^{i\theta_n},
% \end{align}
% % 
% where $n\equal\{1,2\}$ is the number of input photons and 
% % 
% \begin{align}\eqlab{complex overlaps xi}
%     \braket{\psi_{\rm{out}}^{(1)}}{ \mathcal{G}_{\rm{out}}} &= \int_0^T \!\!\!  \xi_{\rm{out}}(t) \mathcal{G}(t\minus \Tout)^* dt \\
%     \braket{\psi_{\rm{out}}^{(2)}}{ \mathcal{G}_{\rm{out}}} &= \nn\\
%     &\hspace{-0.6cm} \int_0^T \!\!\! \int_0^T \!\!\!  \xi_{\rm{out}}(\tau, t) \mathcal{G}(t\minus\Tout)^* \mathcal{G}(\tau\minus\Tout)^*d\tau dt .
% \end{align}
% % 
% The fidelity, $F_n$, should generally be as close to unity as possible and the condition $\theta_2- 2\theta_1\equal \pi$ applies to a controlled-phase gate.
% indicates
% where 
% % 
% \begin{align}\eqlab{overlap phase definition}
%     \theta_{nm} = \arg\!\Big( \braket{\psi_{\rm{out}} }{\mathcal{G}_{\rm{out}}^{(11)}}  \Big)  - 2\arg\!\Big( \braket{\psi_{\rm{out}} }{\mathcal{G}_{\rm{out}}^{(10)}}  \Big).
% \end{align}
%

%----------------------------------------------------------------
\subsection{Material with a Third-order Nonlinearity}\seclab{chi3 interaction main}
%----------------------------------------------------------------
Only modes $a$ and $b$ are needed in the case of a $\chi^{\scriptscriptstyle (3)}$ material. The Hamiltonian corresponding to photon-photon interactions is
\begin{align}\eqlab{H nonlinear chi3}
    \hat{H}_{\!\chi^{(3)}} \!=\! \hbar\chi_3 \Big[ \hat{a}^\dagger\hat{a}\hat{b}^\dagger\hat{b} + \frac{\big( \hat{a}^\dagger\hat{a}\minus 1\big)\hat{a}^\dagger\hat{a} + \big(\hat{b}^\dagger\hat{b}\minus 1\big)\hat{b}^\dagger\hat{b}}{4} \Big].
\end{align}
The corresponding unitary time-evolution operator is
\begin{multline}\eqlab{U nonlinear chi3}
    \hat{U}_{\!\chi^{(3)}} = -i\Delta t\chi_3 \hat{b}^\dagger\hat{b}\hat{a}^\dagger\hat{a} ~-\\
    i\frac14\Delta t\chi_3\Big[ \big( \hat{b}^\dagger\hat{b}- 1\big)\hat{b}^\dagger\hat{b} + \big( \hat{a}^\dagger\hat{a}-1\big)\hat{a}^\dagger\hat{a} \Big]\!.
\end{multline} 
Only states with two photons in the system are affected, so that
\begin{subequations}  \eqlab{U nonlinear chi3 on state}
\begin{align}
    \hat{U}_{\!\chi^{(3)}} \ket{20} &= -i\chi_3\Delta t\frac14 \big(2-1\big)2\ket{20} = -i\frac{\chi_3}{2}\Delta t\ket{20} \\
    \hat{U}_{\!\chi^{(3)}} \ket{11} &= -i\chi_3\Delta t\ket{11} \\
    \hat{U}_{\!\chi^{(3)}} \ket{02} &= -i\chi_3\Delta t\frac14 \big(2-1\big)2\ket{02} = -i\frac{\chi_3}{2}\Delta t\ket{02} .
\end{align}
\end{subequations}
The equations of motion for the corresponding coefficients in~\eqref{ODEs two photon terms main} are therefore modified as
\begin{subequations}\eqlab{ODEs two photon terms main chi3} 
\begin{align}
    \!\dot{\psi}_{20}  &\!= -\Big(i2\delta_a+\Gamma \plus i\frac{\chi_3}{2}\plus i4|\Lambda|\Big)\psi_{20} \minus i\sqrt{2}\Lambda^{\!*}\psi_{11} ~+\nn\\
    &\hspace{5.3cm} \sqrt{2\gamma}\psi_{10}^{\rm{ii}}\xi_{\rm{in}} \\
    % \!\dot{\psi}_{20}  &\!=\! -\big(\Gamma \plus i\chi_3\plus i4|\Lambda|\big)\psi_{20} \minus i\sqrt{2}\Lambda^{\!*}\psi_{11} \plus \sqrt{2\gamma}\psi_{10}^{\rm{ii}}\xi^{\rm{in}} \\
    \!\dot{\psi}_{11}  &\!= -\Big(i(\delta_a+\delta_b)+\frac{\Gamma+\gamma_L}{2}+i\chi_3+ i4|\Lambda|\Big)\psi_{11} ~-\nn\\
    &\hspace{2.2cm} i\sqrt{2}\Big[\Lambda\psi_{20} \plus \Lambda^{\!*}\psi_{02}\Big] \plus \sqrt{\gamma}\psi_{01}^{\rm{ii}}\xi_{\rm{in}} \\
    \!\dot{\psi}_{02}  &\!= \!-\Big(i2\delta_b\plus\gamma_L\plus i\frac{\chi_3}{2}\plus i4|\Lambda|\Big)\psi_{02} - i\sqrt{2}\Lambda\psi_{11} . \eqlab{ODEs two photon terms main chi3 psi_02}
\end{align}  
\end{subequations}
It is seen from~\eqref{ODEs two photon terms main chi3 psi_02} that the amplitude of the state $\ket{02}$ acquires a phase proportional to $\chi_3/2$, which the amplitude of the state $\ket{01}$ in~\eqref{eoms two modes one photon ii 01 main} does not. By a careful choice of storage time, $T_{\rm{store}}$, one may achieve the condition $\Delta\theta\equal \theta_2-2\theta_1 \equal \pi$, where $\theta_n$ is the phase in~\eqref{complex overlaps Fidelity}.~\figref{nonlinear phase and fidelity chi3} plots the phase difference as a function of storage time for a range of different nonlinear coupling coefficients, $\chi_3$. 
\begin{figure}[!h] 
\centering
   \includegraphics[angle=0,origin=c,width=7.5cm] {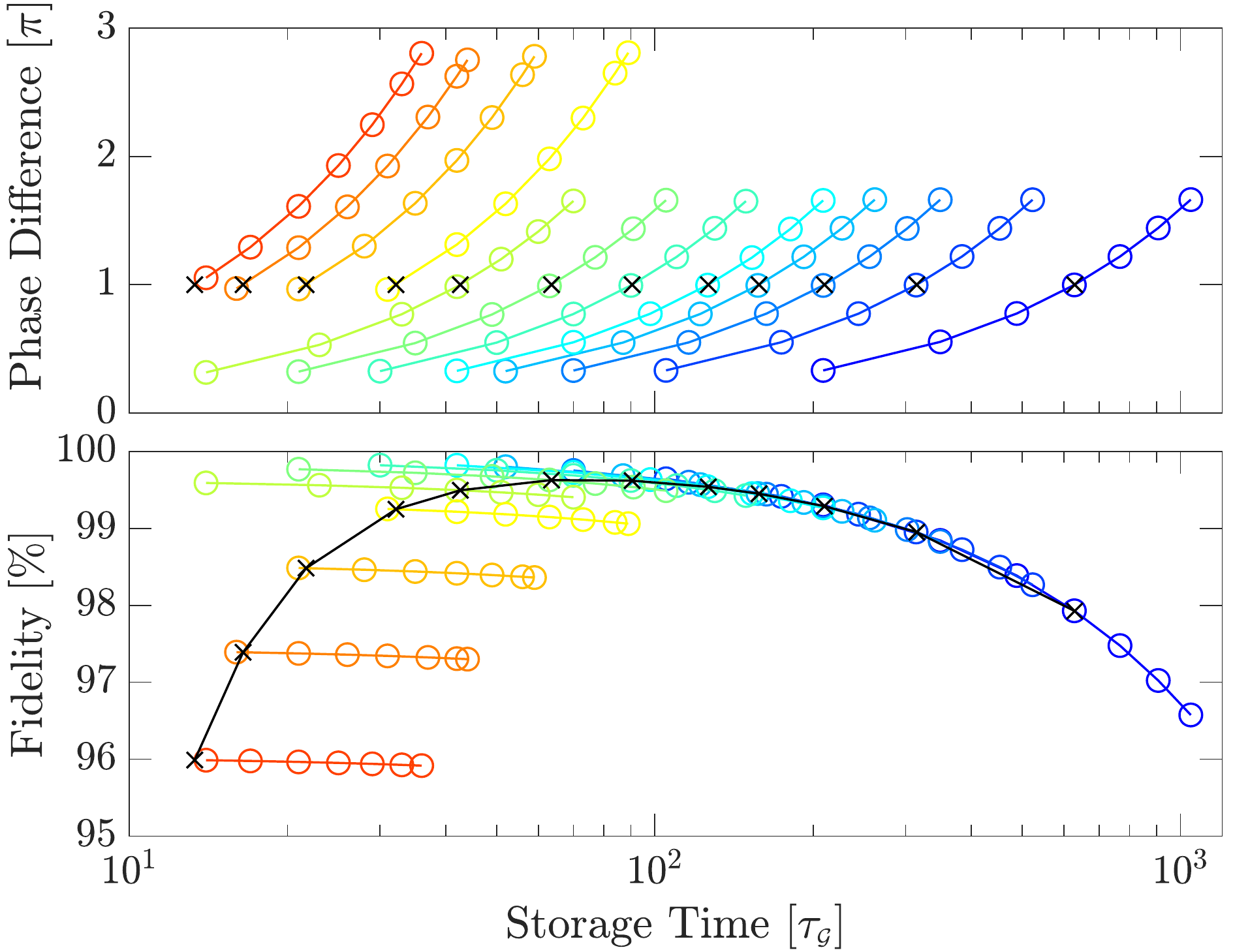}
%   \hspace{1cm}
%   \includegraphics[angle=0,origin=c,width=7cm] {./z_SPM_emission_tauenv1_gamL1e-5.pdf}
\caption{Nonlinear phase difference, $\Delta\theta$, and fidelity, $F_2$, as a function of storage time for different values of the nonlinear coupling rate, $\chi_3$, ranging from $0.01\OmG$ (blue) to $0.5\OmG$ (red). The black line shows the fidelity corresponding to $\Delta\theta\equal \pi$. Parameters: $\gamma\equal 30\OmG$, $\gamma_L\equal 10^{-5}\OmG$, $\tauenv\equal \tauin$.   }
\figlab{nonlinear phase and fidelity chi3}
\end{figure} 
It shows how the phase condition, $\Delta\theta\equal\pi$, may be met using a smaller nonlinearity and larger storage time (blue curve) or a larger nonlinearity and smaller storage time (red curve).~\figref{nonlinear phase and fidelity chi3} also plots the corresponding fidelity, $F_2$, which appears to reach an optimum for $T_{\rm{store}}\approx 100\tauG$. The fidelity degrades when increasing $\chi_3$ because the solutions for the control fields were found assuming a single photon input and photon-photon interactions during the absorption and emission process renders the control fields sub-optimal. The fidelity also degrades if $\chi_3$ is decreased too much because losses increase with increased storage time. 
% Therefore, an intermediate value of $T_{\rm{store}}$ is optimal.

%----------------------------------------------------------------
\subsection{Material with a Second-order Nonlinearity}\seclab{chi2 interaction main}
%----------------------------------------------------------------
For materials exhibiting a $\chi^{\scriptscriptstyle (2)}$ nonlinearity, we explore the process of second-harmonic-generation where $\omega_c\equal 2\omega_b$. With the introduction of mode $c$, the system states are written as $\ket{n_a n_b n_c}\equiv \ket{n_a}\ket{n_b}\ket{n_c}$ with $n_a,~n_b$, and $n_c$ representing the number of photons in each mode. The Hamiltonian describing the interaction is given in~\eqref{H nonlinear chi2 a}. The corresponding unitary time-evolution operator is
\begin{align}\eqlab{U nonlinear chi2}
    \hat{U}_{\rm{SHG}} = -i\chi_2 \Delta t\Big(\hat{c}\hat{b}^\dagger\hat{b}^\dagger + \hat{c}^\dagger\hat{b}\hat{b} \Big).
\end{align}
From~\eqref{U nonlinear chi2} we see that it only causes a coupling between states $\ket{020}$ and $\ket{001}$
\begin{subequations}\eqlab{U nonlinear chi2 action}
\begin{align}
    \hat{U}_{\rm{SHG}} \ket{020} &=  -i\chi_2 \Delta t \sqrt{2} \ket{001}\\
    \hat{U}_{\rm{SHG}} \ket{001} &=  -i\chi_2 \Delta t \sqrt{2} \ket{020}.
\end{align}
\end{subequations}
The equations of motion for coefficients corresponding to two photons in the system are then
\begin{subequations}\eqlab{ODEs two photon terms main chi3 b} 
\begin{align}
    \dot{\psi}_{200}  &= \!- \!\big(i2\delta_a\plus \Gamma\big) \psi_{200} \minus i\sqrt{2}\Lambda^{\!*}\psi_{110} \plus  \sqrt{2\gamma}\psi_{100}^{\rm{ii}}\xi_{\rm{in}} \\
    \dot{\psi}_{110}  &= -\Big(i(\delta_a \plus \delta_b)\plus \frac{\Gamma+\gamma_L}{2}\Big)\psi_{110} - i\sqrt{2}\Lambda\psi_{200} ~-\nn\\
    &\hspace{3cm} i\sqrt{2}\Lambda^{\!*}\psi_{020} + \sqrt{\gamma}\psi_{01}^{\rm{ii}}\xi_{\rm{in}} \\
    \dot{\psi}_{020}  &= -\big(i2\delta_b \plus \gamma_L\big) \psi_{020} \minus i\sqrt{2}\Lambda\psi_{110} \minus i\sqrt{2}\chi_2\psi_{001} \\
    \dot{\psi}_{001}  &= -\Big(i\delta_c+ \frac{\gamma_L}{2}\Big)\psi_{001}  -i\sqrt{2}\chi_2\psi_{020}. 
\end{align}  
\end{subequations}
It is the fact that SHG requires two input photons that enables the phase condition $\Delta\theta\equal \pi$ to be fulfilled. To understand why, consider the case in which the storage time is adjusted such that a single Rabi-flip between states $\ket{020}$ and $\ket{001}$ occur. An example is shown in~\figref{example chi2}. Occupation probabilities of the system states are found in~\appref{two photon dynamics two cavity modes} and plotted as a function of time in~\figref{example chi2}a. It shows how the photons are transferred from state $\ket{020}$ to $\ket{001}$ and back via SHG. The phase of $\psi_{020}(t)$ jumps by $\pi$ as its amplitude becomes zero in the middle of the storage interval (red curve in~\figref{example chi2}b).
\begin{figure}[!h] 
\centering
   \includegraphics[angle=0,origin=c,width=8.5cm] {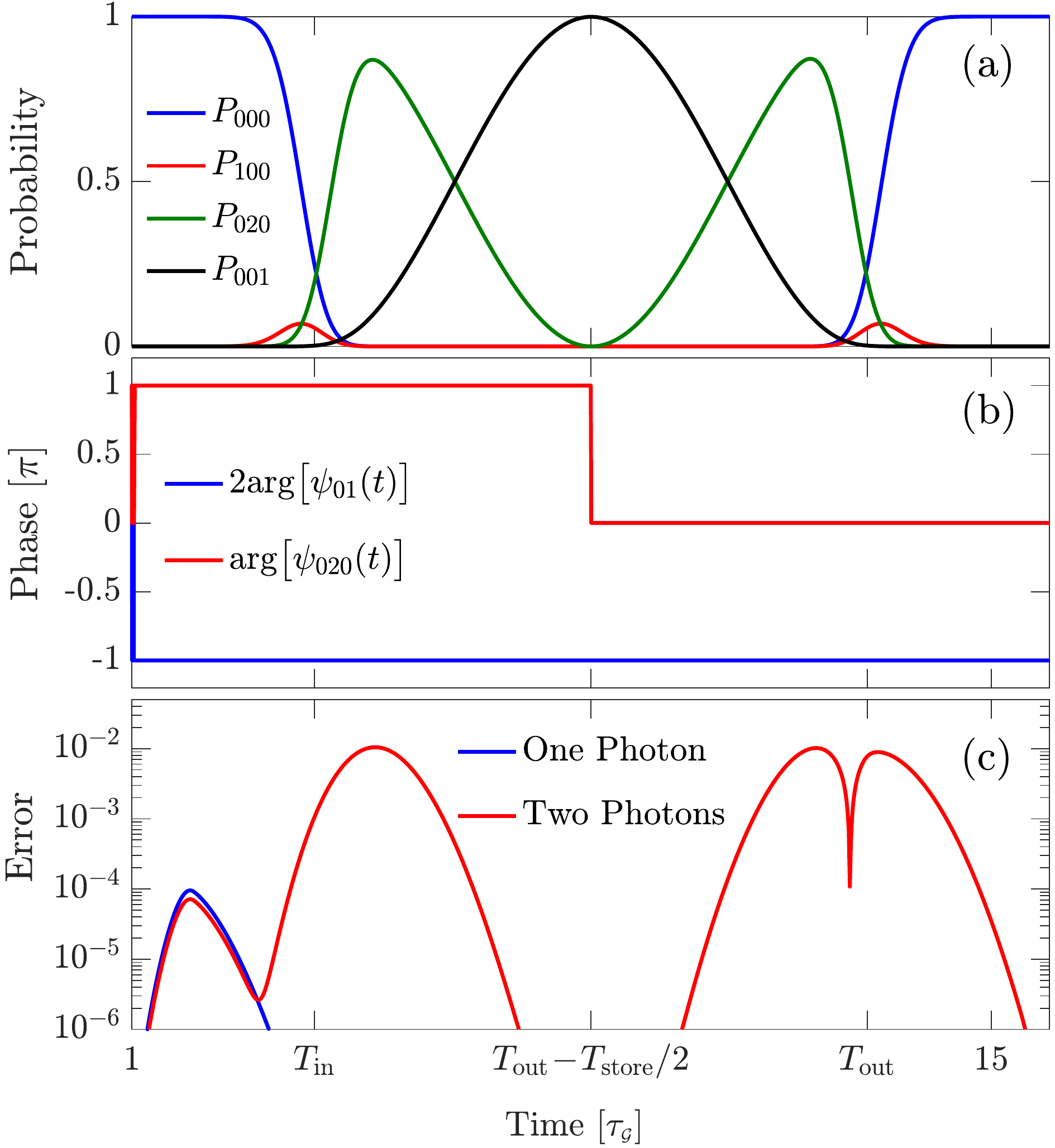}
\caption{(a) Occupation probabilities of system states as a function of time. (b) Phase of the coefficient corresponding to state $\ket{01}$ (blue) and $\ket{020}$ (red). (c) Error measured as the absolute distance from a Gaussian, $|\xi_{\rm{out}}(t)\!-\! \sqrt{\eta}\mathcal{G}(t\!-\!T_{\rm{out}})|$ (blue) and $|\xi_{\rm{out}}(\tau, T_{\rm{out}}) \!+\! \eta\mathcal{G}(\tau\!-\!T_{\rm{out}})\mathcal{G}(0)|$ (red). Parameters: $\gamma\equal 6\OmG$, $\gamma_L\equal 1.5\!\times\!10^{-4}\OmG$, $\tauenv\equal \tauG$, $\eta\equal 0.9963$.   }
\figlab{example chi2}
\end{figure} 
The phase of $\psi_{01}(t)$ (blue curve in~\figref{example chi2}b) remains constant since a single photon cannot undergo SHG. The relevant phase difference, $\Delta\theta$, is therefore seen to be exactly $\pi$.~\figref{example chi2}c shows the error in the output wave packet for both single- and two-photon inputs. Only a negligible error is observed for the single-photon input whereas the two-photon error is more pronounced leading to a fidelity of $F_2\equal 99.1\%$ for this example. Similar to the case of a $\chi^{(3)}$ material, the fidelity of two-photon outputs are degraded by the photon-photon interaction occurring during the absorption and emission process, which is not accounted for in the solution of the control fields.

%----------------------------------------------------------------
\subsection{Interaction with a Two-Level Emitter}\seclab{TLE interaction main}
%----------------------------------------------------------------
We investigate the use of atom-like two-level emitters because their nonlinearity is much stronger than the non-resonant nonlinearities considered above. To ensure complete absorption of incoming photons, the TLE should not be coupled to mode $b$ since we expect the nonlinear interaction during absorption and emission to be prohibitively strong. Instead, we use a tertiary mode, $c$, such that $\omega_c\minus \omega_b \sim \omega_b \minus \omega_a$ is in the GHz range. 
% Mode $c$ is coupled to the TLE with coupling rate $g$ and the coupling between modes $b$ and $c$ is controllable via a classical control field $\Pi(t)$. 
We envision a control scheme where a first control pulse, $\Lambda_i(t)$, is used to absorb incoming photons into mode $b$. Subsequently, a second control pulse, $\Pi(t)$, couples modes $b$ and $c$. Finally, a third control pulse, $\Lambda_o(t)$, couples the photons back into the waveguide through mode $a$. The first and last stage of this control protocol is therefore still described by the equations of motion in~\secref{two modes two photons}.
With the introduction of cavity mode $c$ and the TLE, states with two photons in the system are: $\ket{100}\ket{e}$, $\ket{010}\ket{e}$, $\ket{001}\ket{e}$, $\ket{200}\ket{g}$, $\ket{020}\ket{g}$, $\ket{002}\ket{g}$, $\ket{110}\ket{g}$, $\ket{101}\ket{g}$, and $\ket{011}\ket{g}$. 

During the second stage of the protocol, mode $a$ is empty so we introduce new coefficients, $\phi_{n_b n_c g}(t)$ and $\phi_{n_b n_c e}(t)$, corresponding to states $\ket{0n_b n_c}\ket{g}$ and $\ket{0n_b n_c}\ket{e}$. The dynamics is governed by the following equations of motion
\begin{subequations}\eqlab{ODEs two photon terms TLE main} 
\begin{align}
    \dot{\phi}_{20g}  &=  \!-\!\big(i2\delta_b\plus \gamma_L \plus i\chi_3\plus i4|\Pi|\big) \phi_{20g}  \minus i\sqrt{2}\Pi^*\phi_{11g} \\
    \dot{\phi}_{11g}  &=  -\big[i(\delta_b+\delta_c)+ \gamma_L + i\chi_3+ i4|\Pi|\big] \phi_{11g}  ~-\nn\\
    &\hspace{1.8cm} i\sqrt{2}\Pi\phi_{20g}  - i\sqrt{2}\Pi^*\phi_{02g} -  ig \phi_{10e} \\
    \dot{\phi}_{02g}  &= -\big(i2\delta_c+\gamma_L + i\chi_3+ i4|\Pi|\big) \phi_{02g} ~-\nn\\
    &\hspace{3.4cm} i\sqrt{2}\Pi\phi_{11g} - i\sqrt{2}g\phi_{01e} \\
    \dot{\phi}_{10e}  &= -\Big(i(\delta_b\plus\delta_e) + \frac{\gamma_e+\gamma_L}{2}+ i2|\Pi|\Big) \phi_{10e} ~-\nn\\
    &\hspace{4.0cm} i\Pi^*\phi_{01e} - ig^*\phi_{11g} \\
    \dot{\phi}_{01e}  &= -\Big(i(\delta_c\plus\delta_e) +\frac{\gamma_e+\gamma_L}{2}+ i2|\Pi|\Big) \phi_{01e} ~-\nn\\
    &\hspace{3.7cm} i\Pi\phi_{10e} - i\sqrt{2}g^*\phi_{02g} .
\end{align}  
\end{subequations}
Note that the dynamics is also changed for single-photon inputs, which have the following equations of motion
\begin{subequations}\eqlab{ODEs one photon terms TLE main} 
\begin{align}
    \dot{\phi}_{10g}  &=  -\Big(i\delta_b+\frac{\gamma_L}{2} + i2|\Pi|\Big) \phi_{10g}  -i\Pi^*\phi_{01g} \\
    \dot{\phi}_{01g}  &=  -\Big(i\delta_c\plus \frac{\gamma_L}{2} \plus  i2|\Pi|\Big) \phi_{01g}  \minus i\Pi\phi_{10g}  \minus ig \phi_{00e}\\
    % &\hspace{3cm} i\sqrt{2}\Pi^*\phi_{02g} -  ig \phi_{10e} \\
    % \dot{\phi}_{10g}  &= -\Big(\gamma_e+\frac{\gamma_L}{2}+ i2|\Pi|\Big) \phi_{10e} ~-\nn\\
    % &\hspace{3.6cm} i\Pi^*\phi_{01e} - ig\phi_{11g} \\
    % \dot{\phi}_{01g}  &= -\big(\gamma_L + i\mu+ i4|\Pi|\big) \phi_{02g} ~-\nn\\
    % &\hspace{2.8cm} i\sqrt{2}\Pi\phi_{11g} - i\sqrt{2}g\phi_{01e} \\
    \dot{\phi}_{00e}  &= -\Big(i\delta_e+\frac{\gamma_e}{2} \Big)\phi_{00e} - i g^*\phi_{01g}.
\end{align}  
\end{subequations}
Many interesting properties of the nonlinear interaction may be investigated using~\eqsref{ODEs two photon terms TLE main}{ODEs one photon terms TLE main} but here we again consider the implementation of a c-phase gate. With the protocol described above, the conditions for a successful gate operation are: 1) The occupation probability of mode $b$ must equal one for both single- and two-photon inputs after the application of $\Pi(t)$. 2) The phase difference must be $\arg[\phi_{20g}(T_{\Pi})] \minus 2\arg[\phi_{10g}(T_{\Pi})]\equal \pi$, where $\Pi(t)$ is non-zero only in the interval $t\!\in\![2\Tin, T_{\Pi}]$. We numerically optimize the control function $\Pi(t)$ to fulfill these conditions. 
\begin{figure}[!h] 
\centering
   \includegraphics[angle=0,origin=c,width=8.5cm] {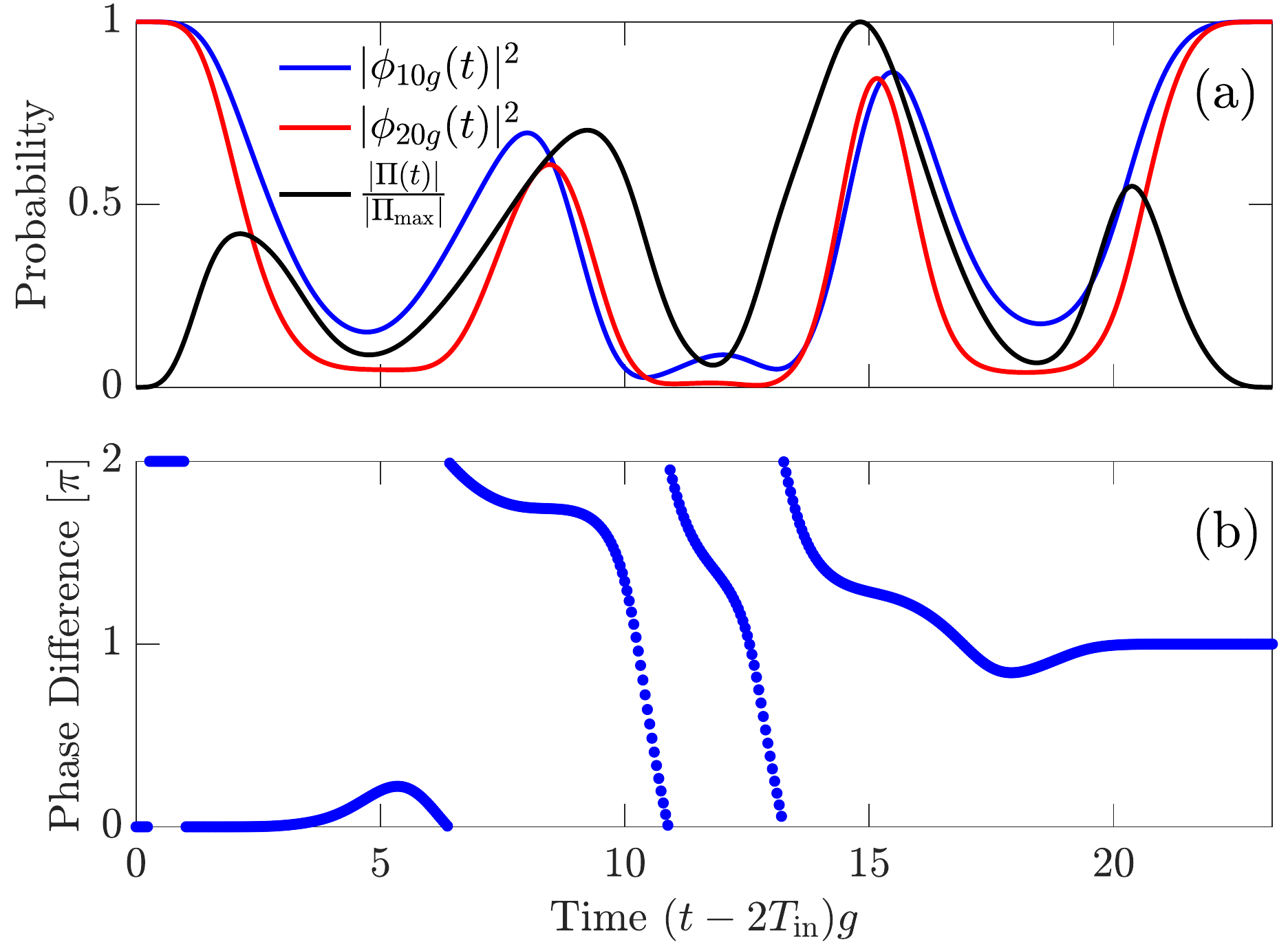}
\caption{Time evolution of the second stage of the control protocol. (a) Probability that all incoming photons occupy mode $b$ for one- (blue) and two-photon (red) inputs. The control function is also plotted (scaled to a maximum of 1). (b) Phase difference $\arg[\phi_{20g}(t\minus 2\Tin)] \minus 2\arg[\phi_{10g}(t\minus 2\Tin)]$ as a function of time.  Parameters: $\gamma_L\equal 0$.}
\figlab{example TLE}
\end{figure} 
An example of the resulting dynamics is shown in~\figref{example TLE}. It shows how the conditions above may be met using a control function plotted in~\figref{example TLE}a.\\  

Here, we considered the host crystal containing the TLE to be a third-order nonlinear material. Many types of TLEs are sensitive to electric fields, which could become problematic if the control field originated from and applied RF field. The optical control fields would not interact with the TLE as they would be very far off-resonant. However, it would be interesting to consider the TLE coupled to mode $b$ and whether an RF control field, $\Lambda(t)$, would be strong enough to effectively detune the TLE and mode $b$ during absorption and emission via an AC Stark shift of the TLE transition energy. This would reduce the effective nonlinear coupling between the photons during absorption and emission and could potentially eliminate the need for mode $c$ and increase the gate operation speed.

An alternative protocol would still use optical control fields for $\Lambda(t)$, to load the photons into mode $b$. The TLE would be coupled to mode $b$, but its transition energy, $\omega_e(t)$, would be controllable via an electrical control field that again tunes the TLE in- and out of resonance with mode $b$ via the AC Stark shifts. During the absorption and emission, the detuning would be large to eliminate any nonlinear interaction, while a similar numerical optimization technique could be used to determine the temporal shape of the electrical control field to implement the c-phase gate.

Note that the three-stage control protocol avoids any error due to nonlinear interactions between the photons during absorption and emission. The fidelity of a c-phase gate with a TLE nonlinearity is therefore only limited by loss when no decoherence mechanisms are included in the model. A similar extension of the control protocol could be applied to the case of second-order nonlinearities by introducing a fourth mode, $b'$, coupled to mode $c$ via SHG. A second control field, $\Pi(t)$, coupling modes $b$ and $b'$ would then effectively turn on the nonlinearity after the photons were coupled into mode $b$.

\section{Discussion} \seclab{discussion}
%----------------------------------------------------------------
Our simulation results illustrate that, within the limitations of our model, it is possible to absorb and emit photons with Gaussian wave packets into- and out of a dynamically coupled cavity. We also show that high fidelity c-phase gates may be implemented using such structures with three different types of nonlinearity. These fidelities were obtained while excluding certain sources of error from our analysis including noise-photons being injected from the loss channel at finite temperatures and decoherence of the TLE. 

We analyzed the interaction with two-level emitters in the context of two-qubit gates, but we expect dynamically coupled cavities to provide performance improvements in other applications as well. For instance, perfect state transfer between photonic qubits and solid-state matter qubits has been proposed using classical control fields coupling the energy levels of the matter qubit~\cite{Cirac1996}. There is a strong analogy between that method and dynamically coupled cavities, however, we expect it to be easier to engineer the photonic- rather than the atomic  degrees of freedom in practical implementations.

\begin{acknowledgments}
The authors thank Joshua Combes for many useful discussions. M.H acknowledges funding from MITRE Corporation.
\end{acknowledgments}

\onecolumngrid 
\appendix

%%%%%%%%%%%%%%%%%%%%%%%%%%%%%%%%%%%%%%%%%%%%%%%%%%%%%%%%%%%%%%%%%%%%%%%%%%
\section{Rotating Frame \applab{rotating frame}}
The Hamiltonian of the three cavity modes, four pump fields, and the TLE is $\hat{H}$, where
\begin{multline}\eqlab{Hamiltonian original frame app}
    \frac{\hat{H}}{\hbar} = \omega_a \hat{a}^\dagger\hat{a} +  \omega_b \hat{b}^\dagger\hat{b}  +  \omega_c \hat{c}^\dagger\hat{c} + \omega_p \hat{p}^\dagger\hat{p} + \omega_1 \hat{p}_1^\dagger\hat{p}_1 + \omega_2 \hat{p}_2^\dagger\hat{p}_2 + \omega_3 \hat{p}_3^\dagger\hat{p}_3 + \omega_e \hat{\sigma}_z +
    i \sqrt{\frac{\gamma}{\Delta t}} \Big( \hat{a}^\dagger \hat{w}_n - \hat{a}\hat{w}_n^\dagger \Big)
      ~+\\ \omega_w\!\sum_{k=1}^N \hat{w}_k^\dagger\hat{w}_k +\chi_2 \Big(\hat{p}^\dagger \hat{a}^\dagger\hat{b} + \hat{p}\hat{b}^\dagger\hat{a}\Big) +
    \chi_3 \Big(\hat{p}_1^\dagger\hat{p}_2 \hat{a}^\dagger\hat{b} + \hat{p}_2^\dagger\hat{p}_1 \hat{b}^\dagger\hat{a}  \Big) + \chi_3 \Big(\hat{p}_1^\dagger\hat{p}_3 \hat{b}^\dagger\hat{c} + \hat{p}_3^\dagger\hat{p}_1 \hat{c}^\dagger\hat{b}  \Big).
\end{multline} 
(Since we wish merely to provide an example, we have left out the cross-phase modulation, self-phase modulation, and second harmonic generation from the Hamiltonian.) We wish to move into the interaction picture, placing the evolution generated by the Hamiltonian $H_0$ into the operators, where  
\begin{multline}\eqlab{rotating frame unitary}
    \frac{H_0}{\hbar} = \omega_{w}\hat{a}^\dagger\hat{a} + (\omega_{b}\minus\delta_b)\hat{b}^\dagger\hat{b}  + (\omega_{c}\minus\delta_c)\hat{c}^\dagger\hat{c}  + \omega_{c}\hat{\sigma}_z + \omega_p \hat{p}^\dagger\hat{p} + \omega_1 \hat{p}_1^\dagger\hat{p}_1 +
    \omega_2 \hat{p}_2^\dagger\hat{p}_2 +\omega_3 \hat{p}_3^\dagger\hat{p}_3 +\omega_w\!\sum_{k=1}^N \hat{w}_k^\dagger\hat{w}_k . 
\end{multline} 
Under this Hamiltonian the evolution of the operators is obtained merely by multiplying them by time-dependent exponentials. Denoting the interaction-picture operators by upper-case letters, we have $\hat{A} = \hat{a} e^{-i\omega_w t}$, $\hat{B}  = \hat{b} e^{-i(\omega_{b}\minus\delta_b) t}$, $\hat{C}  = \hat{c} e^{-i(\omega_{c}\minus\delta_c) t}$, $\hat{\Sigma}_z  = \hat{\sigma}_z e^{-i\omega_{e} t}$, $\hat{P} = \hat{p} e^{-i\omega_p t}$, $\hat{P}_j = \hat{p}_j e^{-i\omega_j t}$ ($j=1,2,3$), $\hat{W}_k = \hat{w}_k e^{-i\omega_w t_k}$. Since we have removed this ``rotating" evolution from the state of the system, we refer to the interaction picture as being in a ``rotating frame". 

The evolution of the state of the system is now given by an effective interaction Hamiltonian, usually referred to as the ``interaction Hamiltonian in the interaction picture", which is given by 
% % 
% \begin{multline}\eqlab{rotating frame unitary}
%     \hat{U}_{\rm{rot}} = \exp\!\Big[ -i\omega_{w}t\hat{A}^\dagger\hat{A} -i(\omega_{b}\minus\delta_b)t\hat{B}^\dagger\hat{B}  -i(\omega_{c}\minus\delta_c)t\hat{C}^\dagger\hat{C}  -i\omega_{c}t\hat{\Sigma}_z -i\omega_p t\hat{P}^\dagger\hat{P} -i\omega_1 t\hat{P}_1^\dagger\hat{P}_1 ~-\\
%     i\omega_2 t\hat{P}_2^\dagger\hat{P}_2 -i\omega_3 t\hat{P}_3^\dagger\hat{P}_3 -i\omega_w t\!\sum_{k=1}^N \hat{W}_k^\dagger\hat{W}_k \Big] 
% \end{multline} 
% %
% such that the Hamiltonian in the rotating frame is~\cite{Kok_book}
% 
\begin{align}\eqlab{Hamiltonian rotating frame definition}
    \hat{H}_{\rm{I}}(t) = \hat{U}(\hat{H} - \hat{H}_0)\hat{U}^\dagger %+ i \hbar \hat{U} \frac{\partial \hat{U}^\dagger}{\partial t} 
\end{align} 
in which $\hat{U} = e^{-i\hat{H}_0t/\hbar}$. Since the right-hand side of the above equation is merely the Hamiltonian $\hat{H} - \hat{H_0}$ evolved in the interaction picture, we obtain $\hat{H}_{\rm{I}}(t)$ merely by replacing the Schr\"{o}dinger picture operators in $\hat{H} - \hat{H_0}$ with their interaction picture counterparts given above. While in general $\hat{H}_{\rm{I}}(t)$ will be time-dependent, if we choose the detuning parameters, $\delta_a$ through $\delta_{e}$, to account for the detunings between the various modes and the TLE, we obtain a time-independent interaction picture Hamiltonian, namely 
\begin{multline}\eqlab{Hamiltonian rotating frame app}
    \hat{H}_{\rm{rot}} = \hbar\delta_a \hat{a}^\dagger\hat{a} +  \hbar\delta_b \hat{b}^\dagger\hat{b}  +  \hbar\delta_c \hat{c}^\dagger\hat{c} + \hbar\delta_e \hat{\sigma}_z + i\hbar \sqrt{\frac{\gamma}{\Delta t}} \Big( \hat{a}^\dagger \hat{w}_n - \hat{a}\hat{w}_n^\dagger \Big) + \hbar\chi_2 \Big(\hat{p}^\dagger \hat{a}^\dagger\hat{b} + \hat{p}\hat{b}^\dagger\hat{a}\Big) ~+\\
    \hbar\chi_3 \Big(\hat{p}_1^\dagger\hat{p}_2 \hat{a}^\dagger\hat{b} + \hat{p}_2^\dagger\hat{p}_1 \hat{b}^\dagger\hat{a}  \Big) + \hbar\chi_3 \Big(\hat{p}_1^\dagger\hat{p}_3 \hat{b}^\dagger\hat{c} + \hat{p}_3^\dagger\hat{p}_1 \hat{c}^\dagger\hat{b}  \Big) . 
\end{multline} 
For the scenario in which the non-linearity is provided by the TLE, the various detunings are chosen to 
satisfy 
% 
% \begin{subequations}\eqlab{detunings TLE}
% \begin{align}
%     \qquad \delta_a &\equiv \omega_a - \omega_w \\
%      \delta_b &\equiv \delta_\Lambda + \delta_a \\
%      \delta_c &\equiv \delta_\Pi + \delta_b \\
%      \delta_e &\equiv \omega_e - \omega_c \\
%      \delta_\Lambda &\equiv (\omega_2 - \omega_1) - (\omega_a-\omega_b) \\
%      \delta_\Pi &\equiv (\omega_3 - \omega_1) - (\omega_b-\omega_c).
% \end{align} 
% \end{subequations}
\begin{align}
\left.\begin{array}{rcl}
         \qquad \delta_a &\equiv& \omega_a - \omega_w \\
     \delta_b &\equiv& \delta_\Lambda + \delta_a \\
     \delta_c &\equiv& \delta_\Pi + \delta_b \\
     \delta_e &\equiv& \omega_e - \omega_c \\
     \delta_\Lambda &\equiv& (\omega_2 - \omega_1) - (\omega_a-\omega_b) \\
     \delta_\Pi &\equiv& (\omega_3 - \omega_1) - (\omega_b-\omega_c)
\end{array} \right\} \;\;\; \mbox{TLE nonlinearity .} \eqlab{detunings TLE}
\end{align} 
Here we have chosen $\delta_b$ to remove the oscillating exponential factor in the FWM term corresponding to the control field $\Lambda(t)$:
\begin{align}\eqlab{rotating frame FWM Lambda term}
     \hat{P}_2^\dagger\hat{P}_1 \hat{B}^\dagger\hat{A} &= \hat{p}_2^\dagger\hat{p}_1 \hat{b}^\dagger\hat{a}\exp\big[ (\omega_2 - \omega_1) + \omega_b - \delta_b - (\omega_a-\delta_a) \big]~\Rightarrow \nn\\
     \delta_b & = (\omega_2 - \omega_1) -(\omega_a - \omega_b) + \delta_a \equiv \delta_\Lambda + \delta_a,  
\end{align} 
where we have defined $\delta_\Lambda$, which describes energy mismatch in the FWM process that couples modes $a$ and $b$. Similarly, we choose $\delta_c$ to remove any exponential factor on the FWM term corresponding to the control field $\Pi(t)$ 
\begin{align}\eqlab{rotating frame FWM Pi term}
     \hat{P}_3^\dagger\hat{P}_1 \hat{C}^\dagger\hat{B} &= \hat{p}_3^\dagger\hat{p}_1 \hat{c}^\dagger\hat{b}\exp\big[ (\omega_3 - \omega_1) - (\omega_b - \delta_b) + (\omega_c-\delta_c) \big]~\Rightarrow \nn\\
     \delta_c & = (\omega_3 - \omega_1) - (\omega_b - \omega_c) + \delta_b \equiv \delta_\Pi + \delta_b,  
\end{align} 
where we defined $\delta_\Pi$, which describes energy mismatch in the FWM process that couples modes $b$ and $c$.

In a $\chi^{(2)}$ material, where there is no control field $\Pi(t)$, we instead define the detunings as
% 
% \begin{subequations}\eqlab{detunings chi-2}
% \begin{align}
%      \chi^{(2)} ~\text{material} \qquad  \delta_a &\equiv \omega_a - \omega_w \\
%      \delta_b &\equiv \delta_\Lambda + \delta_a \\
%      \delta_c &\equiv \omega_c  - 2\omega_b \\
%      \delta_\Lambda &\equiv \omega_p - (\omega_b-\omega_a),
% \end{align} 
% \end{subequations}
\begin{align}
\left. \begin{array}{rcl}
       \delta_a &\equiv& \omega_a - \omega_w \\
     \delta_b &\equiv& \delta_\Lambda + \delta_a \\
     \delta_c &\equiv& \omega_c  - 2\omega_b \\
     \delta_\Lambda &\equiv& \omega_p - (\omega_b-\omega_a) 
\end{array} \right\} \;\;\; \mbox{$\chi^{(2)}$ material}, 
     \eqlab{detunings chi-2}
\end{align} 
where $\delta_c$ now describes energy mismatch in the second harmonic generation process.

% \input{./app_nonlinear_coupling_coefficients}

%%%%%%%%%%%%%%%%%%%%%%%%%%%%%%%%%%%%%%%%%%%%%%%%%%%%%%%%%%%%%%%%%%%%%%%%%%
\section{Dynamics with One Cavity Mode and One Input Photon\applab{one photon dynamics one cavity mode}}
%%%%%%%%%%%%%%%%%%%%%%%%%%%%%%%%%%%%%%%%%%%%%%%%%%%%%%%%%%%%%%%%%%%%%%%%%%
Before the dynamics begins, the state is
\begin{align}\eqlab{psi_0 one mode one photon}
    \ket{\psi_0}=  \sum_{k=1}^N \xi^{\rm{in}}_k  \sqrt{\Delta t} \ket{0} \ket{1_k}, 
\end{align} 
where $\ket{0} \ket{1_k}$ is the state with one photon in bin $k$ and no photons in the system. The state after each time step is found using the time evolution operator
\begin{align}\eqlab{time step evolution}
   \ket{\psi_{n+1}} = \hat{U}_{n+1} \ket{\psi_n}.
\end{align}
After the first time step, the state is therefore
\begin{align}\eqlab{psi_1 one mode one photon}
    \ket{\psi_1}=  \sum_{k=1}^N \xi^{\rm{in}}_k  \sqrt{\Delta t} \ket{0} \ket{1_k} + \sqrt{\gamma}\xi^{\rm{in}}_1  \Delta t \ket{1} \ket{\emptyset} \equiv \sum_{k=1}^N \xi^{\rm{in}}_k  \sqrt{\Delta t} \ket{0} \ket{1_k} + \psi_1(1) \ket{1} \ket{\emptyset},
\end{align} 
where the second term correspond to a photon in bin 1 being absorbed into the cavity mode. After the second step, the state is
\begin{align}\eqlab{psi_2 one mode one photon a}
    \ket{\psi_2}=  \sum_{k=1}^N \xi^{\rm{in}}_k  \sqrt{\Delta t} \ket{0} \ket{1_k} -\sqrt{\gamma}\psi_1(1)\sqrt{\Delta t}\ket{0}\ket{\mathbf{1}_2} + \Big[ \Big(1 -i\delta_a\Delta t - \frac{\Gamma}{2}\Delta t \Big)\psi_1(1) + \sqrt{\gamma}\xi^{\rm{in}}_2\Delta t \Big]\ket{1} \ket{\emptyset},
\end{align} 
where the second term corresponds to a photon being emitted by the cavity into bin 2 on the output side (note that we use boldface notation to distinguish input from output photons). The third term contains a contribution from the identity operator, a decay term, as well as a feeding term corresponding to absorption of a photon from the waveguide in bin 2. At this point, we split the sum over $k$ into $k>n$ corresponding to the photon being on the input side (see~\figref{discrete time illustration}) and $k\leq n$ corresponding to the photon being on the output side
\begin{align}\eqlab{psi_2 one mode one photon b}
    \ket{\psi_2}= \sum_{k=3}^N \xi^{\rm{in}}_k  \sqrt{\Delta t} \ket{0} \ket{1_k}  + \xi^{\rm{out}}_1 \sqrt{\Delta t}\ket{0} \ket{\mathbf{1}_1} +
    \Big[\xi^{\rm{in}}_2-\sqrt{\gamma}\psi_1(1) \Big]\sqrt{\Delta t}\ket{0} \ket{\mathbf{1}_2} +\psi_1(2) \ket{1} \ket{\emptyset}.
\end{align} 
\eqsref{psi_2 one mode one photon a}{psi_2 one mode one photon b} contain all types of states and we can use them to identify the update rules
\begin{align}\eqlab{discrete update rules one mode one photon}
    \psi_1(n+1) &= \psi_1(n) + \Big[\Big(-i\delta_a-\frac{\Gamma}{2}\Big)\psi_1(n)  + \sqrt{\gamma}\xi^{\rm{in}}_{n+1}\Big] \Delta t ~~\Rightarrow \\ \frac{\psi_1(n+1)-\psi_1(n)}{\Delta t} &= -\Big(i\delta_a + \frac{\Gamma}{2}\Big) \psi_1(n)  + \sqrt{\gamma}\xi^{\rm{in}}_{n+1} \\
    \xi^{\rm{out}}_n    &= \xi^{\rm{in}}_n - \sqrt{\gamma}\psi_1(n-1).
\end{align} 
We may now take the continuum limit, $\Delta t\rightarrow 0$, to obtain the equation of motion and input-output relation
\begin{subequations}\eqlab{eom one mode one photon}
\begin{align}
   \dot{\psi}_1(t) &= \Big(-i\delta_a -\frac{\Gamma}{2}\Big)\psi_1(t) + \sqrt{\gamma}\xi_{\rm{in}}(t) \\
   \xi_{\rm{out}}(t)    &= \xi_{\rm{in}}(t) - \sqrt{\gamma} \psi_1(t) .
\end{align}
\end{subequations}
% 

%%%%%%%%%%%%%%%%%%%%%%%%%%%%%%%%%%%%%%%%%%%%%%%%%%%%%%%%%%%%%%%%%%%%%%%%%%
\section{Dynamics with Two Cavity Modes and One Input Photon\applab{one photon dynamics two cavity modes}}
%%%%%%%%%%%%%%%%%%%%%%%%%%%%%%%%%%%%%%%%%%%%%%%%%%%%%%%%%%%%%%%%%%%%%%%%%%
Before the dynamics begins, the state is
\begin{align}\eqlab{psi_0 two modes one photon}
    \ket{\psi_0}=  \sum_{k=1}^N \xi^{\rm{in}}_k  \sqrt{\Delta t} \ket{00} \ket{1_k}, 
\end{align} 
where $\ket{00}$ is the state with no photons in either mode $a$ or $b$. After step one, the state is
\begin{align}\eqlab{psi_1 two modes one photon}
    \ket{\psi_1}=  \ket{\psi_0} + \sqrt{\gamma}\xi^{\rm{in}}_1  \Delta t \ket{10} \ket{\emptyset} \equiv \ket{\psi_0} + \psi_{10}(1) \ket{10} \ket{\emptyset},
\end{align} 
where we defined the amplitude for the state with one photon in mode $a$ and no photons in mode $b$, $\psi_{10}$. After step two, the state is
\begin{multline}\eqlab{psi_2 two modes one photon}
    \ket{\psi_2}=  \sum_{k=3}^N \xi^{\rm{in}}_k  \sqrt{\Delta t} \ket{00} \ket{1_k} -\sqrt{\gamma}\psi_{10}(1)\sqrt{\Delta t}\ket{00} \ket{\mathbf{1}_1} + \Big[\xi^{\rm{in}}(2)-\sqrt{\gamma}\psi_{10}(1) \Big]\sqrt{\Delta t}\ket{00} \ket{\mathbf{1}_2} ~+\\
    \Big[ \Big(1-i\delta_a\Delta t-\frac{\Gamma}{2}\Delta t -i2|\Lambda_2|\Delta t\Big)\psi_{10}(1) + \sqrt{\gamma}\xi^{\rm{in}}_2   \Big] \ket{10} \ket{\emptyset}  -i\Lambda_2\psi_{10}(1)\Delta t \ket{01} \ket{\emptyset} ~\equiv \\
    \sum_{k=3}^N \xi^{\rm{in}}_k  \sqrt{\Delta t} \ket{00} \ket{1_k} + \sum_{k=1}^2 \xi^{\rm{out}}_k  \sqrt{\Delta t} \ket{00} \ket{\mathbf{1}_k} + \psi_{10}\ket{10} \ket{\emptyset} + \psi_{01}\ket{01} \ket{\emptyset}.
\end{multline} 
After step three, the state is
\begin{multline}\eqlab{psi_3 two modes one photon}
    \ket{\psi_3}=  \sum_{k=4}^N \xi^{\rm{in}}_k  \sqrt{\Delta t} \ket{00} \ket{1_k} \plus \sum_{k=1}^3 \xi^{\rm{out}}_k  \sqrt{\Delta t} \ket{00} \ket{\mathbf{1}_k} \plus \Big[ \Big(1 \minus  i\delta_b\Delta t  \minus \frac{\gamma_L}{2}\Delta t\minus i2|\Lambda_3|\Delta t \Big) \psi_{01}(2) \minus i\Lambda_3\psi_{10}(2)\Delta t  \Big]\ket{01} \ket{\emptyset} ~+\\
    \Big[ \Big(1-i\delta_a\Delta t-\frac{\Gamma}{2}\Delta t - i2|\Lambda_3|\Delta t\Big)\psi_{10}(1) -i\Lambda_3^*\psi_{01}(2)\Delta t + \sqrt{\gamma}\xi^{\rm{in}}_3    \Big]\ket{10} \ket{\emptyset} .
\end{multline} 
\eqref{psi_3 two modes one photon} contains all the possible dynamics and we can use it to read off the update rules
\begin{align}\eqlab{update rules two modes one photon}
    \frac{\psi_{10}(n+1) - \psi_{10}(n)}{\Delta t} &=  \Big(-i\delta_a-\frac{\Gamma}{2}-i2|\Lambda_{n+1}| \Big)\psi_{10}(n) -i\Lambda_{n+1}^*\psi_{01}(n) + \sqrt{\gamma}\xi^{\rm{in}}_{n+1} \\
    \frac{\psi_{01}(n+1) - \psi_{01}(n)}{\Delta t} &= \Big(-i\delta_b-\frac{\gamma_L}{2}-i2|\Lambda_{n+1}| \Big)\psi_{01}(n)  -i\Lambda_{n+1}\psi_{10}(n) \\
     \xi^{\rm{out}}_n    &= \xi^{\rm{in}}_n - \sqrt{\gamma}\psi_{10}(n-1).
\end{align} 
In the continuum limit, we have the ODEs and input-output relation
\begin{subequations}\eqlab{eoms two modes one photon}
\begin{align}
    \dot{\psi}_{10}(t)  &=  -\Big(i\delta_a+\frac{\Gamma}{2}+i2|\Lambda(t)| \Big)\psi_{10}(n) -i\Lambda(t)^{\!*}\psi_{01}(t) + \sqrt{\gamma}\xi_{\rm{in}}(t) \\
    \dot{\psi}_{01}(t)  &= -\Big(i\delta_b +\frac{\gamma_L}{2} + i2|\Lambda(t)|\Big) \psi_{01}(t)  -i\Lambda(t)\psi_{10}(t) \\
     \xi_{\rm{out}}(t)  &= \xi_{\rm{in}}(t) - \sqrt{\gamma}\psi_{10}(t).
\end{align} 
\end{subequations}
%

%%%%%%%%%%%%%%%%%%%%%%%%%%%%%%%%%%%%%%%%%%%%%%%%%%%%%%%%%%%%%%%%%%%%%%%%%%
\section{Dynamics with Two Cavity Modes and Two Input Photons\applab{two photon dynamics two cavity modes}}
%%%%%%%%%%%%%%%%%%%%%%%%%%%%%%%%%%%%%%%%%%%%%%%%%%%%%%%%%%%%%%%%%%%%%%%%%%
For identical input photons, the input state is
\begin{align}\eqlab{psi_0 two modes two photons}
    \ket{\psi_0}=  \sqrt{2}\sum_{j=1}^N\sum_{k>j}^N \xi^{\rm{in}}_j \xi^{\rm{in}}_k  \Delta t \ket{00} \ket{ 1_j 1_k}. 
\end{align} 
Let us show that the state in~\eqref{psi_0 two modes two photons} is normalized. In the continuum limit, it corresponds to 
\begin{align}\eqlab{psi_0 two modes two photons continuum}
    \ket{\psi}=  \sqrt{2}\int_{0}^{T} dt_j\int_{t_j}^{T}dt_k \xi(t_j) \xi(t_k)   \ket{00} \ket{ 1_j 1_k},
\end{align} 
where we omitted the ${}^{\rm{in}}$ superscripts. Let us calculate its norm
\begin{subequations}
\begin{align}\eqlab{psi_0 two modes two photons continuum normalization}
    \expect{\psi\big|\psi} &=  2\int_{0}^{T} dt_j'\int_{t_j'}^{T}dt_k' \int_{0}^{T} dt_j\int_{t_j}^{T}dt_k \xi^*(t_{j'})  \xi(t_j) \xi^*(t_{k'})\xi(t_k)   \expect{1_j'\big| 1_j} \expect{1_k'\big| 1_k} ~\Rightarrow\\ 
    % \expect{\psi\big|\psi} &=  2\int_{0}^{T} dt_j'\int_{t_j'}^{T}dt_k' \int_{0}^{T} dt_j\int_{t_j}^{T}dt_k \xi^*(t_{j'})  \xi(t_j) \xi^*(t_{k'})\xi(t_k)   \delta(t_{j'}-t_j) \delta(t_k-t_{k'}) ~\Rightarrow\\
    \expect{\psi\big|\psi} &= 2 \int_{0}^{T} dt_j \big|\xi(t_j)\big|^2  \int_{t_j}^{T}dt_k  \big|\xi(t_j)\big|^2 = 2\int_{0}^{T} dt_j \big|\xi(t_j)\big|^2 \bigg[ \int_{0}^{T} dt_k  \big|\xi(t_k)\big|^2 -    \int_{0}^{t_j}dt_k  \big|\xi(t_k)\big|^2 \bigg] ~\Rightarrow \\
    \expect{\psi\big|\psi} &=  2\int_{0}^{T} dt_j \big|\xi(t_k)\big|^2 \bigg[ 1 - \int_{0}^{t_j}dt_k  \big|\xi(t_k)\big|^2 \bigg] = 2 - 2\int_{0}^{T} dt_j |\xi(t_j)\big|^2 \int_{0}^{t_j}dt_k  \big|\xi(t_k)\big|^2 ~\Rightarrow\\
    \expect{\psi\big|\psi} &=  2 - 2 \int_{0}^{T} dt_j \dot{\Xi}(t_j) \Xi(t_j) = 2 - 2\int_{0}^{T}dt_j \frac{d}{dt_j}\Big(\frac12 \Xi^2(t_j)\Big) = 2-\Big[\Xi(T) - \Xi(0) \Big] = 2 - (1-0) = 1,
\end{align} 
\end{subequations}
where $\dot{\Xi}(t_j)\equal |\xi(t_j)|^2$.\\

To begin with, we follow the dynamics of states with one photon in the system and one photon on the input side
\begin{align}\eqlab{psi_10ii definition}  
    \ket{\psi_n} =  \psi_{10}^{\rm{ii}}(n) \sum_{k>n}^N \xi^{\rm{in}}_k \sqrt{\Delta t} \ket{10}\ket{ 1_k} + \psi_{01}^{\rm{ii}}(n) \sum_{k>n}^N \xi^{\rm{in}}_k \sqrt{\Delta t} \ket{01}\ket{ 1_k} +\ldots 
\end{align} 
The superscript "ii" signifies that the equation of motion for $\psi_{10}^{\rm{ii}}(t)$ is driven by two photons on the input side. As in~\appsref{one photon dynamics one cavity mode}{one photon dynamics two cavity modes} we follow the evolution of these states through the first time steps in order to identify the update rules. After the first step, we have
\begin{align}\eqlab{psi_1 two modes two photons ii}  
    \ket{\psi_1} = \sqrt{2}\xi^{\rm{in}}_1\sqrt{\gamma}\Delta t \sum_{k>1}^N \xi^{\rm{in}}_k \sqrt{\Delta t} \ket{10}\ket{ 1_k} +\ldots =  \psi_{10}^{\rm{ii}}(1)\!\sum_{k>1}^N \xi^{\rm{in}}_k \sqrt{\Delta t}\ket{10}\ket{1_k} +\ldots,
\end{align} 
After step 2, we have
\begin{multline}\eqlab{psi_2 two modes two photons ii}  
    \ket{\psi_2} = \Big[\Big(1-i\delta_a\Delta t-\frac{\Gamma}{2}\Delta t - i2|\Lambda_2|\Delta t\Big)\psi_{10}^{\rm{ii}}(1) + \sqrt{2}\xi^{\rm{in}}_2\sqrt{\gamma}\Delta t \Big]\psi_{10}^{\rm{ii}}(1)\!\sum_{k>2}^N \xi^{\rm{in}}_k \sqrt{\Delta t} \ket{10}\ket{1_k} ~-\\
    i\Lambda_2 \psi_{10}^{\rm{ii}}(1)\!\sum_{k>2}^N \xi^{\rm{in}}_k \sqrt{\Delta t} \ket{01}\ket{1_k} + \ldots = \psi_{10}^{\rm{ii}}(2)\!\sum_{k>2}^N \xi^{\rm{in}}_k \sqrt{\Delta t}\ket{10}\ket{1_k} + \psi_{01}^{\rm{ii}}(2)\!\sum_{k>2}^N \xi^{\rm{in}}_k \sqrt{\Delta t}\ket{01}\ket{1_k} + \ldots
\end{multline} 
After step 3, all the possible interactions linking $\psi_{10}^{\rm{ii}}$ and $\psi_{01}^{\rm{ii}}$ are included 
\begin{multline}\eqlab{psi_3 two modes two photons ii}  
    \ket{\psi_3} =  \Big[\Big(1-i\delta_a\Delta t-\frac{\Gamma}{2}\Delta t - i2|\Lambda_3|\Delta t\Big)\psi_{10}^{\rm{ii}}(2) + \sqrt{2}\xi^{\rm{in}}_3\sqrt{\gamma}\Delta t - i\Lambda_3^* \psi_{01}^{\rm{ii}}(2)\Delta t \Big] \sum_{k>3}^N \!\xi^{\rm{in}}_k\sqrt{\Delta t} \ket{10}\ket{ 1_k}   ~+\\
     \Big[\big(1-i\delta_b\Delta t-\frac{\gamma_L}{2}\Delta t - i2|\Lambda_3|\Delta t\big)\psi_{01}^{\rm{ii}}(2) - i\Lambda_3 \psi_{10}^{\rm{ii}}(2) \Delta t\Big] \sum_{k>3}^N \!\xi^{\rm{in}}_k\sqrt{\Delta t} \ket{01}\ket{ 1_k}   + \ldots 
\end{multline} 
From~\eqref{psi_3 two modes two photons ii} we identify the equations of motion in the continuum limit
\begin{subequations}\eqlab{eoms two modes two photons ii} 
\begin{align}
    \dot{\psi}_{10}^{\rm{ii}}(t) &= -\Big(i\delta_a-\frac{\Gamma}{2} + i2|\Lambda(t)|\Big)\psi_{10}^{\rm{ii}}(t) - i\Lambda(t)^{\!*}\psi_{01}^{\rm{ii}}(t) + \sqrt{2\gamma}\xi_{\rm{in}}(t) \eqlab{eoms two modes two photons ii 10} \\
    \dot{\psi}_{01}^{\rm{ii}}(t) &= -\Big(i\delta_b-\frac{\gamma_L}{2}+ i2|\Lambda(t)|\Big)\psi_{01}^{\rm{ii}}(t) - i\Lambda(t)\psi_{10}^{\rm{ii}}(t) \eqlab{eoms two modes two photons ii 01} .
\end{align} 
\end{subequations}

Next, we consider states with two photons in the system
\begin{align}\eqlab{psi_20 psi_11 psi_02 definition}  
    \ket{\psi_n} =  \psi_{20}(n) \ket{20}\ket{\emptyset} + \psi_{11}(n) \ket{11}\ket{\emptyset} + \psi_{02}(n) \ket{02}\ket{\emptyset} +\ldots 
\end{align} 
These states first appear after step 2
\begin{align}\eqlab{psi_2 two modes two photons 20 11 02}  
    \ket{\psi_2} &= \sqrt{2\gamma}\psi_{10}^{\rm{ii}}(1) \Delta t \ket{20}\ket{\emptyset} + \sqrt{\gamma} \psi_{01}^{\rm{ii}}(1)\xi^{\rm{in}}_2 \Delta t \ket{11}\ket{\emptyset} + \ldots = \psi_{20}(2) \ket{20}\ket{\emptyset} + \psi_{11}(2) \ket{11}\ket{\emptyset} + \ldots,
\end{align} 
where the factor of $\sqrt{2}$ in the first term comes from $\hat{a}^\dagger$ acting on $\ket{1_a}$. After step 3, we have 
\begin{multline}\eqlab{psi_3 two modes two photons 20 11 02}  
    \ket{\psi_3} = \Big[ \Big(1- i2\delta_a\Delta t-\Gamma\Delta t -i4|\Lambda_3|\Delta t\Big)\psi_{20}(2)  -i\sqrt{2}\Lambda^*_3 \psi_{11}(2)\Delta t +  \psi_{10}^{\rm{ii}}(2)\xi^{\rm{in}}_3 \sqrt{2\gamma} \Delta t \Big] \ket{20}\ket{\emptyset} ~+\\
    \Big[\Big(1 \minus i(\delta_a\plus\delta_b)\Delta t\minus \frac{\Gamma\plus\gamma_L}{2}\Delta t \minus i4|\Lambda_3|\Delta t \Big)\psi_{11}(2)  \minus i\sqrt{2}\Delta t \Lambda_3\psi_{20}(2) +\sqrt{\gamma}\psi_{01}^{\rm{ii}}(2)\xi^{\rm{in}}_3 \Delta t \Big] \ket{11}\ket{\emptyset} ~-\\
    i\sqrt{2}\Lambda_3 \psi_{11}(2)\Delta t \ket{02}\ket{\emptyset} + \ldots
\end{multline} 
%
% Similarly, step 4 introduces the amplitude corresponding to both photons being in mode $A$
% % 
% \begin{align}\eqlab{psi_4 two modes two photons 20 11 02}  
%     \ket{\psi_4} &= -i\sqrt{2}\Lambda_4 \psi_{11}(3)\Delta t \ket{02}\ket{\emptyset}  + \ldots
% \end{align} 
% %
After step 4, all the dynamics describing the states with two photons in the system is present
\begin{multline}\eqlab{psi_4 two modes two photons 20 11 02}
    \ket{\psi_4} = \Big[ \Big(1-i2\delta_a\Delta t-\Gamma\Delta t -i4|\Lambda_4|\Delta t\Big)\psi_{20}(3)  -i\sqrt{2}\Lambda^*_4 \psi_{11}(3)\Delta t +  \psi_{10}^{\rm{ii}}(3)\xi^{\rm{in}}_4 \sqrt{2\gamma} \Delta t \Big] \ket{20}\ket{\emptyset} ~+\\
    \Big[ \Big(1 \minus i(\delta_a\plus\delta_b)\Delta t\plus \frac{\Gamma\plus\gamma_L}{2}\Delta t \minus  i4|\Lambda_4|\Delta t \Big)\psi_{11}(3)  \minus i\sqrt{2}\Delta t \Big( \Lambda_4\psi_{20}(3) \plus \Lambda_4^*\psi_{02}(3)\Big) \plus \psi_{01}^{\rm{ii}}(3)\xi^{\rm{in}}_4\sqrt{\gamma}\Delta t \Big] \ket{11}\ket{\emptyset} ~+\\
    \Big[\Big(1-i2\delta_b-\gamma_L\Delta t -i4|\Lambda_4|\Delta t\Big)\psi_{02}(3) -i\sqrt{2}\Lambda_4 \psi_{11}(3) \Big]\ket{02}\ket{\emptyset} + \ldots
\end{multline}   
We identify the equations of motion in the continuum limit
\begin{subequations}\eqlab{ODEs two photon terms}
\begin{align} 
    \dot{\psi}_{20}(t)  &= -\big(i2\delta_a+\Gamma+i4|\Lambda(t)|\big)\psi_{20}(t) - i\sqrt{2}\Lambda(t)^{\!*}\psi_{11}(t) + \sqrt{2\gamma}\psi_{10}^{\rm{ii}}(t)\xi_{\rm{in}}(t) \\
    \dot{\psi}_{11}(t)  &= -\Big(i(\delta_a+\delta_b)+\frac{\Gamma+\gamma_L}{2}+i4|\Lambda(t)|\Big)\psi_{11}(t) - i\sqrt{2}\Lambda(t)\psi_{20}(t) - i\sqrt{2}\Lambda(t)^{\!*}\psi_{02}(t) + \sqrt{\gamma}\psi_{01}^{\rm{ii}}(t)\xi_{\rm{in}}(t) \\
    \dot{\psi}_{02}(t)  &= -\big(i2\delta_b+\gamma_L+i4|\Lambda(t)|\big)\psi_{02}(t) - i\sqrt{2}\Lambda(t)\psi_{11}(t) . 
\end{align}  
\end{subequations}

Next, we consider states with one photon on the input- and one on the output side. There are two paths resulting in this state (see~\figref{two photons two modes paths}). One, a photon is coupled into the waveguide from the system while the other photon remains on the input side. Two, one of the two input photons passes by the system without interacting. If this occurs in bin $m$, the contribution to the state is  
\begin{align}\eqlab{psi_00 definition}  
    \ket{\psi_m} =  \Big[-\sqrt{\gamma}\psi_{10}^{\rm{ii}}(m) + \sqrt{2}\xi^{\rm{in}}_m \Big] \sum_{k>m}^N  \xi^{\rm{in}}_k \Delta t  \ket{00} \ket{1_k \mathbf{1}_m} +\ldots = \psi_{00}(m)\sum_{k>m}^N  \xi^{\rm{in}}_k \Delta t  \ket{00} \ket{1_k \mathbf{1}_m} 
\end{align} 
If the photon remaining on the input side is absorbed, it gives rise to states with one photon in the system and one on the output side
\begin{align}\eqlab{psi_10 psi_01 i definition}  
    \ket{\psi_n} =  \psi_{00}(m) \Big[ \psi_{10}^{\rm{i}}(m,n) \sqrt{\Delta t}  \ket{10} \ket{\mathbf{1}_m} + \psi_{01}^{\rm{i}}(m,n) \sqrt{\Delta t}  \ket{01} \ket{\mathbf{1}_m} \Big] + \ldots ,
\end{align} 
where we factored out $\psi_{00}(m)$ to obtain equations of motion for $\psi_{10}^{\rm{i}}(\tau, t)$ and $\psi_{01}^{\rm{i}}(\tau, t)$ that are similar to~\eqref{eom one mode one photon}. These amplitudes are functions of two times, where $\tau\equal t_m$ describes the time the dynamics was initialized by the formation of the state $\ket{1_k \mathbf{1}_m}$. The superscript "i" signifies that the equations of motion for $\psi_{10}^{\rm{i}}(\tau, t)$ and $\psi_{01}^{\rm{i}}(\tau, t)$ are driven by one photon on the input side. Let us again follow the evolution of~\eqref{psi_00 definition} for a few time steps to determine the equations of motion for $\psi_{10}^{\rm{i}}(\tau, t)$ and $\psi_{01}^{\rm{i}}(\tau, t)$. At step $n+1$, we have 
% In the second step, the absorbed photon may be re-emitted 
% % 
% \begin{align}\eqlab{psi_2 two modes two photons 00}  
%     \ket{\psi_{m+1}} = - \sqrt{\gamma}\psi_{10}^{\rm{ii}}(1)\sum_{k>2}^N  \xi^{\rm{in}}_k \Delta t  \ket{00} \ket{1_k \mathbf{1}_2}  + \ldots 
%     \equiv \psi_{00}(2,2) \sum_{k>2}^N \xi^{\rm{in}}_k \Delta t  \ket{00} \ket{1_k \mathbf{1}_2},
% \end{align} 
% %
% where boldface is used to indicate that the photon is on the output side.
% 
\begin{multline}\eqlab{psi_m+1 2 2 i}  
    \ket{\psi_{n+1}} = \Big[ \Big(1 - i\delta_a\Delta t- \frac{\Gamma}{2}\Delta t - i2|\Lambda_{n+1}|\Delta t\Big)\psi_{10}^{\rm{i}}(m,n) -i\Lambda_{n+1}^*\psi_{01}(m,n)\Delta t + \sqrt{\gamma}\xi^{\rm{in}}_{n+1}\Delta t \Big] \sqrt{\Delta t}\ket{10} \ket{\mathbf{1}_m} ~+ \\
    \Big[ \Big(1 - i\delta_b\Delta t - \frac{\gamma_L}{2}\Delta t - i2|\Lambda_{n+1}|\Delta t\Big)\psi_{01}^{\rm{i}}(m,n) -i\Lambda_{n+1}\psi_{10}(m,n)\Delta t\Big] \sqrt{\Delta t}\ket{01} \ket{\mathbf{1}_m} + \ldots ~=\\
    \psi_{00}(m)\Big[ \psi_{10}^{\rm{i}}(m,n+1)  \sqrt{\Delta t}\ket{10} \ket{\mathbf{1}_m} + \psi_{01}^{\rm{i}}(m,n+1)  \sqrt{\Delta t}\ket{01} \ket{\mathbf{1}_m}\Big] + \ldots
\end{multline} 
From~\eqref{psi_m+1 2 2 i} we obtain the equations of motion 
\begin{subequations}\eqlab{eoms two modes two photons i}
\begin{align}
    \dot{\psi}_{10}^{\rm{i}}(\tau, t)  &=  -\Big(i\delta_a+\frac{\Gamma}{2}+i2|\Lambda(t)| \Big)\psi_{10}^{\rm{i}}(\tau, t) -i\Lambda(t)^{\!*}\psi_{01}^{\rm{i}}(\tau, t) + \sqrt{\gamma}\xi_{\rm{in}}(t) \\
    \dot{\psi}_{01}^{\rm{i}}(\tau, t)  &= -\Big(i\delta_b +\frac{\gamma_L}{2} + i2|\Lambda(t)|\Big) \psi_{01}(\tau, t)  -i\Lambda(t)\psi_{10}^{\rm{i}}(\tau, t) .
\end{align} 
\end{subequations}
Comparing~\eqsref{eom one mode one photon}{eoms two modes two photons i} we see that there is an additional factor of $\sqrt{2}$ on the driving term $\sqrt{\gamma}\xi_{\rm{in}}(t)$ in~\eqref{eom one mode one photon} because it is driven by two photons as opposed to one in~\eqref{eoms two modes two photons i}. The initial condition for~\eqref{eoms two modes two photons i} is $\psi_{10}^{\rm{i}}(\tau,\tau)\equal 0$ and $\psi_{01}^{\rm{i}}(\tau,\tau)\equal 0$ because the system started out in the state $\ket{00}$ in~\eqref{psi_00 definition}.\\

Finally, we need to consider states with one photon in the system and one photon on the output side
\begin{align}\eqlab{psi_10 psi_01 o definition}  
    \ket{\psi_n} =  \psi_{10}^{\rm{o}}(m,n) \sqrt{\Delta t}  \ket{10} \ket{\mathbf{1}_m} + \psi_{01}^{\rm{o}}(m,n) \sqrt{\Delta t}  \ket{01} \ket{\mathbf{1}_m} + \ldots
\end{align} 
There are four different paths leading to this state. One (Two), a photon couples into the waveguide while the state of the system is $\ket{20}$ ($\ket{11}$). Three (Four), the photon on the input side passes by the system without interacting while the system is in the state $\ket{10}$ ($\ket{01}$). If this occurs in bin $m$, the contribution to the state is 
\begin{multline}\eqlab{psi_10 psi_01 o initial condition}
    \ket{\psi_m} = \Big[ -\sqrt{2\gamma} \psi_{20}(m)  + \psi_{10}^{\rm{ii}}(m)\xi^{\rm{in}}_m  \Big] \sqrt{\Delta t} \ket{10}\ket{\mathbf{1}_m}   + \Big[  - \sqrt{\gamma}\psi_{11}(m) + \psi_{01}^{\rm{ii}}(m)\xi^{\rm{in}}_m \Big] \sqrt{\Delta t} \ket{01}\ket{\mathbf{1}_m} +\ldots ~=\\
    \psi_{10}^{\rm{o}}(m,m) \sqrt{\Delta t} \ket{10}\ket{\mathbf{1}_m} + \psi_{01}^{\rm{o}}(m,m) \sqrt{\Delta t} \ket{01}\ket{\mathbf{1}_m} + \ldots
\end{multline} 
At time $t_{m+1}$, the state is
\begin{multline}\eqlab{psi_m+1 2 2 o} 
    \ket{\psi_{m+1}} = \Big[ \Big(1-i\delta_a\Delta t-\frac{\Gamma}{2}\Delta t - i2|\Lambda_{m+1}|\Delta t\Big)\psi_{10}^{\rm{o}}(m,m) -i\Lambda_{m+1}^* \psi_{01}^{\rm{o}}(m,m) \Delta t\Big] \sqrt{\gamma\Delta t}\ket{10}\ket{\mathbf{1}_m} ~+\\
   \Big[ \Big(1-i\delta_b\Delta t-\frac{\gamma_L}{2}\Delta t - i2|\Lambda_{m+1}|\Delta t\Big)\psi_{01}^{\rm{o}}(m,m) -i\Lambda_{m+1} \psi_{10}^{\rm{o}}(m,m) \Delta t\Big] \sqrt{\gamma\Delta t}\ket{01}\ket{\mathbf{1}_m} + \ldots
    % \ket{\psi_5} \equiv -\psi_{10}^{\rm{o}}(4,5) \sqrt{\gamma\Delta t}\ket{10}\ket{\mathbf{1}_4} -\psi_{01}^{\rm{o}}(4,5) \sqrt{\gamma\Delta t}\ket{01}\ket{\mathbf{1}_4}+ \ldots 
\end{multline}   
From~\eqref{psi_m+1 2 2 o} we identify the equations of motion 
\begin{subequations}\eqlab{eoms two modes two photons o} 
\begin{align}
    \dot{\psi}_{10}^{\rm{o}}(\tau, t) &= -\Big(i\delta_a+\frac{\Gamma}{2} + i2|\Lambda(t)|\Big)\psi_{10}^{\rm{o}}(\tau, t) - i\Lambda(t)^{\!*}\psi_{01}^{\rm{i}}(\tau, t) \eqlab{SPM 2 photons ODE psi_10^o} \\
    \dot{\psi}_{01}^{\rm{o}}(\tau, t) &= -\Big(i\delta_b+\frac{\gamma_L}{2}+ i2|\Lambda(t)|\Big)\psi_{01}^{\rm{o}}(\tau, t) - i\Lambda(t)\psi_{10}^{\rm{o}}(\tau, t). \eqlab{SPM 2 photons ODE psi_01^o} 
\end{align} 
\end{subequations}
\eqref{eoms two modes two photons o} must be solved for two sets of initial conditions corresponding to the first ($\psi_{10}^{\rm{i}}(\tau, \tau) \equal 1$ and $\psi_{01}^{\rm{i}}(\tau, \tau)\equal 0$) and second ($\psi_{10}^{\rm{i}}(\tau, \tau) \equal 0$ and $\psi_{01}^{\rm{i}}(\tau, \tau)\equal 1$) term in~\eqref{psi_10 psi_01 o definition}, respectively. We introduce functions $L_{10}(\tau,t)$, $L_{01}(\tau,t)$, $M_{10}(\tau,t)$, and $M_{01}(\tau,t)$, where $L$ correspond to $\psi^{\rm{o}}$ with the first initial condition and $M$ correspond to $\psi^{\rm{o}}$ with the second initial condition.\\

The final step is to identify all terms of the output state using~\figref{two photons two modes paths} and the derivations above. From~\eqref{psi_00 definition} we have the contributions
\begin{align}\eqlab{output two modes two photons a}
    \xi_{\rm{out}}(\tau, t)     = -\sqrt{\gamma}\psi_{10}^{\rm{i}}(\tau, t) \Big[-\sqrt{\gamma}\psi_{10}^{\rm{ii}}(\tau) + \sqrt{2}\xi_{\rm{in}}(\tau) \Big] + \ldots
\end{align} 
From~\eqref{psi_10 psi_01 o initial condition} we have the contributions
\begin{align}\eqlab{output two modes two photons b}
    \xi_{\rm{out}}(\tau, t)     = -\sqrt{\gamma}L_{10}(\tau, t) \Big[ -\sqrt{2\gamma} \psi_{20}(\tau)  + \psi_{10}^{\rm{ii}}(\tau)\xi_{\rm{in}}(\tau)  \Big] -\sqrt{\gamma}M_{10}(\tau, t)\Big[  - \sqrt{\gamma}\psi_{11}(\tau) + \psi_{01}^{\rm{ii}}(\tau)\xi_{\rm{in}}(\tau) \Big] +  \ldots
\end{align} 
The remaining contributions to the output state come from both photons passing by the system without interacting and decay from system state $\ket{10}$ followed by the second input photon passing by the system
\begin{align}\eqlab{output two modes two photons c}
    \xi_{\rm{out}}(\tau, t)     = \Big[\sqrt{2}\xi_{\rm{in}}(\tau) - \sqrt{\gamma}\psi_{10}^{\rm{ii}}(\tau)    \Big]\xi_{\rm{in}}(t) + \ldots
\end{align} 
If we define the output state as 
\begin{align}\eqlab{output state two modes two photons} 
    \ket{\psi_{\rm{out}}} \equiv \int_0^T \!\!d\tau \int_0^T  \!\!dt\xi_{\rm{out}}(\tau, t) \hat{a}^\dagger(\tau)\hat{a}^\dagger(t) \ket{\emptyset},
\end{align}
then the output wave packet is
\begin{multline}\eqlab{output wave packet} 
    \xi_{\rm{out}}(\tau, t) \equiv \xi_{\rm{in}}(\tau)\xi_{\rm{in}}(t) + \frac{1}{\sqrt{2}}\Big[ \sqrt{2}\gamma\psi_{20}(\tau) L_{10}(\tau, t) + \gamma\psi_{11}(\tau) M_{10}(\tau, t) -\sqrt{\gamma}\psi_{10}^{\rm{ii}}(\tau)\xi_{\rm{in}}(\tau)L_{10}(\tau, t)  ~-\\ \sqrt{\gamma}\psi_{01}^{\rm{ii}}(\tau)\xi_{\rm{in}}(\tau)M_{10}(\tau, t) +
    \gamma\psi_{10}^{\rm{ii}}(\tau)\psi_{10}^{\rm{i}}(\tau, t) -\sqrt{\gamma}\psi_{10}^{\rm{ii}}(\tau)\xi_{\rm{in}}(t) -\sqrt{2\gamma}\xi_{\rm{in}}(\tau)\psi_{10}^{\rm{i}}(\tau, t)  \Big], ~~\tau\leq t,
\end{multline}
and $\xi_{\rm{out}}(\tau, t) = \xi^{\rm{out}}(t, \tau)$. The factor of $1/\sqrt{2}$ comes from the integrals in~\eqref{output state two modes two photons} spanning the entire time interval, whereas the terms in Eqs. (\ref{eq:output two modes two photons a})-(\ref{eq:output two modes two photons c}) were derived using the definition in~\eqref{psi_0 two modes two photons}, where each state appears only once in the summations.\\

The probability of finding the system in a state with $n_a$ photons in mode $a$ and $n_b$ photons in mode $b$ at time $t_n$ is found from the expectation value
\begin{align} 
   P_{lm}(t_n) = \big\langle \psi_n\big| \Big( \ket{n_a n_b}\bra{n_a n_b}\otimes \hat{\mathbb{I}}_{\rm{field}} \Big) \big| \psi_n\big\rangle  = \sum_{j,k=1}^N \big| \bra{1_j 1_k}\bra{n_a n_b} \psi_n\rangle \big|^2, ~~\text{with} ~~ \hat{\mathbb{I}}_{\rm{field}} = \sum_{j,k=1}^N \ket{1_j 1_k} \bra{1_j 1_k}.
\end{align} 
It is instructive to use~\figref{two photons two modes paths} to keep track off all paths when evaluating the overlap $\langle 1_j 1_k | \langle n_a n_b | \psi_n\rangle$. For $n_a\equal n_b\equal 0$, we see that there are contributions from the two paths leading to states with one photon on the input- and one on the output side as well as contributions from both photons being on the output side. The first contribution is
\begin{multline}\eqlab{P_00 a} 
    \big|\expect{00|00}\big|^2 \!\sum_{j',k'=1}^N \sum_{m=1}^n \big| \expect{1_{j'} 1_{k'} | \psi_m } \big|^2=  \!\!\sum_{j',k'=1}^N \sum_{k>n}^N \sum_{m=1}^n \Delta t\Big| \xi^{\rm{in}}_k \big[ \sqrt{2}\xi^{\rm{in}}_m -\sqrt{\gamma}\psi_{10}^{\rm{ii}}(m)  \big]\Big|^2 \big|\expect{1_{j'} 1_{k'} | 1_k \mathbf{1}_m} \big|^2 ~=\\
    \sum_{k>n}^N \sum_{m=1}^n \Delta t\Big| \xi^{\rm{in}}_k \big[ \sqrt{2}\xi^{\rm{in}}_m -\sqrt{\gamma}\psi_{10}^{\rm{ii}}(m)  \big]\Big|^2. 
\end{multline}
The state $\ket{\psi_m}$ is from~\eqref{psi_00 definition} and the summation over $m$ is included since the photon on the output side could have made it there in any bin prior to $t_n$. Similarly, the contribution from the output state is
\begin{align}\eqlab{P_00 b} 
    \big|\expect{00|00}\big|^2 \!\sum_{j',k'=1}^N \sum_{m'=1}^n \sum_{m=1}^n \!\Delta t\big| \xi^{\rm{out}}_{m'm}\big|^2 \big|\expect{1_{j'} 1_{k'} | \mathbf{1}_{m'} \mathbf{1}_m} \big|^2  =\sum_{m'=1}^n \sum_{m=1}^n \!\Delta t\big| \xi^{\rm{out}}_{m'm}\big|^2 .
\end{align}
Adding the contributions from~\eqsref{P_00 a}{P_00 b} and taking the continuum limit, we get
\begin{align}\eqlab{P_00 c} 
    P_{00}(t_n) = \int_{t_n} ^T \!\!|\xi^{\rm{in}}(s)|^2 ds \!\int_0^{t_n} \!\Big|\sqrt{2}\xi_{\rm{in}}(\tau) - \gamma\psi_{10}^{\rm{ii}}(\tau)\Big|^2\!d\tau  + \int_0^{t_n}\!\!\int_0^{t_n}|\xi_{\rm{out}}(\tau, s)|^2 dsd\tau .
\end{align}
There are 7 different paths leading to the system state $\ket{10}$ and the probability is
\begin{multline}\eqlab{P_10} 
    P_{10}(t_n) = |\psi_{10}^{\rm{ii}}(t_n)|^2 \!\int_{t_n}^T \!\!|\xi^{\rm{in}}(s)|^2 ds  +
    \int_0^{t_n} \!\Big|\psi_{10}^{\rm{ii}}(\tau)\xi_{\rm{in}}(\tau)L_{10}(\tau, t_n) + \psi_{01}^{\rm{ii}}(\tau)\xi_{\rm{in}}(\tau)M_{10}(\tau, t_n)  ~-\\
    \sqrt{2\gamma}\psi_{20}(\tau)L_{10}(\tau, t_n) - \sqrt{\gamma}\psi_{11}(\tau)M_{10}(\tau, t_n) -
    \sqrt{\gamma}\psi_{10}^{\rm{ii}}(\tau)\psi_{10}^{\rm{i}}(\tau, t_n) + \sqrt{2}\xi_{\rm{in}}(\tau)\psi_{10}^{\rm{i}}(\tau, t_n) \Big|^2 d\tau .
\end{multline}
Similarly, the probability of the system state $\ket{01}$ is
\begin{multline}\eqlab{P_01} 
    P_{01}(t_n) = |\psi_{01}^{\rm{ii}}(t_n)|^2 \!\int_{t_n}^T \!\!|\xi^{\rm{in}}(s)|^2 ds  +
    \int_0^{t_n} \!\Big|\psi_{10}^{\rm{ii}}(\tau)\xi_{\rm{in}}(\tau)L_{01}(\tau, t_n) + \psi_{01}^{\rm{ii}}(\tau)\xi_{\rm{in}}(\tau)M_{01}(\tau, t_n)  ~-\\
    \sqrt{2\gamma}\psi_{20}(\tau)L_{01}(\tau, t_n) - \sqrt{\gamma}\psi_{11}(\tau)M_{01}(\tau, t_n) -
    \sqrt{\gamma}\psi_{10}^{\rm{ii}}(\tau)\psi_{01}^{\rm{i}}(\tau, t_n) + \sqrt{2}\xi_{\rm{in}}(\tau)\psi_{01}^{\rm{i}}(\tau, t_n) \Big|^2 d\tau .
\end{multline}
The probability distributions for states with two photons in the system are simply
\begin{align}\eqlab{P_20 P_11 P_02} 
    P_{20}(t_n) =  \big|\psi_{20}(t_n)\big|^2, ~~P_{11}(t_n) =  \big|\psi_{11}(t_n)\big|^2, ~~P_{02}(t_n) =  \big|\psi_{02}(t)\big|^2 .
\end{align}
\section{Absorption of Photon Wavepacket}\applab{app absorption}
%%%%%%%%%%%%%%%%%%%%%%%%%%%%%%%%%%%%%%%%%%%%%%%%%%%%%%%%%%%%%%%%%%%%%%%%%%%%%%%%%%%%%%%%
We write the driving function as $\Lambda(t)\equiv |\Lambda(t)|\exp[i\phi(t)]$ and our goal is to determine the amplitude, $|\Lambda(t)|$, and phase, $\phi(t)$, such that an incoming photon in the wave packet $\xi_{\rm{in}}(t)$ is fully absorbed into mode $b$. The equations of motion are written in~\eqref{eoms two modes one photon}, but we repeat them here for easy reference
\begin{subequations}\eqlab{eoms two modes one photon absorption}
\begin{align}
    \dot{\psi}_{10}(t)  &=  -\Big(i\delta_a+\frac{\Gamma}{2}+i2|\Lambda(t)| \Big)\psi_{10}(t) -i|\Lambda(t)|e^{-i\phi(t)}\psi_{01}(t) + \sqrt{\gamma}\xi(t) \eqlab{EOM cp as}\\
    \dot{\psi}_{01}(t)  &= -\Big(i\delta_b+\frac{\gamma_L}{2}+i2|\Lambda(t)|\Big)\psi_{01}(t)  -i|\Lambda(t)|e^{i\phi(t)}\psi_{10}(t) \eqlab{EOM cp ai+}\\
     \xi^{\rm{out}}(t)  &= \xi(t) - \sqrt{\gamma}\psi_{10}(t).
\end{align} 
\end{subequations}
Note that we have omitted the subscript of $\xi_{\rm{in}}(t)$ in~\eqref{eoms two modes one photon absorption} for notational convenience.
% % 
% \begin{subequations}\eqlab{eom cross phase}
% \begin{align}
% 	\dot{\psi}_{10}(t)	    &= \bigg(- \frac{\Gamma}{2} -i2 |\Lambda|\bigg)\psi_{10} - i |\Lambda|e^{-i\phi}\psi_{01}  + \sqrt{\gamma}\xi^{\rm{in}} \eqlab{EOM cp as}\\
% 	\dot{\psi}_{01}(t) 	&= \bigg(-\frac{\gamsub{L}}{2} - i2 |\Lambda|\bigg)\psi_{01} - i |\Lambda|e^{i\phi} \psi_{10} \eqlab{EOM cp ai+}\\	
%     \xi^{\rm{out}}    &= \xi^{\rm{in}} - \sqrt{\gamma}\psi_{10},
% \end{align} 
% \end{subequations}
% % 
Absorbing the incoming pulse implies $\xi_{\rm{out}}\equal 0$ and therefore $\psi_{10}\equal \xi_{\rm{in}}/\sqrt{\gamma}$. Substituting this into~\eqref{EOM cp ai+} and re-arranging terms yields
\begin{align}\eqlab{EOM cp ai+ 2}
	\frac{d}{dt} \Big(\psi_{01}(t) e^{-Q(t)} \Big)e^{Q(t)}	= \frac{-i}{\sqrt{\gamma}} |\Lambda(t)| e^{i\phi(t)} \xi(t) ~\Rightarrow ~~ \psi_{01}(t)	= \frac{-i}{\sqrt{\gamma}} e^{Q(t)}  \int_0^t\! e^{-Q(s)}  |\Lambda(s)|e^{i\phi(s)} \xi(s) ds ,
\end{align} 
where we defined the functions
\begin{align}\eqlab{P definition}
	Q(t) 	= -iP(t) - i\Big(\delta_b + \frac{\gamsub{L}}{2}\Big) t , \;\;\;  P(t)    = 2\!\int_0^t \!|\Lambda(s)|ds.
\end{align} 
Substituting $\psi_{10}\equal \xi/\sqrt{\gamma}$ into~\eqref{EOM cp as} yields
\begin{align}\eqlab{EOM cp as 2a}
	 \frac{(\gamma-\gamsub{L})}{2}\xi(t) - \dot{\xi}(t) - i\big(\delta_a + 2 |\Lambda(t)|\big)\xi(t) 	&=  i |\Lambda(t)| e^{-i\phi(t)}\sqrt{\gamma}\psi_{01}(t).
\end{align} 
Multiplying~\eqref{EOM cp as 2a} by $\xi(t)^{\!*}\!\exp(\gamsub{L} t)$ and defining real functions $f_i$ and $g_i$, we find
\begin{align}\eqlab{EOM cp as 3}
	f_i(t) + i g_i(t)	&=   |\Lambda(t)|e^{-i\phi(t)}\xi(t)^{\!*} e^{(-i\delta_b+\frac{\gamsub{L}}{2})t}e^{-iP(t)}  \int_0^t\! e^{(i\delta_b+ \frac{\gamsub{L}}{2})s}e^{iP(s)}  |\Lambda(s)|e^{i\phi(s)} \xi(s)ds,
\end{align} 
with 
\begin{subequations}\eqlab{fi gi definition}
\begin{align}
	 f_i(t) &= \Big(\frac{\gamma-\gamsub{L}}{2} \xi(t) - \dot{\xi}(t)\Big)\xi(t)^{\!*} e^{\gamsub{L}t} \eqlab{fi definition}\\
     g_i(t) &= -\big(\delta_a + 2 |\Lambda(t)|\big) |\xi(t)|^2 e^{\gamsub{L}t}. \eqlab{gi definition}
\end{align} 
\end{subequations}
Note that~\eqref{fi definition} assumes an input wavepacket without chirp, $\frac{d}{dt}[\arg{\xi(t)}]\equal 0$. The RHS of~\eqref{EOM cp as 3} can be written as
\begin{align}\eqlab{complex product}
	 \big[x(t)-iy(t)\big] \int_0^t\! \big[x(s)+iy(s)\big]ds =  x(t)\int_0^t\!x(s)ds + y(t)\int_0^t\! y(s)ds + i\Big(x(t) \int_0^t\! y(s)ds - y(t)\int_0^t\! x(s)ds\Big),
\end{align} 
where 
\begin{subequations}\eqlab{x and y chi-3}
\begin{align}
	 x(t)   &=   |\Lambda(t)||\xi(t)| \exp(\gamsub{L}t/2) \cos\!\big[\phi(t)+\delta_b t +P(t) +\arg(\xi)\big]\\
	 y(t)   &=  |\Lambda(t)||\xi(t)| \exp(\gamsub{L}t/2)\sin\!\big[\phi(t)+\delta_b t+P(t)+\arg(\xi)\big].
\end{align} 
\end{subequations}
By defining the functions
\begin{align}\eqlab{aux functions}
	 X(t)=\int_0^t\!x(s)ds = R(t)\cos\!\big[\theta(t)\big], ~ Y(t)=\int_0^t\!y(s)ds = R(t)\sin\!\big[\theta(t)\big],
\end{align} 
\eqref{EOM cp as 3} can be split into real and imaginary parts
\begin{align}\eqlab{EOM cp as 4a}
	 f_i = \dot{X}X + \dot{Y}Y, \qquad 	 g_i = \dot{X}Y - \dot{Y}X.
\end{align} 
Using the definition in~\eqref{aux functions}, we have
\begin{multline}\eqlab{RHS real}
	 f_i = \dot{X}X + \dot{Y}Y = \big[\dot{R}\cos(\theta) - R\sin(\theta)\dot{\theta}\big]R\cos(\theta) + \big[\dot{R}\sin(\theta) + R\cos(\theta)\dot{\theta} \big]R\sin(\theta) = \dot{R}R = \frac12 \frac{d}{dt}\Big( R^2\Big),     
\end{multline} 
which has the solution
\begin{align}\eqlab{R sol}
	 R(t) = \sqrt{ 2\int_0^t\! f_i(s)ds}.
\end{align} 
Similarly, 
\begin{align}\eqlab{RHS imag}
	 g_i = \dot{X}Y - \dot{Y}X = \big[\dot{R}\cos(\theta) - R\sin(\theta)\dot{\theta}\big]R\sin(\theta) - \big[\dot{R}\sin(\theta) + R\cos(\theta)\dot{\theta} \big]R\cos(\theta) = -R^2\dot{\theta}.
\end{align} 
Using the result in~\eqref{R sol}, the solution for $\theta$ is 
\begin{align}\eqlab{theta sol}
	 \theta(t) = -\frac12 \int_0^t\!\! \frac{g_i(s)}{ \int_0^s\! f_i(z)dz}ds.
\end{align} 
To find the solution for $|\Lambda(t)|$ we evaluate $x^2+y^2\equal |\Lambda|^2 |\xi|^2 \exp(\gamsub{L} t)$ using the results above
\begin{multline}\eqlab{absolute value absorption}
 |\Lambda|^2 |\xi|^2e^{\gamma_L t} =  \dot{X}^2 + \dot{Y}^2 = \big[\dot{R}\cos(\theta) - R\sin(\theta)\dot{\theta}\big]^2 + \big[\dot{R}\sin(\theta) + R\cos(\theta)\dot{\theta} \big]^2 = \dot{R}^2 + R^2\dot{\theta}^2 = \frac{1}{2\int\!f_i} \big( g_i^2 + f_i^2\big) .    
\end{multline} 
Inserting the definition of $g_i$ from~\eqref{gi definition} yields
\begin{multline}\eqlab{p sol} 
	 |\Lambda|^2 |\xi^2| \exp(\gamsub{L} t) = \frac{1}{2\mathcal{F}_i} \Big[ \big(\delta_a + 2 |\Lambda|\big)^2 \exp(2\gamsub{L} t)|\xi|^4 + f_i^2\Big] ~\Rightarrow \\ 
     |\Lambda(t)| = \frac{2\delta_a\xi^4 e^{\gamma_L t} \pm \sqrt{2} e^{-\frac{\gamma_L}{2} t}|\xi| \sqrt{f_i^2\big( \mathcal{F}_i  - 2 \xi^2e^{\gamma_L t} \big) + \delta_a^2\xi^4 \mathcal{F}_i e^{2\gamma_L t} }  }{2 \xi^2\big[ \mathcal{F}_i - 2\xi^2 e^{\gamma_L t}   \big]},
\end{multline} 
where $\mathcal{F}_i(t)$ is the anti-derivative of $f_i(t)$. If $\delta_a\equal 0$, the solution is
\begin{align}\eqlab{LAM delta_a = 0}
	 |\Lambda(t)| = \frac{|f_i(t)| e^{-\gamsub{L}t/2}}{\sqrt{2} |\xi(t)|}\frac{1}{\sqrt{ \mathcal{F}_i  - 2|\xi(t)|^2e^{\gamsub{L}t} } }.
\end{align}
Knowing $|\Lambda(t)|$ means $g_i$ is a known function and $x$ and $y$ may be evaluated using $\theta$ from~\eqref{theta sol}. Then, the phase $\phi$ is
\begin{align}\eqlab{phi sol}
	 \phi(t) &= -\delta_b t - 2\int_0^t \!|\Lambda(s)|ds - \arg(\xi) + \tan^{-1}\!\bigg( \frac{y(t)}{x(t)}\bigg) .
\end{align}
% 
% To obtain $x$ note that 
% % 
% \begin{align} 
% 	x & = \dot{X} = \dot{R}\cos(\theta) - R\sin(\theta)\dot\theta = \frac{f_i \cos\theta + g_i \sin\theta}{\sqrt{2 \int f_i}}  .
% \end{align} 
% %
To obtain $x$ and $y$, note that 
\begin{align} 
	x & = \dot{X} = \dot{R}\cos(\theta) - R\sin(\theta)\dot\theta = \frac{f_i \cos(\theta) + g_i \sin(\theta)}{\sqrt{ 2\int\!f_i}} \\ 
	y & = \dot{Y} = \dot{R}\sin(\theta) + R\cos(\theta)\dot\theta = \frac{f_i \sin(\theta) - g_i \cos(\theta)}{\sqrt{ 2\int\!f_i}}.
\end{align} 
%

%
%%%%%%%%%%%%%%%%%%%%%%%%%%%%%%%%%%%%%%%%%%%%%%%%%%%%%%%%%%%%%%%%%%%%%%%%%%%%%%%%%%%%%%%%
\subsection{When Does a Solution Exist?}\applab{when a solution exists}
%%%%%%%%%%%%%%%%%%%%%%%%%%%%%%%%%%%%%%%%%%%%%%%%%%%%%%%%%%%%%%%%%%%%%%%%%%%%%%%%%%%%%%%%
From~\eqsref{fi definition}{p sol} it is seen that $|\Lambda(t)|$ is only a real finite function if (assuming $\xi$ is real and there is no loss, $\gamsub{L}\equal 0$)
\begin{align}\eqlab{p sol condition}
2 \int_0^t \Big( \frac{\gamma}{2} \xi^2(s)ds - \!\xi(s)\dot{\xi}(s)\Big) ds  - 4\xi^2(t) > 0   \quad \Rightarrow \nn \\
\frac{\gamma}{2} \int_0^t\!\xi^2(s)ds - \int_0^t\! \frac12 \frac{d}{ds} \Big(\xi^2(s) \Big)ds  - 2\xi^2(t) > 0 \quad \Rightarrow \quad \xi^2(t) < \frac{\gamma}{5} \int_0^t\!\xi^2(s)ds . 
\end{align} 
A general identity holds for inequalities of the type in~\eqref{p sol condition}~\cite{Gronwall1919}  
\begin{align}\eqlab{Gronwall ineq}
\dot{u}(t) \leq \beta(t) u(t) \quad \Rightarrow \quad u(t) \leq u(a) \exp\Big(\int_a^t \!\!\beta(s)ds \Big) .
\end{align} 
Comparing~\eqref{Gronwall ineq} to~\eqref{p sol condition} shows that
\begin{align}\eqlab{Gronwall ineq example}
u(t) \leq u(0) \exp\Big(\frac{\gamma}{5} t \Big), \qquad u(t) \equiv \int_0^t \!\xi^2(s)ds . 
\end{align} 
Since $u(0)$ should equal zero, we see that this cannot be fulfilled. If $t\equal 0$ is excluded from the interval over which the solution must be valid, then $u(0^+)$ can be made arbitrarily small and~\eqref{Gronwall ineq example} provides a bound on what the rising edge of the wave packet can look like. However, since $u(T)\equal 1$ in order for the input quantum state to be normalized, we see that the wave packet length increases as $u(0^+)$ decreases. In physical terms, a finite length wave packet cannot be fully absorbed into a resonator without letting the coupling rate, $\gamma$, tend to infinity, if only for an infinitely short time. This is because the exponential decay out of the resonator only asymptotically approaches a state where the entire cavity population has coupled into the waveguide. 

\section{Emission of Photon Wavepacket}\applab{app emission}
%%%%%%%%%%%%%%%%%%%%%%%%%%%%%%%%%%%%%%%%%%%%%%%%%%%%%%%%%%%%%%%%%%%%%%%%%%%%%%%%%%%%%%%%
Without any driving field, the equations of motion are
\begin{subequations}\eqlab{eoms two modes one photon emission}
\begin{align}
    \dot{\psi}_{10}(t)  &=  -\Big(i\delta_a+\frac{\Gamma}{2}+i2|\Lambda(t)| \Big)\psi_{10}(t) -i|\Lambda(t)|e^{-i\phi(t)}\psi_{01}(t)  \eqlab{EOM cp as UL}\\
    \dot{\psi}_{01}(t)  &= -\Big(i\delta_b+\frac{\gamma_L}{2}+i2|\Lambda(t)|\Big)\psi_{01}(t)  -i|\Lambda(t)|e^{i\phi(t)}\psi_{10}(t) \eqlab{EOM cp ai+ UL}\\
     \xi(t)  &= - \sqrt{\gamma}\psi_{10}(t) ,
\end{align} 
\end{subequations}
where we dropped the subscript on $\xi_{\rm{out}}$ for notational convenience.
% % 
% \begin{subequations}\eqlab{eom cross phase UL}
% \begin{align}
% 	\dot{\psi}_X(t)	    &= \bigg(- \frac{\Gamma}{2} -i2 |\Lambda|\bigg)\psi_{10} - i |\Lambda|e^{-i\phi}\psi_{01}  \eqlab{EOM cp as UL}\\
% 	\dot{\psi}_A(t) 	&= \bigg(-\frac{\gamma_L}{2} - i2 |\Lambda|\bigg)\psi_{01} - i |\Lambda|e^{i\phi} \psi_{10} \eqlab{EOM cp ai+ UL}\\	
%     \xi                 &= \sqrt{\gamma_{X}}\psi_{10},
% \end{align} 
% \end{subequations}
% % 
% where we again dropped sub- and superscripts on $\xi_A^{\rm{out}}$ for notational convenience.
Substituting in $\psi_{10}\equal -\xi/\sqrt{\gamma}$, we have
\begin{align}
	\dot{\xi}    &= -\bigg(\frac{\Gamma}{2} +i\big(\delta_a + 2 |\Lambda|\big)\bigg)\xi + i |\Lambda|e^{-i\phi}\sqrt{\gamma}\psi_{01}  \eqlab{EOM cp as UL 2}\\
	\dot{\psi}_{01}(t) 	&= -\bigg(i\delta_b+\frac{\gamma_L}{2} + i2 |\Lambda|\bigg)\psi_{01} + i\frac{ |\Lambda|e^{i\phi}}{\sqrt{\gamma}} \xi . \eqlab{EOM cp ai+ UL 2}
\end{align} 
Using the same functions $P(t)$ and $Q(t)$ as in~\appref{app absorption}, \eqref{EOM cp ai+ UL 2} can be solved 
\begin{align} \eqlab{sol Ai+ 2 UL}
	\frac{d}{dt}\Big( \psi_{01}(t) e^{-Q(t)} \Big)e^{Q(t)} 	&=  i\frac{ |\Lambda(t)| e^{i\phi(t)}}{\sqrt{\gamma}} \xi(t)   \; \Rightarrow \; \psi_{01}(t)e^{-Q(t)} - \psi_{01}(0) = \frac{i}{\sqrt{\gamma}}\int_0^t  |\Lambda(s)| e^{i\phi(s)} \xi(s)e^{-Q(s)} ds \; \Rightarrow \nn\\
	\psi_{01}(t) &= e^{Q(t)} \Big[\psi_{01}(0) + \frac{i}{\sqrt{\gamma}}\int_0^t  |\Lambda(s)| e^{i\phi(s)} \xi(s)e^{-Q(s)} ds   \Big] .
\end{align} 
Comparing~\eqsref{sol Ai+ 2 UL}{EOM cp as UL 2} we see that
\begin{align} 
	\dot{\xi}(t) + \frac{\Gamma}{2}\xi(t) + i\big(\delta_a+2 |\Lambda(t)|\big)\xi(t)	&= i |\Lambda(t)| e^{-i\phi(t)}\sqrt{\gamma}e^{Q(t)}\Big[\psi_{01}(0)   + \frac{i}{\sqrt{\gamma}}\int_0^t  |\Lambda(s)| e^{i\phi(s)} \xi(s)e^{-Q(s)} ds   \Big].
\end{align}
Multiplying on both sides by $-\xi^* \exp[\gamma_Lt]$ yields
\begin{multline}\eqlab{eq UL re im}
-\Big(\dot{\xi}(t) + \frac{\Gamma}{2}\xi(t)\Big)\xi(t)^{\!*}e^{\gamma_Lt}  
- i\big(\delta_a+2 |\Lambda(t)|\big)|\xi(t)|^2e^{\gamma_Lt}	= \\
-i |\Lambda(t)| e^{-i\phi(t)}\xi(t)^{\!*} e^{(-i\delta_b+\frac{\gamma_L}{2})t}e^{-iP(t)}\Big[\psi_{01}(0)\sqrt{\gamma}  +\int_0^t i  |\Lambda(s)| e^{i\phi(s)} \xi(s)e^{(i\delta_b+\frac{\gamma_L}{2})s}e^{iP(s)} ds   \Big]
\end{multline} 
Let us assume that $\psi_{01}(0)$ is complex-valued with a phase $\theta_0$. Then,~\eqref{eq UL re im} can be rewritten as
\begin{align}
	 {\rm{LHS}} &= -i |\Lambda(t)| e^{-i\phi(t)}\xi(t)^{\!*} e^{(-i\delta_b+\frac{\gamma_L}{2})t}e^{-iP(t)}\Big[|\psi_{01}(0)|e^{i\theta_0}\sqrt{\gamma}  +\int_0^t i  |\Lambda(s)| e^{i\phi(s)} \xi(s)e^{(i\delta_b+\frac{\gamma_L}{2})s}e^{iP(s)} ds   \Big] \nn\\
	 {\rm{LHS}} \!\times\! e^{-i\theta_0} &= -i |\Lambda(t)| e^{-i\phi(t)}\xi(t)^{\!*} e^{(-i\delta_b+\frac{\gamma_L}{2})t}e^{-iP(t)}\Big[|\psi_{01}(0)|\sqrt{\gamma}  +\int_0^t i  |\Lambda(s)| e^{i\phi(s)} \big(\xi(s)e^{-i\theta_0}\big)e^{(i\delta_b+\frac{\gamma_L}{2})s}e^{iP(s)} ds   \Big] \nn\\
	 {\rm{LHS}}  &= -i |\Lambda(t)| e^{-i\phi(t)}\big(\xi(t)^{\!*}e^{i\theta_0}\big) e^{(-i\delta_b+\frac{\gamma_L}{2})t}e^{-iP}\Big[|\psi_{01}(0)|\sqrt{\gamma}  +\int_0^t i  |\Lambda(s)| e^{i\phi(s)} \big(\xi(s)e^{-i\theta_0}\big)e^{(i\delta_b+\frac{\gamma_L}{2})s}e^{iP(s)} ds   \Big] . \eqlab{eq UL re im 2}
\end{align} 
\eqref{eq UL re im 2} may be written as
\begin{multline}\eqlab{complex product UL}
	 -f_o + ig_o = (x-iy) \Big( C + \int_0^t\! \big[ x(s)+iy(s)\big]ds \Big) ~=\\
	 x\Big[C+\int_0^t\!x(s)ds\Big] + y\int_0^t\! y(s)ds + i\Bigg(x\int_0^t\! y(s)ds - y\Big[C+\int_0^t\! x(s)ds\Big]\Bigg), 
\end{multline} 
where 
\begin{subequations}\eqlab{C, x, y defs}
\begin{align}
C       &= |\psi_{01}(0)|\sqrt{\gamma}\\
x       &=  -|\Lambda(t)| |\xi(t)| \exp(\gamma_Lt/2) \sin\!\big[\phi(t)+\delta_b t + P(t)+\arg(\xi) - \theta_0\big] \\
y       &=  |\Lambda(t)| |\xi(t)| \exp(\gamma_Lt/2)\cos\!\big[\phi(t)+\delta_b t + P(t)+\arg(\xi) - \theta_0\big] \\
f_o     &= \Big(\dot{\xi}(t) + \frac{\Gamma}{2}\xi(t)\Big)\xi(t)^{\!*}e^{\gamma_Lt} \eqlab{fo definition}\\
g_o     &= -\big(\delta_a + 2 |\Lambda(t)|\big) |\xi(t)|^2 e^{\gamma_Lt}.  \eqlab{go definition}
\end{align}
\end{subequations}
%
% To identify the real and imaginary parts of the LHS of~\eqref{eq UL re im} we write the input field in polar form, $\xi(t)\equiv r_{\rm{out}}(t)\exp(i\phi_{\rm{out}}(t))$. The derivative is
% % 
% \begin{align}\eqlab{dot S out}
% 	 \dot{\xi} = \big(\dot{r}_{\rm{out}} + i\dot{\phi}_{\rm{out}} r_{\rm{out}}\big) e^{i\phi_{\rm{out}}}
% \end{align} 
% % 
% Then the functions $f_o$ and $g_o$ are  
% % 
% \begin{align}\eqlab{fo go definition}
% % f_o     &= -\big(\dot{r}_{\rm{out}}r_{\rm{out}}  +  \frac{\gamma}{2}r_{\rm{out}}^2\big)e^{\gamma_Lt}  	\\
% f_o     &= \Big(\dot{\xi}(t) + \frac{\gamma}{2}\xi(t)\Big)\xi(t)^{\!*}e^{\gamma_Lt} \\
% g_o     &= -2 |\Lambda(t)| |\xi(t)|^2 e^{\gamma_Lt}.
% \end{align} 
% % 
Let us define the functions
\begin{align}\eqlab{aux functions UL}
	 X(t)=C+\int_0^t\!x(s)ds = R(t)\cos\!\big[\theta(t)\big], ~~ Y(t)=\int_0^t\!y(s)ds = R(t)\sin\!\big[\theta(t)\big].
\end{align} 
Equating real and imaginary parts of~\eqref{complex product UL} yields
\begin{align}\eqlab{EOM cp as 4}
	 -f_o(t) = \dot{X}(t)X(t) + \dot{Y}(t)Y(t), \qquad 	 g_o(t) = \dot{X}(t)Y(t) - \dot{Y}(t)X(t),
\end{align} 
where $x(t)\equal\dot{X}(t)$ and $y(t)\equal\dot{Y}(t)$. Using the definition in~\eqref{aux functions UL}, we have
\begin{align}\eqlab{RHS real UL}
	 -f_o = \dot{X}X + \dot{Y}Y &= \big[\dot{R}\cos(\theta) - R\sin(\theta)\dot{\theta}\big]R\cos(\theta) + \big[\dot{R}\sin(\theta) + R\cos(\theta)\dot{\theta} \big]R\sin(\theta) = \dot{R}R = \frac12 \frac{d}{dt}\Big( R^2\Big) \; \Rightarrow \nn\\
	 R(t)^2 - R(0)^2 &= -\int_0^t 2 f_o(s)ds .
\end{align} 
Since $R^2\equal X^2+Y^2$, we have $R(0)^2 \equal C^2$ and therefore
\begin{align}\eqlab{R sol UL}
	 R(t) = \sqrt{ C^2-2\!\int_0^t\! f_o(s)ds \Big)}.
\end{align} 
Similarly, 
\begin{align}\eqlab{RHS imag UL}
	 g_o = \dot{X}Y - \dot{Y}X = \big[\dot{R}\cos(\theta) - R\sin(\theta)\dot{\theta}\big]R\sin(\theta) - \big[\dot{R}\sin(\theta) + R\cos(\theta)\dot{\theta} \big]R\cos(\theta) = -R^2\dot{\theta}.
\end{align} 
Using the result in~\eqref{R sol UL} and the initial condition $\theta(0)\equal 0$, the solution for $\theta$ is 
\begin{align}\eqlab{theta sol UL}
	 \theta(t) = -\int_0^t\!\! \frac{g_o(s)}{ C^2 - 2\!\int_0^s\! f_o(z)dz} ds.
\end{align} 
To find the solution for $|\Lambda(t)|$ we evaluate $x^2+y^2\equal  |\Lambda|^2 |\xi|^2 \exp(\gamma_L t)$ using the results above
\begin{multline}\eqlab{absolute value}
 |\Lambda|^2 |\xi|^2 e^{\gamma_L t} =  \dot{X}^2 \plus \dot{Y}^2 = \big[\dot{R}\cos(\theta) \minus R\sin(\theta)\dot{\theta}\big]^2 \plus \big[\dot{R}\sin(\theta) \plus R\cos(\theta)\dot{\theta} \big]^2 = \dot{R}^2 \plus R^2\dot{\theta}^2 =  \frac{g_o^2 + f_o^2}{C^2-2\!\int\!f_o}  .    
\end{multline} 
Inserting the definition of $g_o$ from~\eqref{go definition} yields
\begin{multline}\eqlab{Lambda sol UL} 
	  |\Lambda|^2 |\xi|^2 \exp(\gamma_L t) = \frac{1}{C^2-2\mathcal{F}_o} \Big[ \big(\delta_a+2 |\Lambda|\big)^2 \exp(2\gamma_L t)|\xi|^4 + f_o^2\Big] ~\Rightarrow \\
	  |\Lambda(t)| = e^{-\gamma_L t} \frac{2\delta_a|\xi|^3 e^{2\gamma_L t} \pm \sqrt{e^{\gamma_L t} f_o^2 \big( C^2 - 2\mathcal{F}_o - 4 e^{2\gamma_L t} \xi^2\big) + \delta_a^2\xi^4 \big(C^2- 2\mathcal{F}_o\big) e^{3\gamma_L t} }  }{ |\xi| \big[ C^2 -  2\mathcal{F}_o - 4\xi^2 e^{\gamma_L t}   \big]},
    %   |\Lambda(t)|  &= \frac{|f_o| \exp(-\gamma_Lt/2)}{ |\xi|}\frac{1}{\sqrt{ C^2 - 2\int_0^t\!f_o(s)ds - 4|\xi|^2e^{\gamma_Lt} } } .
\end{multline} 
where $\mathcal{F}_o(t)$ is the anti-derivative of $f_o(t)$. If $\delta_a\equal 0$, the solution is
\begin{align}\eqlab{LAMo sol delta_a=0}
	 |\Lambda(t)|  &= \frac{|f_o| \exp(-\gamma_Lt/2)}{ |\xi|}\frac{1}{\sqrt{ C^2 - 2\mathcal{F}_o - 4|\xi|^2e^{\gamma_Lt} } } .
\end{align} 
Knowing $|\Lambda(t)|$ means $g_o$ is a known function and $x$ and $y$ may be evaluated using $\theta$ from~\eqref{theta sol UL}. Then, the phase $\phi$ is
\begin{align}\eqlab{phi sol UL}
	 \phi(t) &= -\delta_b t - 2\int_0^t |\Lambda(s)|ds  - \arg(\xi) +\theta_0 + \tan^{-1}\!\bigg( \frac{-x(t)}{y(t)}\bigg) .
\end{align} 
To obtain $x$ and $y$, note that 
\begin{align} 
	x & = \dot{X} = \dot{R}\cos(\theta) - R\sin(\theta)\dot\theta = \frac{-f_o \cos(\theta) + g_o \sin(\theta)}{\sqrt{ C^2 -  2\int\!f_o}} \\ 
	y & = \dot{Y} = \dot{R}\sin(\theta) + R\cos(\theta)\dot\theta = \frac{-f_o \sin(\theta) - g_o \cos(\theta)}{\sqrt{ C^2 -  2\int\!f_o}}.
\end{align} 
%

%%%%%%%%%%%%%%%%%%%%%%%%%%%%%%%%%%%%%%%%%%%%%%%%%%%%%%%%%%%%%%%%%%%%%%%%%%%%%%%%%%%%%%%%
\subsection{Gaussian Wave Packet}\applab{Gaussian emission Chi3 app}
%%%%%%%%%%%%%%%%%%%%%%%%%%%%%%%%%%%%%%%%%%%%%%%%%%%%%%%%%%%%%%%%%%%%%%%%%%%%%%%%%%%%%%%%
The Gaussian wave packet is
\begin{align}\eqlab{Gaussian app}
\xi_{\rm{in}}(t) = \mathcal{G}(t-\Tin)  =  \sqrt{\frac{2}{\tauin}} \left(\frac{\text{ln}(2)}{\pi}\right)^{\!\frac{1}{4}} \exp\!\left(\!-2\text{ln}(2)\frac{(t-\Tin)^2}{\tauin^2} \right) , 
\end{align} 
where $|\mathcal{G}(t)|^2$ has a full width at half maximum (FWHM) temporal width $\tauin$, spectral width $\OmG\equal 4\text{ln}(2)/\tauG$, and integrates to 1 (over the infinite interval from $-\infty$ to $\infty$). As discussed in~\appref{when a solution exists}, it is not possible to fully absorb this wave packet and this issue manifests in the denominator of~\eqref{p sol} being imaginary during the rising edge of the Gaussian where 
\begin{align}\eqlab{LAMi real} 
	  2\int_0^t\!f_i(s)ds - 4|\xi_{\rm{in}}(t)|^2e^{\gamsub{L}t} \leq 0.
\end{align} 
$\Lambda_i$ diverges at the cross-point determined by an equality in~\eqref{LAMi real}. This is illustrated in~\figref{smoothing functions app}.
% As in~\appref{Gaussian absorption Chi3 app}, we consider a Gaussian wave packet given by~\eqref{Gaussian app}.
% 
\begin{figure}[!h] 
\centering
   \includegraphics[angle=0,origin=c,width=7cm] {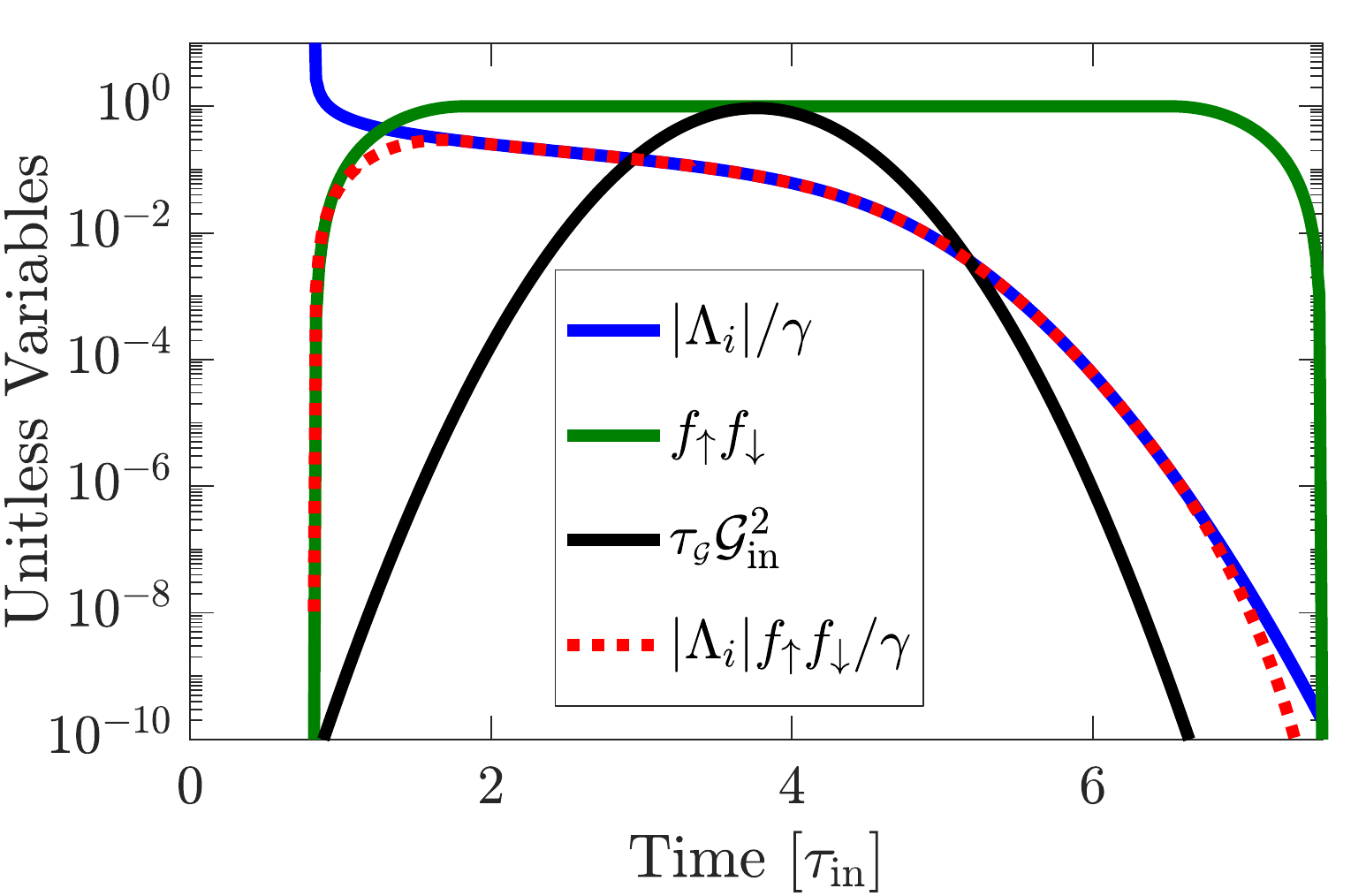}
   \hspace{1cm}
   \includegraphics[angle=0,origin=c,width=7cm] {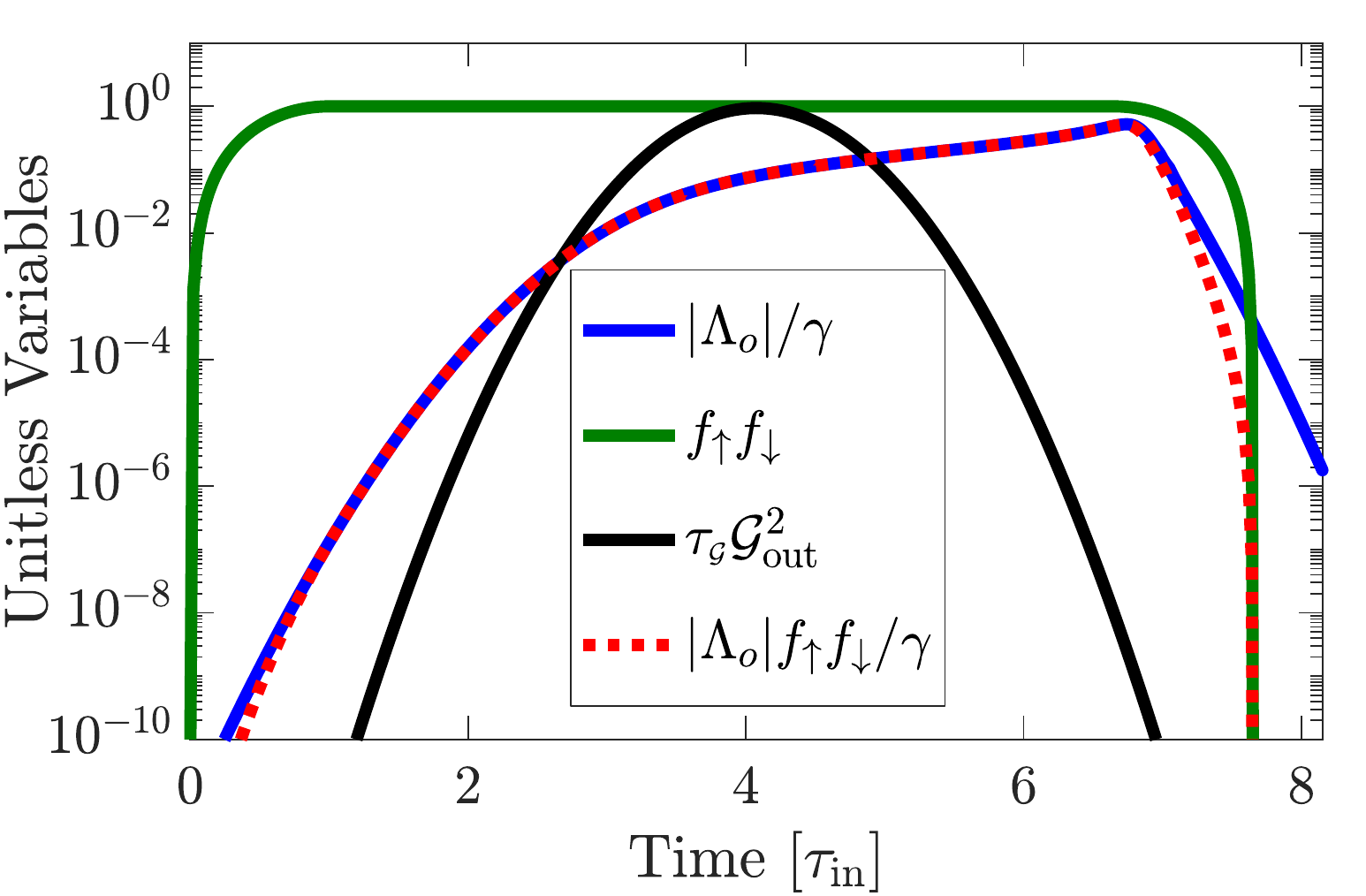}
\caption{Illustration of the solutions for $|\Lambda_{i/o}|$ along with the smoothing functions in~\eqref{f up down i app} that ensures well-behaved control fields. Parameters: (a) $\gamma\equal 30\OmG$, $\gamma_L\equal 0$, $\tauenv\equal \tauin$. (b) $\gamma_L\equal 10^{-5}\OmG$.  }
\figlab{smoothing functions app}
\end{figure} 
To avoid divergences and keep $|\Lambda_i|$ real, we multiply the solution in~\eqref{p sol} by smoothing functions
\begin{align}\eqlab{f up down i app}
	f_{\uparrow}(t) &= \frac{1 \plus \sin\!\big( \frac{\pi t}{\tauenv}\big) }{2}  \Theta\Big(t \plus \frac{\tauenv}{2}    \Big)  \Theta\Big(\frac{\tauenv}{2} \minus t \Big) + \Theta\Big( t\minus\frac{\tau_{\rm{e}}}{2} \Big) \\
	f_{\downarrow}(t) &= \frac{1 \minus \sin\!\big( \frac{\pi t}{\tauenv}\big) }{2}  \Theta\Big(t \plus \frac{\tauenv}{2}    \Big)  \Theta\Big(\frac{\tauenv}{2} \minus t \Big) + \Theta\Big( \minus \frac{\tau_{\rm{e}}}{2} \minus t\Big)
\end{align} 
where $\Theta$ is a step function that equals 1 for positive arguments and 0 for negative arguments. The smoothing functions rise from 0 to 1 ($\uparrow$) or fall from 1 to 0 ($\downarrow$) in the interval $t\!\in\![-\tauenv/2,~ \tauenv/2]$ as half a period of the sine function.

%%%%%%%%%%%%%%%%%%%%%%%%%%%%%%%%%%%%%%%%%%%%%%%%%%%%%%%%%%%%%%%%%%%%%%%%%%%%%%%%%%%%%%%%
\section{Input Pump Fields for Absorption and Emission}\applab{input pumps supp}
%%%%%%%%%%%%%%%%%%%%%%%%%%%%%%%%%%%%%%%%%%%%%%%%%%%%%%%%%%%%%%%%%%%%%%%%%%%%%%%%%%%%%%%%

The resonator modes that couple to the pump fields are identical and the Hamiltonian associated with those modes is 
\begin{align}\eqlab{H pump}
   \hat{H}_n^{\rm{pump}} = i\hbar \sqrt{\frac{\gamma_p}{\Delta t}} \sum_{m=1}^2\! \Big( \hat{p}_m^\dagger \hat{W}_n - \hat{p}_m\hat{W}_n^\dagger \Big) + \hbar \hat{p}_1^\dagger\hat{p}_1\hat{p}_2^\dagger\hat{p}_2 +
   \frac12 \hbar \!\sum_{m=1}^2 \!\Big( \hat{p}_m^\dagger\hat{p}_m-1\Big)\hat{p}_m^\dagger\hat{p}_m.
\end{align}

The temporal shape of the input pump functions can be found by considering their equations of motion
\begin{subequations}\eqlab{eom pumps xpm}
\begin{align}
	\dot{\alpha}_1	&= \Big(- \frac{\Gamma_p}{2} -i\chi_3\big(|\alpha_1|^2+|\alpha_2|^2\big) \Big)\alpha_1    + \sqrt{\gamma_p}\xi_1 \eqlab{eom pumps xpm p1}\\
	\dot{\alpha}_2 	&= \Big(- \frac{\Gamma_p}{2} -i\chi_3\big(|\alpha_1|^2+|\alpha_2|^2 \big)\Big)\alpha_2   + \sqrt{\gamma_p}\xi_2 \eqlab{eom pumps xpm p2}.
\end{align} 
\end{subequations}
From~\secref{main model} we have $\Lambda = \chi_3 \alpha_2^* \alpha_1=|\Lambda|\exp(i\phi)\equal \chi_3 r_{\!\alpha}^2\exp[i(\phi_1-\phi_2)]$ and since we assumed $|\alpha_1|\equal |\alpha_2|$, we can express the complex amplitudes in polar form: $\alpha_1\equal r_{\!\alpha}\exp(i\phi_1)$ and $\alpha_2\equal r_{\!\alpha}\exp(i\phi_2)$, with $\phi\equal \phi_1 - \phi_2$. The goal is to determine the complex-valued input fields, $\xi_1$ and $\xi_2$, such that~\eqref{eom pumps xpm} yields the correct intra-cavity control fields $\alpha_1$ and $\alpha_2$.
% $ \chi_3\alpha_2^*\alpha_1 \equal \Lambda$ and 
% $\chi_3\langle \hat{p}_2^\dagger \hat{p}_2\rangle \equal \chi_3\langle \hat{p}_1^\dagger \hat{p}_1 \rangle \equal |\Lambda_n|$
% where we have assumed their cavity modes are identical and the $3 |\Lambda(t)|$ comes from cross-phase modulation $(2 |\Lambda(t)|)$ and self-phase modulation $( |\Lambda(t)|)$. The definition of $\Lambda(t)$ is 
% 
% \begin{align} \eqlab{kappa def}
% 	\Lambda \equiv \alpha_2^* \alpha_1 ~~\Rightarrow ~~ |\Lambda| = |\alpha_1||\alpha_2|, ~\phi(t) = \arg[\alpha_1] - \arg[\alpha_2].
% \end{align} 
% %
% Note that we have assumed $|A_{P_1}(t)|\equal|A_{P_2}(t)|\equal \sqrt{|\Lambda(t)|}$ in previous sections. 
Let us write the pump fields in polar form: $\xi_n\equal q_n\exp(i\psi_n)$, and substitute into~\eqref{eom pumps xpm}
\begin{align} \eqlab{eom pumps xpm polar}
	\dot{\alpha}_n &= \big( \dot{r}_{\!\alpha} + i\dot{\phi}_{n} r_{\!\alpha} \big) e^{i\phi_n} = \big(- \frac{\Gamma_p}{2} - i2\chi_3 r_{\!\alpha}^2 \big)r_{\!\alpha} e^{i\phi_n}   + \sqrt{\gamma_p} q_n e^{i\psi_n}.
\end{align} 
Separating equations for the real and imaginary parts yields
\begin{subequations}
\begin{align} 
	\dot{r}_{\!\alpha}         &= -\frac{\Gamma_p}{2}r_{\!\alpha} + \sqrt{\gamma_p}q_n\cos(\psi_n - \phi_n) \eqlab{eom pumps re}\\
    \dot{\phi}_{n}  &= -2\chi_3 r_{\!\alpha}^2 + \sqrt{\gamma_p}\frac{q_n}{r_{\!\alpha}}\sin(\psi_n - \phi_n). \eqlab{eom pumps im}
\end{align} 
\end{subequations}
Let us guess that $q_1\equal q_2\equal q$ and $\psi_1-\phi_{1} \equal -(\psi_2-\phi_{2})$. Since $\phi\equal \phi_1 - \phi_2$, we have  
\begin{align} \eqlab{dot phi}
	\dot{\phi}  = \dot{\phi}_1 - \dot{\phi}_2 = \sqrt{\gamma_p}\frac{q}{r_{\!\alpha}} \Big[\sin(\psi_1-\phi_1) -  \sin(\psi_2-\phi_2) \Big] = 2 \sqrt{\gamma_p}\frac{q}{r_{\!\alpha}} \sin(\psi_1-\phi_1)
\end{align} 
Re-arranging~\eqsref{eom pumps re}{dot phi}, we have
\begin{align} \eqlab{psi_1 - phi_1 sol}
	\frac12 \frac{r_{\!\alpha} \dot{\phi}}{\big(\dot{r}_{\!\alpha} + \frac{\Gamma_p}{2} r_{\!\alpha}\big)} =  \tan(\psi_1-\phi_1) , ~~\Rightarrow ~~ \psi_1-\phi_1 = \arctan\!\Bigg[  \frac12 \frac{r_{\!\alpha} \dot{\phi}}{\big(\dot{r}_{\!\alpha} + \frac{\Gamma_p}{2} r_{\!\alpha}\big)}  \Bigg].
\end{align} 
Using the identity $\cos[\arctan(x)]\equal 1/\sqrt{1+x^2}$, we find $q$ from~\eqref{eom pumps re}
\begin{align} \eqlab{solution q}
	\dot{r}_{\!\alpha} + \frac{\Gamma_p}{2} r_{\!\alpha}  = \sqrt{\gamma_p}q \frac{1}{\sqrt{ 1 + \frac14 \Big( \frac{r_{\!\alpha} \dot{\phi}}{\dot{r}_{\!\alpha} + \frac{\Gamma_p}{2} r_{\!\alpha}} \Big)^2}}, ~~\Rightarrow~~ q = \frac{1}{\sqrt{\gamma_p}} \sqrt{ \Big(\dot{r}_{\!\alpha} + \frac{\Gamma_p}{2}r\Big)^2 + \frac{\dot{\phi}^2 r_{\!\alpha}^2}{4} }.
\end{align} 
Using the identity $\sin[\arctan(x)]\equal x/\sqrt{1+x^2}$, we may insert~\eqref{psi_1 - phi_1 sol} into~\eqref{eom pumps im} to obtain
\begin{subequations}\eqlab{dot{phi_n} sol}
\begin{align} 
    \dot{\phi}_{1}  &= -2\chi_3 r_{\!\alpha}^2 + \sqrt{\gamma_p}\frac{q}{r_{\!\alpha}} \Big[\frac12 \frac{r_{\!\alpha} \dot{\phi}}{\big(\dot{r}_{\!\alpha} + \frac{\Gamma_p}{2} r_{\!\alpha}\big)} \Big]\frac{1}{\sqrt{ 1 + \frac14 \Big( \frac{r_{\!\alpha} \dot{\phi}}{\dot{r}_{\!\alpha} + \frac{\Gamma_p}{2} r_{\!\alpha}} \Big)^2}} = -2\chi_3 r_{\!\alpha}^2 + \frac{\dot{\phi}}{2}\\
    \dot{\phi}_{2}  &= -2\chi_3 r_{\!\alpha}^2 - \frac{\dot{\phi}}{2}.
\end{align} 
\end{subequations}
Integrating~\eqref{dot{phi_n} sol}, and inserting into~\eqref{psi_1 - phi_1 sol}, we find
% If we guess that $q_1\equal q_2\equal q$ and $\psi_1-\phi_{1} \equal -(\psi_2-\phi_{2})$, then
% % 
% \begin{align} \eqlab{eom pumps xpm phase}
%   \dot{\phi} &= \dot{\phi}_{1} - \dot{\phi}_{2} =  -2\sqrt{\gamma_p}\frac{q}{r_{\!\alpha}}\sin(\psi_1 - \phi_{1}) ~\Rightarrow \psi_1 - \phi_{1} = -\sin^{-1}\!\Big(  \frac{\dot{\phi} r_{\!\alpha}}{2\sqrt{\gamma_p} q } \Big).
% \end{align} 
% 
% Inserting this into~\eqref{eom pumps xpm re} yields
% % 
% \begin{align} \eqlab{eom pumps xpm re 2}
% 	\dot{r}_{\!\alpha} + \frac{\Gamma_p}{2}r_{\!\alpha} &= \sqrt{\gamma_p}q \sqrt{ 1 - \frac{\dot{\phi}^2 r_{\!\alpha}^2}{4\gamma_p q^2 } } ~\Rightarrow~ q = \frac{1}{\sqrt{\Gamma_p}} \sqrt{ \Big(\dot{r}_{\!\alpha} + \frac{\Gamma_p}{2}r\Big)^2 + \frac{\dot{\phi}^2 r_{\!\alpha}^2}{4} }.
% \end{align} 
% % 
% Finding $\phi_n$ is done by substituting~\eqref{eom pumps xpm phase} into~\eqref{eom pumps xpm im}
% 
\begin{subequations} \eqlab{eom pumps xpm im 2}
\begin{align} 
    \psi_{1}(t) &= -2\int_0^t\! |\Lambda(s)|ds + \frac{\phi(t)}{2} + \arctan\!\Bigg[  \frac12 \frac{r_{\!\alpha} \dot{\phi}}{\big(\dot{r}_{\!\alpha} + \frac{\Gamma_p}{2} r_{\!\alpha}\big)}  \Bigg] \\ 
     \psi_{2}(t) &= -2\int_0^t\! |\Lambda(s)|ds - \frac{\phi(t)}{2} - \arctan\!\Bigg[  \frac12 \frac{r_{\!\alpha} \dot{\phi}}{\big(\dot{r}_{\!\alpha} + \frac{\Gamma_p}{2} r_{\!\alpha}\big)}  \Bigg].
\end{align} 
\end{subequations}
% 
% The solution for $\psi_n$ is given by substituting the results in~\eqref{eom pumps xpm im 2} into~\eqref{eom pumps xpm phase}.

% \input{./app_nonlinear_dynamics}

% \include{./kurt_derivation}
% \include{./absorbing_photons}

% To start the appendixes, use the \verb+\appendix+ command.

% \section{A little more on appendixes}
% \subsection{\label{app:subsec}A subsection in an appendix}

% \nocite{*}

%\bibliography{Mendeley}% Produces the bibliography via BibTeX.
%

\end{document}